\documentclass[orivec,envcountsect,envcountsame]{llncs}
 
\usepackage{xargs}
\usepackage{xstring}

\usepackage{mathtools}
\usepackage{amsmath}
\usepackage{rotating}

\usepackage{amssymb}
\usepackage{enumerate}              \usepackage{etex,etoolbox}          \usepackage{graphicx}               

\usepackage{hyperref}
\usepackage[capitalise]{cleveref}\crefname{fact}{Fact}{fact}\crefname{lstlisting}{Algorithm}{Algorithm}

\usepackage{xifthen}                \usepackage{appendix}               

\usepackage{stmaryrd}

\usepackage[usenames,dvipsnames,svgnames,table]{xcolor} \usepackage{tikz}                   \usetikzlibrary{shadows,arrows,shapes,automata,positioning,decorations.markings}

\newif\ifhideproofs
\hideproofsfalse

\usepackage{mathtools} \usepackage{amsbsy}

\DeclareMathSizes{10}{10}{6}{6}

\tikzset{
  cnode/.style={
    shape=circle,
    minimum size = 0mm,
    inner sep = 1pt,
    font=\tiny,
    draw
  },
  carrow/.style={
    ->,
    shorten >=1pt,
    >=stealth',
    auto,
    draw,
    sloped
  }
}

\newif\ifemi

\DeclareGraphicsExtensions{.png,.PNG,.pdf,.PDF,.jpg,.mps,.jpeg,.jbig2,.jb2,.JPG,.JPEG,.JBIG2,.JB2}

\usepackage{bm}
\usepackage{xspace}

\usepackage{xifthen}        \newcommand{\ifempty}[3]{\ifthenelse{\isempty{#1}}{#2}{#3}}

\newcommand{\mkfun}[4][\colorFun]{
  \newcommand{#2}[1][#4]{
    {#1\textsf{#3}}
    \ifempty{##1}{}{
      ({##1})}
  }
}

\newcommand{\mkuop}[4][\colorFun]{
  \newcommand{#2}[1][#4]{
    {#1\textsf{#3}}
    \ifempty{##1}{}{
      \, {##1}}
  }
}

\newcommand{\hidden}[1]{}

\newcommand{\cf}[2]{
  \fontsize{#1}{#1}{\selectfont{#2}}
}
\ifemi
\usepackage{showlabels}

\newcommand{\emi}[2]{
  \marginpar{\fcolorbox{red}{shadecolor}{\cf{#1}{{#2}}}}
}
\newcommand{\emic}[2]{\par
  \fcolorbox{red}{shadecolor}{\parbox{\linewidth}{ 
      \color{gray}
      \begin{description}
      \item[{\color{blue} #2}]{\sf #1}
      \end{description}}}
}
\else
\newcommand{\emi}[2]{}
\newcommand{\emic}[2]{}{}
\fi

\newcommand{\sst}{\;\big|\;}
\newcommand{\qst}{\;\colon\;}

\newcommand{\conf}[1]{\ensuremath{\langle {#1} \rangle}}

\newcommand{\qqand}{\qquad\text{and}\qquad}
\newcommand{\qand}{\quad\text{and}\quad}

\newcommand{\rmkend}{\finex}

\newcommand{\quo}[1]{\lq\lq {#1}\rq\rq}
\def\finex{{\unskip\nobreak\hfil
\penalty50\hskip1em\null\nobreak\hfil{\Large $\diamond$}
\parfillskip=0pt\finalhyphendemerits=0\endgraf}}

\definecolor{shadecolor}{rgb}{1,0.99,0.9}
\definecolor{bg}{rgb}{0.95,0.95,0.95}

\newcommand{\rel}{{\mathcal R}}

\newcommand{\Dual}[1]{\overline{#1}}

\newcommand{\chora}[1][A]{\mathsf{C{#1}}}
\newcommand{\sync}[3]{#1^{\ptp[#2] \bowtie \ptp[#3]}}
\newcommand{\compose}[4]{#1\,\!_{\ptp[#3] \bowtie \ptp[#4]}\! #2}
\newcommand{\restr}[2]{{{#1}_{\mid{#2}}}}
\newcommand{\proj}[2]{#1\!\downarrow_{{\ptp[#2]}}}
\newcommand{\proji}[4]{{#1}\!\downarrow_{{\ptp[#2]}}^{{\ptp[#3]} \bowtie {\ptp[#4]}}}
\newcommand{\arro}[1]{\xrightarrow{#1}}
\newcommand{\lang}[1]{\mathcal{L}(#1)}
\newcommand{\ssem}[1]{{\llbracket #1 \rrbracket}}
\newcommand{\close}[1]{close( #1 )}
\newcommand{\comm}{\sim}
\newcommand{\pstate}[2]{\vec #1(#2)}
 \usepackage[procnames]{listings}
\usepackage{mfirstuc}

\newcommandx{\gspider}[2][1=T_1,2=T_2,usedefault=@]{\text{spider}(\ifempty{#1}{\_}{#1}, \ifempty{#2}{\_}{#2})}
\newcommandx{\glink}[3][1=\aG,2=T_1,3=T_2,usedefault=@]{\text{link}(\ifempty{#1}{\_}{#1}, \ifempty{#2}{\_}{#2},  \ifempty{#3}{\_}{#3})}
\newcommandx{\gintcomp}[2][1=I,2=J,usedefault=@]{\stackrel{\p[#1],\p[#2]}{\bowtie}}

\newcommand{\iH}{\p[h]\xspace}
\newcommand{\iK}{\p[k]\xspace}
\newcommand{\iI}{\p[i]\xspace}
\newcommand{\iJ}{\p[j]\xspace}

\newcommand{\fillcolor}{orange!5}
\def\colorPtp{\color{blue}}
\def\colorFun{\color{NavyBlue}}

\def\colorOp{\color{OliveGreen}}
\def\colorNode{\color{cyan}}
\def\colorR{\color{OliveGreen}}
\def\colorE{\color{orange}}

\def\colorMsg{\color{BrickRed}}

\newcommand{\msg}[1][m]{\mathsf{\colorMsg{#1}}}
\newcommand{\msgset}{\mathcal{\colorMsg M}}

\newcommand{\lset}{\mathcal{L}}
\newcommand{\lint}{\lset_{\text{int}}}
\newcommand{\lact}{\lset_{\text{act}}}
\newcommand{\lio}{\lset_{\text{i/o}}}

\newcommand{\sset}{\mathcal{S}}

\newcommand{\ptp}[1][A]{\ensuremath{\mathsf{\colorPtp{\capitalisewords{#1}}}}}

\newcommand{\p}{\ptp}
\newcommand{\q}{{\ptp[B]}}

\newcommandx{\ggcommon}[3][1=\ptp,2={\aH},3={\aH'},usedefault=@]{f_{#1}}
\newcommandx{\opair}[2][1={\ae},2={\ae'},usedefault=@]{\conf{{#1},{#2}}}
\newcommandx{\hopair}[2][1={\aE},2={\aE'},usedefault=@]{\llparenthesis\, {#1},{#2}\, \rrparenthesis}
\newcommandx{\wf}[2][1={\aG},2={\aG'},usedefault=@]{wf({#1}, {#2})}
\newcommandx{\wb}[2][1={\aG},2={\aG'},usedefault=@]{wb({#1}, {#2})}
\newcommandx{\ws}[2][1={\aG},2={\aG'},usedefault=@]{ws({#1}, {#2})}
\newcommandx{\widx}[2][1={\aW},2={i},usedefault=@]{{#1}[{#2}]}
\newcommandx{\outop}[2][1=\gname,2={}]{{\colorOp{!}}^{{#1}{#2}}}
\newcommandx{\inop}[2][1=\gname,2={}]{{\colorOp{?}}^{{#1}{#2}}}
\newcommandx{\aout}[5][1={\p},2={\q},3={},4=m,5={},usedefault=@]{
  \achan[#1][#2] \outop[{#3}] {\msg[#4]}{#5}
}
\newcommandx{\ain}[5][1={\p},2={\q},3={},4=m,5={},usedefault=@]{
  \achan[#1][#2] \inop[{#3}] {\msg[#4]}{#5}
}
\newcommandx{\adep}[1][1={}]{
  \conf{ \aout[@][@][@][@][{#1}], \ain[@][@][@][@][{#1}]}
}

\newcommandx{\hproj}[2][1=\aH, 2=\ptp, usedefault=@]{
  \ifempty{#1}{}{{#1}}\ifempty{#2}{}{{^{\scriptscriptstyle @{#2}}}}
}
\newcommandx{\eproj}[2][1=\aE,2=\ptp, usedefault=@]{
  {{#1}}\ifempty{#2}{}{{^{\scriptscriptstyle @{#2}}}}
}

\newcommandx{\mklooptwo}[4][1=.5,2=1.5]{ \node[ogate,above = #1 of {#3}] (entry#3) {};
  \pgfgetlastxy \xentry \yentry;
  \pgfmathtruncatemacro{\xentryrounded}{\xentry};
\path (#4);
   \pgfgetlastxy \xexit \yexit;
  \pgfmathtruncatemacro{\xexitrounded}{\xexit};
  \path[line] (entry#3) -- (#3);
  \pgfmathsetmacro\tmpdiff{abs(\xentryrounded - \xexitrounded)}
  \path[line] (#4) -|  ($(#4)+(\tmpdiff,0)+(#2,0)$) |- (entry#3);
}

\newcommand{\apom}{r}

\newcommand{\aR}[1][R]{{\colorR{#1}}}

\newcommandx{\detM}[1][1=\aCM,usedefault=@]{\Delta({#1})}

\newcommand{\gsubs}[2]{^{#1} / _{#2}}
\newcommandx{\gsubst}[3][1=\aM,2=q,3=q',usedefault=@]{
  \left \{\gsubs{#3}{#2} \right \}#1
}

\newcommand{\II}{\mathsf{\color{blue} I}}
\newcommand{\JJ}{\mathsf{\color{blue}J}}
\newcommand{\HH}{\mathsf{\color{blue}H}}
\newcommand{\KK}{\mathsf{\color{blue}K}}

\newcommand{\Set}[1]{\{\,#1\,\}}

\newcommandx{\cm}[2][1=\ptp, 2=\aM]{{#2}_{#1}}
\newcommandx{\achan}[2][1=A,2=B,usedefault=@]{{\ptp[#1]\,\ptp[#2]}}
\newcommand{\ptpset}{\mathcal{\colorPtp{P}}}

\newcommand{\oact}{\outop[]}
\newcommand{\iact}{\inop[]}
\newcommand{\tset}{\to}

\newcommandx{\cauttr}[5][1=p,2={\gint[][A][@][B]}, 3=q, 4=H, 5=K, usedefault=@]{
  {#1} \xrightarrow[{\ptp[#4]\ \ptp[#5]}]{#2} {#3}
}
\newcommand{\RS}[1][]{\mathcal{R}({#1})}

\newcommand{\trans}[2][{}]{\,\xrightarrow{#2}_{#1}\,}

\newcommandx{\acfsmout}[3][1=A,2=B,3=m,usedefault=@]{\achan[{#1}][{#2}] \oact {\msg[{#3}]}}
\newcommandx{\acfsmin}[3][1=A,2=B,3=m,usedefault=@]{\achan[{#1}][{#2}] \iact {\msg[{#3}]}}
\newcommandx{\fsaout}[2][1={\p},2={},usedefault=@]{
  \ptp[#1] \ \outop[]\ \msg[{#2}]
}
\newcommandx{\fsain}[2][1={\p},2={},usedefault=@]{
  \ptp[#1] \ \inop[]\ \msg[{#2}]
}

\makeatletter
\newcommand{\linenumfontsize}{\@setfontsize{\linenumfontsize}{3pt}{3pt}}
\makeatother

\newcommand{\aG}{\mathsf{G}}

\newcommand{\gseqop}{{\colorOp ;}\,}
\newcommand{\gparop}{{\colorOp \ |\ }}
\newcommand{\gchoop}{{\colorOp \ +\ }}
\newcommand{\grecop}{{\colorOp *}}
\newcommand{\grecopp}{{\colorOp{@}}}
\newcommand{\gname}[1][i]{{\colorNode{\scriptstyle\textsf{#1}}}}
\newcommandx{\nmerge}[2][1={i},2={},usedefault=@]{
  \ifempty{#2}{
    \ifempty{#1}{\mu}{\gname[-{#1}]}
  }{-{#2}}
}

\mkfun{\esbj}{sbj}{\ae}

\makeatletter \@ifclassloaded{exam-paper}{}{\makeatletter \@ifclassloaded{test}{}\makeatother }
\makeatother

\newcommandx{\gnode}[2][1=i,2=\gint,usedefault=@]{
  {\ifempty{#1}{}{\colorNode{\gname[{#1}].}}} {#2}
}

\newcommandx{\gint}[4][1={i},2=A,3=\msg,4=B,usedefault=@]{
  \ptp[#2] {\colorOp \xrightarrow{\scriptscriptstyle\gname[#1]}} \ptp[#4] \colon {\msg[{#3}]}
}
\newcommandx{\gout}[4][1=\gname,2=\ptp,3=\msg,4={\ptp[C]},usedefault=@]{
  \achan[{#2}][{#4}] {\colorOp {\colorOp{!}}} {\msg[{#3}]}
}
\newcommandx{\gin}[4][1={},2=\ptp,3=\msg,4={\ptp[C]},usedefault=@]{
  \achan[{#2}][{#4}] {\colorOp {\colorOp{?}}} {\msg[{#3}]}
}
\newcommandx{\gseq}[3][1=i,2={\aG},3={\aG'},usedefault=@]{
  \gnode[{#1}][{#2} \gseqop {#3}]
}
\newcommandx{\gpar}[3][1={},2={\aG},3={\aG'},usedefault=@]{
  \gnode[{#1}][\ifempty{#1}{{#2} \gparop {#3}}{({#2} \gparop {#3})}]
}
\newcommandx{\gcho}[3][1={},2={\aG},3={\aG'},usedefault=@]{
  \gnode[{#1}][\ifempty{#1}{{#2} \gchoop {#3}}{\big({#2} \gchoop {#3}\big)}]
}
\newcommandx{\gchov}[3][1={},2={\aG},3={\aG'},usedefault=@]{
  \gnode[{#1}][\left(
  \begin{array}l
    \ifempty{#1}{{#2} \\ \gchoop \\ {#3}}{\!\!{#2} \\ \gchoop \\ {#3}}
  \end{array}\right)
  ]
}
\newcommandx{\grec}[3][1={},2={\aG},3={\p},usedefault=@]{
  \gnode[{#1}][\ifempty{#1}{\grecop {#2} \grecopp {#3}}{\big(\grecop {#2} \grecopp {#3}\big)}]
}

\newcommand{\getcentroid}[2]{
    \coordinate (tmpgatecoord) at (0,0);
    \foreach \n [count=\i] in {#1}{
      \path (\n);
      \coordinate (tmpgatecoord) at ($(tmpgatecoord) + (\n)$);
      \coordinate (#2) at ($1/\i*(tmpgatecoord)$);
}
}

\tikzset{
  src/.style={draw,circle,fill=white,
    minimum size=2mm,
    inner sep=0pt},
  sink/.style={draw,circle,double,fill=white,
    minimum size=1.5mm,
    inner sep=0pt},
  node/.style={draw,circle,fill=black,
    minimum size=2mm,
    inner sep=0pt},
block/.style = {rectangle, draw=gray, align=center, fill=orange!25, rounded corners=0.1cm,
    minimum size=5mm, inner sep=2pt},
  prenode/.style = {minimum size=9pt,inner sep=2pt, font=\Large},
bblock/.style = {rectangle, draw=blue!50, opacity=.5, line width=1pt, align=center, fill=white, rounded corners=0.1cm,
    minimum size=7mm, inner sep=2pt},
  prenode/.style = {minimum size=9pt,inner sep=2pt, font=\Large},
agate/.style={draw, rectangle,
    minimum size=3mm,
    inner sep=0pt,
    fill=orange!25,
    postaction={path picture={\draw[red]
        ([yshift=\gatedistanceinand]path picture bounding box.south) --
        ([yshift=-\gatedistanceinand]path picture bounding box.north) ;}}},
ogate/.style = {
    diamond, draw, fill=orange!25,
    minimum size=4mm,
    inner sep=0pt,
    postaction={path picture={\draw[red]
        ([yshift=\gatedistancein]path picture bounding box.south) -- ([yshift=-\gatedistancein]path picture bounding box.north)
        ([xshift=-\gatedistancein]path picture bounding box.east) -- ([xshift=\gatedistancein]path picture bounding box.west)
        ;}}},
anygate/.style = {circle, draw, fill=white,
    minimum size=4mm,
    inner sep=0pt,
    postaction={path picture={\draw[black]
        ([xshift=-\gatedistancein,yshift=\gatedistancein]path picture bounding box.south east) --
        ([xshift=\gatedistancein,yshift=-\gatedistancein]path picture bounding box.north west)
        ([xshift=-\gatedistancein,yshift=-\gatedistancein]path picture bounding box.north east) --
        ([xshift=\gatedistancein,yshift=\gatedistancein]path picture bounding box.south west)
        ;}}},
elli/.style = {draw,densely dotted,-},
line/.style = {draw,->, rounded corners=0.07cm,>=latex},
  nline/.style = {draw,semithick, ->},
  pline/.style = {draw,->,>=latex},
  node distance=1cm and 0.7cm,
  baseline=(current  bounding  box.center),
}

\tikzset{
  graphaxiom/.style={
    node distance=0.7 and 0.7cm,
    scale=.5,
    transform shape
  }
}

\tikzset{
  hgsem/.style={
    draw,
    node distance=2cm and 1cm,
    transform shape,
    smooth,
    every node/.style = {font=\sffamily\bfseries}
  }
}

\tikzset{
  hgstyle/.style={
    src color={#1},
    tgt color={#1},
    centroid color={#1},
    centroid label={#1},
    centroid name={#1},
    centroid radius={#1},
    centroid ratio={#1},
    xoffset={#1},
    yoffset={#1},
    xsrcoffset={#1},
    ysrcoffset={#1},
    xtgtoffset={#1},
    ytgtoffset={#1},
    font={#1},
    centroid angle={#1},
    centroid tolerance={#1}
  },
src color/.store in = \hgsrccol,
  tgt color/.store in = \hgtgtcol,
  centroid color/.store in =\hgfillcolor,
  centroid label/.store in =\hglabel,
  centroid name/.store in =\hgname,
  centroid radius/.store in = \hgradius,
  centroid ratio/.store in = \hgratio,
  xoffset/.store in =\hgxoffset,
  yoffset/.store in =\hgyoffset,
  xsrcoffset/.store in =\hgxsrcoffset,
  ysrcoffset/.store in =\hgysrcoffset,
  xtgtoffset/.store in =\hgxtgtoffset,
  ytgtoffset/.store in =\hgytgtoffset,
  centroid angle/.store in =\hgangle,
  centroid tolerance/.store in =\hgtolerance,
src color = black,
  tgt color = black,
  centroid color = orange!40,
  centroid label={},
  centroid name={dummycentroid},
  centroid radius = .7pt,
  centroid ratio = .35,
  xoffset = 0,
  yoffset = 0,
  xsrcoffset = 0,
  ysrcoffset = 0,
  xtgtoffset = 0,
  ytgtoffset = 0,
  font=\sffamily\scriptsize,
  centroid angle=0,
  centroid tolerance=10pt
}

\newcommandx{\mkhg}[5][1={},4={},5={},usedefault=@]{
  \begingroup
  \tikzset{#1}
  \StrCount{#2,}{,
  }[\l] \StrCount{#3,}{,}[\m] \ifthenelse{\l = 1 \AND \m = 1}{
    \ifempty{#4}{
      \ifempty{#5}{
        \path[hgsem, ->, >=stealth', shorten >=1pt] (#2) -- (#3);
      }{
        \path[hgsem, ->, >=stealth', shorten >=1pt] (#2) #5 (#3);
      }
    }{
      \ifempty{#5}{
        \path[hgsem, ->, >=stealth', shorten >=1pt, #4] (#2) -- (#3);
      }{
        \path[hgsem, ->, >=stealth', shorten >=1pt, #4] (#2) #5 (#3);
      }
    }
  }{
    \coordinate (srcoffset) at (\hgxsrcoffset,\hgysrcoffset);
    \coordinate (tgtoffset) at (\hgxtgtoffset,\hgytgtoffset);
    \getcentroid{#2}{srccentroid};
    \getcentroid{#3}{tgtcentroid};
    \node[label={left:\hglabel}] (\hgname) at ($(srccentroid)!{1-\hgratio}!\hgangle:(tgtcentroid) + (\hgxoffset,\hgyoffset)$) {};
    \pgfgetlastxy \xc \yc;
    \pgfmathtruncatemacro{\xcontrol}{\xc};
    \pgfmathtruncatemacro{\ycontrol}{\yc};
    \foreach \n in {#2}{
      \path (\n);
      \pgfgetlastxy \xntmp \yntmp;
      \pgfmathtruncatemacro{\xn}{\xntmp};
      \pgfmathtruncatemacro{\yn}{\yntmp};
      \pgfmathsetmacro\xtmpdiff{abs(\xn - \xcontrol + \hgxsrcoffset)};
      \pgfmathsetmacro\ytmpdiff{abs(\yn - \ycontrol + \hgytgtoffset)};
      \ifdim \xtmpdiff pt > \hgtolerance
      \ifempty{#4}{
        \path[hgsem, \hgsrccol] (\n) .. controls ($(srccentroid.center) + (srcoffset)$) .. (\hgname.center);
      }{
        \path[hgsem, \hgsrccol] (\n) .. controls ($(srccentroid.center) + (srcoffset)$) .. (\hgname.center);
      }
      \else
      \ifempty{#4}{
        \path[hgsem, \hgsrccol] (\n) -- (\hgname.center);
      }{
        \path[hgsem, \hgsrccol, #4] (\n) -- (\hgname.center);
      }
      \fi
    }
    \foreach \n in {#3}{
      \path (\n);
      \pgfgetlastxy \xntmp \yntmp;
      \pgfmathtruncatemacro{\xn}{\xntmp};
      \pgfmathtruncatemacro{\yn}{\yntmp};
      \pgfmathsetmacro\xtmpdiff{abs(\xn - \xcontrol)};
      \pgfmathsetmacro\ytmpdiff{abs(\yn - \ycontrol)};
      \ifdim \xtmpdiff pt > \hgtolerance
      \ifempty{#4}{
        \path[hgsem, ->, >=stealth', shorten >=1pt, \hgtgtcol] (\hgname.center) .. controls (tgtcentroid.center) and ($(tgtcentroid.center) + (tgtoffset)$) .. (\n);
      }{
        \path[hgsem, ->, >=stealth', shorten >=1pt, \hgtgtcol,#4] (\hgname.center) .. controls (tgtcentroid.center) and ($(tgtcentroid.center) + (tgtoffset)$) .. (\n);
      }
      \else
      \ifempty{#4}{
        \path[hgsem, ->, >=stealth', shorten >=1pt, \hgtgtcol] (\hgname.center) --  (\n);
      }{
        \path[hgsem, ->, >=stealth', shorten >=1pt, \hgtgtcol] (\hgname.center) --  (\n);
      }
      \fi
    }
    \fill[\hgfillcolor] (\hgname) circle [radius=\hgradius];
  }
  \endgroup
}

\newcommandx{\hgordeq}[1][1={\aH},usedefault=@]{\sqsubseteq_{#1}}
\newcommandx{\gintsem}[4][4=.5]{
  \tikz[hgsem,scale=#4,every node/.style={font=\scriptsize}]{
    \node (out) {$\aout[{#1}][{#2}][][{\msg[{#3}]}]$};
    \node[below = 20pt of out] (in) {$\ain[{#1}][{#2}][][{\msg[{#3}]}]$};
    \mkhg{out}{in};
  }
}

\newcommandx{\gsem}[2][1={\aG},2={},usedefault=@]{[\![ {#1} ]\!]_{#2}}
\newcommandx{\rbot}{\text{undef}}
\newcommandx{\rtrs}[1][1={\aH},usedefault=@]{{#1}^{\star}}
\newcommandx{\gord}[1][1={\aG},usedefault=@]{\leq_{#1}}
\newcommandx{\gordeq}[1][1={\aG},usedefault=@]{\leq_{#1}}
\mkfun{\cause}{cs}{}
\mkfun{\effect}{ef}{}

\mkfun{\participants}{ptps}{}

\newcommandx{\aW}{w}
\newcommandx{\rlang}{\mathcal{L}}
\newcommand{\gfun}[1]{\ensuremath{\mathsf{\colorFun #1}}}
\mkfun{\eact}{\gfun{act}}{}
\mkfun{\enode}{\gfun{cp}}{}
\mkfun{\cp}{\gfun{cp}}{}

\mkuop{\rmax}{\gfun{max}}{\aH}
\mkuop{\rmin}{\gfun{min}}{\aH}
\mkuop{\rMAX}{\gfun{lst}}{\aH}
\mkuop{\rMIN}{\gfun{fst}}{\aH}

\newcommandx{\rseq}[2][1=\aG,2={\aG'},usedefault=@]{\gfun{seq}({#1},{#2})}
\newcommandx{\rpar}[2][1=\aG,2={\aG'},usedefault=@]{\gfun{par}({#1},{#2})}

\newcommandx{\gproj}[2][1=\aG,2=\ptp]{{#1}\!\downarrow_{#2}}
\newcommandx{\gprojPN}[2][1=\aG,2=\ptp]{{#1}\!\downarrow_{#2}^{\mathsf{P\hspace{-0.5pt}N}}}
\newcommandx{\cinit}[1][1={\aQzero},usedefault=@]{{#1}}
\newcommandx{\cfinal}[1][1={q_e},usedefault=@]{{#1}}

\newcommandx{\geproj}[4][1=\aG,2=\ptp,3=\cinit,4=\cfinal,usedefault=@]{
  {#1}\downarrow_{#2}^{{#3},{#4}}
}

\newcommand*{\StrikeThruDistance}{0.15cm}\tikzset{strike thru arrow/.style={
    decoration={markings, mark=at position 0.5 with {
        \draw [blue, thick,-] 
            ++ (-\StrikeThruDistance,-\StrikeThruDistance) 
            -- ( \StrikeThruDistance, \StrikeThruDistance);}
    },
    postaction={decorate},
}}

\newcommandx{\ich}[1][1={\aG},usedefault=@]{{#1}^{\oplus}}
\newcommandx{\ichedges}[2][1={\aG},2={\gname},usedefault=@]{{#1}^{\oplus}({#2})}
\newcommandx{\parts}[1]{2^{#1}}
\newcommandx{\actch}{c}
\newcommandx{\soundactch}[2][1={\aG},2={\actch},usedefault=@]{{#1} \,\circledR\, {#2}}
\newcommandx{\rOnActch}[2][1={\aG},2={\actch},usedefault=@]{{#1} \setminus {#2}}
\newcommandx{\rOnActchClean}[2][1={\aG},2={\actch},usedefault=@]{{#1} \circledR {#2}}
\newcommandx{\rAllEvents}[1][1={\aG},usedefault=@]{\mathit{dom}(#1)}

\newcommand{\AV}{\mathcal{V}}
\newcommand{\aH}{H}

\newcommandx{\hgvertex}[2][1=\al,2=\gname,usedefault=@]{{#1}_{\textcolor{red}{[{#2}]}}}
\newcommand{\aE}{{\colorE E}}
\renewcommand{\ae}[1][e]{{\colorE{#1}}}

\newcommand{\al}[1][l]{{\colorE{#1}}}
\newcommandx{\hyedge}[1]{\{#1\}}

\newcommandx{\rdiv}[2][1=\gcho,2=\ptp,usedefault=@]{
  \gfun{div}_{#2}(#1)
}

\newcommandx{\rrdiv}[5][1={\aG},2={\aG'},3={\AV},4={\AV'},5=\ptp,usedefault=@]{
  \gfun{div}^{#3,#4}_{#5}(#1,#2)
}
\newcommandx{\pdiv}[3][1={\apom_1},2={\apom_2},3={\apom},usedefault=@]{
  \gfun{div}_{#3}(#1,#2)
}
\newcommandx{\pfork}[3][1={\apom_1},2={\apom_2},3={\apom},usedefault=@]{
  \gfun{fork}_{#3}(#1,#2)
}

\newcommandx{\mkint}[6][3=i,4=\p,5=\msg,6=\q,usedefault=@]{
  \node[bblock,{#1}] (#2) {$\gint[#3][#4][#5][#6]$};
}

\newcommandx{\mkgraph}[3][1=.5cm]{
  \node[source,above = #1 of {#2}] (src#2) {};
  \node[sink,below  = #1 of {#3}] (sink#3) {};
  \path[line] (src#2) -- (#2);
  \path[line] (#3) -- (sink#3);
}

\newcommandx{\mkloop}[4][1=.5,2=1.5]{
  \node[ogate,above = #1 of {#3}] (entry#3) {};
  \pgfgetlastxy \xentry \yentry;
  \pgfmathtruncatemacro{\xentryrounded}{\xentry};
  \node[ogate,below  = #1 of {#4}] (exit#4) {};
  \pgfgetlastxy \xexit \yexit;
  \pgfmathtruncatemacro{\xexitrounded}{\xexit};
  \path[line] (entry#3) -- (#3);
  \path[line] (#4) -- (exit#4);
  \pgfmathsetmacro\tmpdiff{abs(\xentryrounded - \xexitrounded)}
  \path[line] (exit#4) -|  ($(exit#4)+(\tmpdiff,0)+(#2,0)$) |- (entry#3);
}

\newcommandx{\mkfork}[4][2=gatenode,3=i,4=.6,usedefault=@]{
  \mkgatebegin{#1}[{\gname[#3]}][agate][#4]{#2}
}

\newcommandx{\mkbranch}[4][2=gatenode,3=i,4=.6,usedefault=@]{
  \mkgatebegin{#1}[{\gname[#3]}][ogate][#4]{#2}
}

\newcommandx{\mkgatebegin}[5][2={},3=ogate,4=.5]{
\coordinate (gatecord) at (0,0);
  \foreach \n [count=\i] in {#1}{
    \pgfgetlastxy \xc \yc;
    \path (\n);
    \pgfgetlastxy \xn \yn;
    \coordinate (gatecord) at ($(gatecord) + (\xn,0)$);
    \coordinate (gatecord) at ($1/\i*(gatecord)$);
    \ifdim \yn < \yc
    \node (max) at (0,\yc) {};
    \else
    \node (max) at (0,\yn) {};
    \fi
  }
  \coordinate (gatecord) at ($(gatecord) + (0,#4) + (max)$);
  \node[#3,label={below:$#2$}] (#5) at (gatecord) {};
  \pgfgetlastxy{\xgate}{\ygate};
  \pgfmathtruncatemacro{\xgateround}{\xgate};
  \StrCount{#1,}{,}[\l] \ifnum \l < 2 {\errmessage{#1 argument should be a comma-separated list of lenght >= 2}}
  \else{
    \foreach \n in {#1}{
      \path (\n);
      \pgfgetlastxy{\xnode}{\ynode};
      \pgfmathtruncatemacro{\xnround}{\xnode};
      \pgfmathsetmacro\tmpdiff{abs(\xnround - \xgateround)}
      \ifdim \tmpdiff pt > 1 pt \path[line] (#5) -| (\n);
      \else
        \path[line] (#5) -- (\n);
      \fi
    }
  }
  \fi
}

\newcommandx{\gorthopath}[4][3=|-,4=-2pt,usedefault=@]{
  \path #1;
  \pgfgetlastxy{\xabove}{\yabove};
  \pgfmathtruncatemacro{\xaboveround}{\xabove};
  \path #2;
  \pgfgetlastxy{\xbelow}{\ybelow};
  \pgfmathtruncatemacro{\xnround}{\xbelow};
  \pgfmathsetmacro\tmpdiff{abs(\xnround - \xaboveround)}
  \ifdim \tmpdiff pt > 1 pt \path[line] #2 #3 #1;
  \else
  \path[line] #2 -- #1;
  \fi
}

\newcommandx{\mkmerge}[4][2=gatenode,3=i,4=0,usedefault=@]{\mkgateend{#1}[{\ifempty{#3}{}{\nmerge[#3]}}][ogate][#4]{#2}}

\newcommandx{\mkjoin}[4][2=gatenode,3=i,4=0,usedefault=@]{\mkgateend{#1}[{\ifempty{#3}{}{\nmerge[#3]}}][agate][#4]{#2}}

\newcommandx{\mkgateend}[5][2={},3=ogate,4=.5,usedefault=@]{
\coordinate (gatecord) at (0,0);
  \foreach \n [count=\i] in {#1}{
    \pgfgetlastxy \xc \yc;
    \path (\n);
    \pgfgetlastxy \xn \yn;
    \coordinate (gatecord) at ($(gatecord) + (\xn,0)$);
    \coordinate (gatecord) at ($1/\i*(gatecord)$);
    \ifdim \yn > \yc
    \node (min) at (0,\yc) {};
    \else
    \node (min) at (0,\yn) {};
    \fi
  }
  \coordinate (gatecord) at ($(gatecord) - (0,#4) + (min)$);
  \node[#3,label={above:$#2$}] (#5) at (gatecord) {};
  \pgfgetlastxy{\xgate}{\ygate};
  \pgfmathtruncatemacro{\xgateround}{\xgate};
  \StrCount{#1,}{,}[\l] \ifnum \l < 2 {\errmessage{#1 argument should be a comma-separated list of lenght >= 2}}
  \else{
    \foreach \n in {#1}{
      \path (\n);
      \pgfgetlastxy{\xnode}{\ynode};
      \pgfmathtruncatemacro{\xnround}{\xnode};
      \pgfmathsetmacro\tmpdiff{abs(\xnround - \xgateround)}
      \ifdim \tmpdiff pt > 1 pt \path[line] (\n) |- (#5);
      \else
        \path[line] (\n) -- (#5);
      \fi
    }
  }
  \fi
}

\newcommand{\gatedistancein}{3pt}
\newcommand{\gatedistanceinand}{2pt}

\usetikzlibrary{
  arrows,
  backgrounds,
  chains,
  calc,
  decorations.markings,decorations.pathreplacing,
  fadings,
  fit,
  patterns,
  petri,
  positioning,
  shadows,
  shapes,automata,shapes.callouts
}

\tikzset{
  src/.style={draw,circle,fill=white,
    minimum size=2mm,
    inner sep=0pt
  },
  sink/.style={draw,circle,double,fill=white,
    minimum size=1.5mm,
    inner sep=0pt
  },
  node/.style={draw,circle,fill=black,
    minimum size=2mm,
    inner sep=0pt
  },
source/.style={draw,circle,fill=white,
    minimum size=3mm,
    inner sep=0pt
  },
  sink/.style={draw,circle,double,fill=white,
    minimum size=3mm,
    inner sep=0pt
  },
block/.style = {rectangle, draw=gray, align=center, fill=orange!25, rounded corners=0.1cm,
    minimum size=5mm, inner sep=2pt},
  prenode/.style = {minimum size=9pt,inner sep=2pt, font=\Large},
bblock/.style = {rectangle, draw=blue!50, opacity=.5, line width=1pt, align=center, fill=white, rounded corners=0.1cm,
    minimum size=7mm, inner sep=2pt},
  prenode/.style = {minimum size=9pt,inner sep=2pt, font=\Large},
agate/.style={draw, rectangle,
    minimum size=3mm,
    inner sep=0pt,
    fill=orange!25,
    postaction={path picture={\draw[red]
        ([yshift=\gatedistanceinand]path picture bounding box.south) --
        ([yshift=-\gatedistanceinand]path picture bounding box.north) ;}}
  },
ogate/.style = {
    diamond, draw, fill=orange!25,
    minimum size=4mm,
    inner sep=0pt,
    postaction={path picture={\draw[red]
        ([yshift=\gatedistancein]path picture bounding box.south) -- ([yshift=-\gatedistancein]path picture bounding box.north)
        ([xshift=-\gatedistancein]path picture bounding box.east) -- ([xshift=\gatedistancein]path picture bounding box.west)
        ;}}},
altogate/.style = {
    diamond, draw,
    minimum size=4mm,
    inner sep=0pt,
    postaction={path picture={\draw
        ([yshift=\gatedistancein]path picture bounding box.south) -- ([yshift=-\gatedistancein]path picture bounding box.north)
        ([xshift=-\gatedistancein]path picture bounding box.east) -- ([xshift=\gatedistancein]path picture bounding box.west)
        ;}}},
  altgate/.style={draw, rectangle,
    minimum size=3mm,
    inner sep=0pt,
    postaction={path picture={\draw
        ([yshift=\gatedistanceinand]path picture bounding box.south) --
        ([yshift=-\gatedistanceinand]path picture bounding box.north) ;}}},
anygate/.style = {circle, draw, fill=white,
    minimum size=4mm,
    inner sep=0pt,
    postaction={path picture={\draw[black]
        ([xshift=-\gatedistancein,yshift=\gatedistancein]path picture bounding box.south east) --
        ([xshift=\gatedistancein,yshift=-\gatedistancein]path picture bounding box.north west)
        ([xshift=-\gatedistancein,yshift=-\gatedistancein]path picture bounding box.north east) --
        ([xshift=\gatedistancein,yshift=\gatedistancein]path picture bounding box.south west)
        ;}}
  },
smallglobal/.style={
        node distance=1cm and 0.8cm, semithick, scale=0.8, every node/.style={transform shape}
  },
elli/.style = {draw,densely dotted,-},
line/.style = {draw,->, rounded corners=0.07cm,>=latex},
  nline/.style = {draw,semithick, ->},
  pline/.style = {draw,->,>=latex},
  node distance=1cm and 0.7cm,
  baseline=(current  bounding  box.center),
  local/.style={rectangle, draw, fill=\fillcolor, drop shadow,
    text centered, rounded corners, minimum height=5em
  },
  bigar/.style={
    draw,very thick, ->
  },
  process/.style={rectangle, draw=gray, fill=\fillcolor, drop shadow,
    text centered, minimum height=5em,text=gray
  },
  choreo/.style={rectangle, draw, fill=\fillcolor, drop shadow,
    text centered, rounded corners, minimum height=5em
  },
mycfsm/.style={
        font=\footnotesize,
        initial where=above,
        ->,>=stealth,auto, node distance=1cm and 1cm,
        scale=1, every node/.style={transform shape},
        every state/.style=inner sep=2pt,
        baseline=(current  bounding  box.center)
  },
  machinecloud/.style={
    cloud, cloud puffs=10, cloud ignores aspect, minimum height=.1cm, minimum width=2cm, draw
  },
  fitting node/.style={
    inner sep=0pt,
    fill=none,
    draw=none,
    reset transform,
    fit={(\pgf@pathminx,\pgf@pathminy) (\pgf@pathmaxx,\pgf@pathmaxy)}
  },
  mypetri/.style={
    font=\footnotesize,
    baseline=(current  bounding  box.center)
  },
  silentrans/.style = {rectangle, draw=black, align=center, fill=black,
    minimum height=1pt,
    minimum width=15pt,
    inner sep=1.5pt
  },
  reset transform/.code={\pgftransformreset},
  tmtape/.style={draw,minimum size=1.2cm}
}

\newcommand{\gunlessop}{\mbox{\colorOp\tiny\tt unless}}

\newcommandx{\gtry}[5][1=\gname,2={\aG_1 \gchoop \cdots \gchoop \aG_n},3=\gin,4=\gout,5={j},usedefault=@]{
  \def\foo{\gtryop\ {#2} \ \gcatchop\ {#3} {\colorOp \Rightarrow} {#4} {\colorOp \bullet} {\gname[{#5}]}}
  \gnode[{#1}][{\ifempty{#1} {\foo } { \big(\foo \big) }}]
}

\newcommandx{\gtrycatch}[4][1=\gname,2={\aG},3=\gin,4={\aG'},usedefault=@]{
  \def\foo{\gtryop\ {#2} \ \gcatchop\ {#3} \gdoop\ {#4}}
  \gnode[{#1}][{\ifempty{#1} {\foo} {\big( \foo \big) }}]
}

\def\colorGuard{\color{cyan}}
\newcommand{\aguard}{{\colorGuard \phi}}
\newcommandx{\agG}[2][1={\aG},2=\aguard]{{#1} \ifempty{#2}{}{\ \gunlessop\ {#2}}}

\newcommandx{\grcho}[4][1=\gname,2={\agG},3={\agG[\aG'][\aguard']},4={\cdots},usedefault=@]{
  \def\foo{{#2} {\ \ifempty{#4}{\gchoop}{\gchoop \ {#4}\  \gchoop}\ } {#3}}
  \gnode[{#1}][\ifempty{#1}{\foo}{\big( \foo \big)}]
}

\newcommandx{\ggprefix}[3][1=\ptp,2={\aR},3={\aR'},usedefault=@]{f_{#1}} \newcommand{\aconfigfn}{\chi}
\newcommand{\aconfig}{\ell}

\newcommand{\lstates}{\statemap}
\newcommandx{\sysconfig}[3][1=\lstates,2=\aconfigfn,3={},usedefault=@]{
  \conf{ {#1},{#2} \ifempty{#3}{}{, #3} }
}
\newcommand{\sysctxfn}[1][]{\gamma_{#1}}
\newcommandx{\sysctx}[2][1=\aQ,2={},usedefault=@]{({#1},\sysctxfn[{#2}])}

\newcommandx{\alog}[4][1=\msg,2=q,3=\gname,4=t,usedefault=@]{\big({#1},{#2},{#3},{#4}\big)}

\newcommand{\aCM}{M}\newcommand{\aM}{\aCM}
\newcommand{\aQ}{Q}
\newcommandx{\aQzero}[1][1=,usedefault=@]{
  {\ifempty{#1}{q_0}{q_{0#1}}}
}
\newcommand{\badbranches}[1][]{\beta\ifempty{#1}{}{\big({#1}\big)}}
\newcommand{\aTrs}{\tset}
\newcommandx{\guardedaction}[2][1=\al,2=\aguard,usedefault=@]{
  {#1} \ifempty{#2}{}{/} {#2}
}
\newcommandx{\atrM}[4][1=q,2=\al,3={\hat q,\hat \al, \aguard},4=q',usedefault=@]{
  {#1} \xrightarrow[{#3}]{\guardedaction[{#2}][]} {{#4}}
}
\newcommandx{\atrS}[5][
  1={\sysconfig[@][@][\badbranches]},
  2=\al,
  3=\aguard,
  4={\sysconfig[\lstates'][\aconfigfn'][\badbranches]},
  5=\sysctx,usedefault=@
]{
  {#1} \xRightarrow{\qquad} {{#4}}
}
\newcommandx{\arevtrS}[2][
  1={\sysconfig[@][@][\badbranches]},
  2={\sysconfig[\lstates'][\aconfigfn'][\badbranches']},
  usedefault=@
]{
  {#1} \rightsquigarrow {#2}
}
\newcommand{\aCS}[1][S]{\mathsf{#1}}

\newcommandx{\enables}[2][1=\aconfigfn,2=\aguard,usedefault=@]{{#1} \vdash {#2}}
\newcommandx{\gprojfn}[5][1=\aG,2=\ptp,3=\cinit,4=\cfinal,5={},usedefault=@]{
  \mathbf{proj}_{#2}({#1},{#3},{#4}\ifempty{#5}{}{,{#5}})
}

\newcommandx{\rbp}[3][1=\aG,2=\aconfigfn,3=\achan,usedefault=@]{\mathtt{RBP}_{{#1},{#2}}\ifempty{#3}{}{\big({#3}\big)}}

\newcommand{\apseudoCFSM}{\mathtt{M}}
\newcommandx{\pseudoseq}[2][1=\apseudoCFSM,2=\apseudoCFSM',usedefault=@]{{#1}  ; {#2}}
\newcommandx{\pseudoCFSM}[4][1=\aQ,2=\aQzero,3=\cfinal,4=\aTrs,usedefault=@]{(#1 \ ; #2 \ ; #3 \ ; #4)}
\newcommandx{\markt}[3][1=\hat{\al},2=\hat{q},3=\aguard,usedefault=@]{\%\big({#1} , {#2}, {#3}\big)}

\newcommandx{\borderfn}[2][1=\aconfig,2=\aloop,usedefault=@]{
  \mathsf{border}_{{#2}}\ifempty{#1}{}{\big({#1}\big)}
}

\newcommandx{\ggvisually}[4][1=5pt,2=15pt,3=5pt,4=5pt,usedefault=@]{
  \def\dist{\hspace{1.0cm}}
  \tikzset{
    mycallout/.style={
      fill=green!10, opacity=.5, overlay, align=center,
      cloud callout, cloud puffs=15, aspect=1.9, cloud ignores aspect, cloud puff arc=100, shading=ball
    }
  }
  $\begin{array}{c@{\dist}c@{\dist}c@{\dist}c@{\dist}c}
\begin{tikzpicture}[node distance=0.9cm and 0.4cm, every node/.style={scale=.7,transform shape}]
       \mkint{}{int}[]
       \mkgraph{int}{int};
       \node[mycallout, above = .3cm of srcint, xshift=1cm, callout absolute pointer={(srcint.east)}] {source node};
       \node[mycallout, below = .3cm of sinkint, xshift=-1cm, callout absolute pointer={(sinkint.west)}] {sink node};
     \end{tikzpicture}
     &
\begin{tikzpicture}[node distance=.9cm and 0.4cm, every node/.style={scale=.7,transform shape}]
       \node[bblock] at (0,0) (g) {$\aG$};
       \node[node, below=of g] (s1) {};
       \node[bblock, below=of s1] (gp) {$\aG'$};
       \path[line,dotted] (g) -- (s1);
       \path[line,dotted] (s1) -- (gp);
     \end{tikzpicture}
     &
\begin{tikzpicture}[node distance=.4cm and 0.4cm, every node/.style={scale=.7,transform shape}]
       \node[bblock] at (-.7,0) (g) {$\aG$};
       \node[bblock] at (.7,0)  (gp) {$\aG'$};
       \node[node, above=of g] (f) {};
       \node[node, below=of g] (j) {};
       \node[node, above=of gp] (fp) {};
       \node[node, below=of gp] (jp) {};
       \path[line,dotted] (f) -- (g);
       \path[line,dotted] (g) -- (j);
       \path[line,dotted] (fp) -- (gp);
       \path[line,dotted] (gp) -- (jp);
       \mkfork{f,fp}[fork][][#1];
       \mkjoin{j,jp}[join][][#2];
       \mkgraph{fork}{join};
       \node[mycallout, above = .3cm of fork, xshift=1cm, callout absolute pointer={(fork.east)}] {fork gate};
       \node[mycallout, above = -.9cm of join, xshift=-1cm, callout absolute pointer={(join.west)}] {join gate};
     \end{tikzpicture}
     &
\begin{tikzpicture}[node distance=.4cm and 0.4cm, every node/.style={scale=.7,transform shape}]
       \node[bblock] at (-.7,0) (g) {$\aG$};
       \node[bblock] at (.7,0)  (gp) {$\aG'$};
       \node[node, above=of g] (f) {};
       \node[node, below=of g] (j) {};
       \node[node, above=of gp] (fp) {};
       \node[node, below=of gp] (jp) {};
       \path[line,dotted] (f) -- (g);
       \path[line,dotted] (g) -- (j);
       \path[line,dotted] (fp) -- (gp);
       \path[line,dotted] (gp) -- (jp);
       \mkbranch{f,fp}[fork][][#3];
       \mkmerge{j,jp}[join][][#4];
       \mkgraph{fork}{join};
       \node[mycallout, above = .3cm of fork, xshift=1cm, callout absolute pointer={(fork.east)}] {branch gate};
       \node[mycallout, above = -.9cm of join, xshift=-1cm, callout absolute pointer={(join.west)}] {merge gate};
     \end{tikzpicture}
     &
\begin{tikzpicture}[node distance=0.4cm and 0.4cm, every node/.style={scale=.7,transform shape}]
       \node[bblock] (g) {$\aG$};
       \node[node, above=.5cm of g] (f) {};
       \node[node, below=.5cm of g] (j) {};
       \path[line,dotted] (f) -- (g);
       \path[line,dotted] (g) -- (j);
       \mkloop[.5][@]{f}{j}[];
       \mkgraph[.4cm]{entryf}{exitj};
       \node[mycallout, above = .2cm of entryf, xshift=1.3cm, callout absolute pointer={(entryf.east)}] {loop entry};
       \node[mycallout, above = -.7cm of exitj, xshift=-1.3cm, callout absolute pointer={(exitj.west)}] {loop exit};
     \end{tikzpicture}
     \\
     \text{\scriptsize interaction}
     &
     \text{\scriptsize sequential}
     &
     \text{\scriptsize parallel}
     &
     \text{\scriptsize branching}
     &
     \text{\scriptsize iteration}
   \end{array}$
}

\title{
  Composition of choreography automata\\
  {\small \sc (Technical Report)}}
	
\author{Franco Barbanera\inst 1, Ivan Lanese\inst 2, Emilio Tuosto\inst 3}
\authorrunning{F. Barbanera et. al.}
\institute{
  Dept. of Mathematics and Computer Science, University of Catania (Italy)
  \and
  Focus Team, University of Bologna/INRIA (Italy)
  \and
  Gran Sasso Science Institute (Italy)
  }

\spnewtheorem{fact}{Fact}{\bfseries}{\itshape}
\begin{document}

\maketitle              

\begin{abstract}
  Coreography automata are an automata-based model of choreographies,
  that we show to be a compositional one.
Choreography automata
  represent global views of choreographies (and rely on the well-known
  model of communicating finite-state machines to model local
  behaviours). The projections of well-formed global views are live as
  well as lock- and deadlock-free.
In the class of choreography automata we  define an internal
  operation of {\em composition}, which connects two global views via roles
  acting as interfaces.
We show that under mild conditions the composition of well-formed
  choreography automata is well-formed.
The composition operation enables for a flexible
  modular mechanism at the design level.
\end{abstract}

\section{Introduction}\label{sec:intro}
Choreographic approaches to the modelling, analysis, and programming
of message-passing applications abound.
Several models based on behavioural types have been proposed to
analyse properties such as liveness or deadlock-freedom (e.g.,
\cite{ScalasY19,honda16jacm,cdyp16} and the
survey~\cite{hlvlcmmprttz16} to mention but few) while other
approaches have considered syntax-free models~\cite{gt18}.
Abstract models have also been advocated to verify and debug
choreographic specifications~\cite{kps19,adsgpt19} using modelling
languages such as BPMN~\cite{BPMN}.
At a programming level, choreographic programming has been
explored in~\cite{lgmz08,mon13}.
The ICT industrial sector is also starting to acknowledge
the potential of choreographies~\cite{bon18,ad16}.

A distinguished feature of choreographies is the coexistence of two
distinct yet related views of a distributed system: the \emph{global}
and the \emph{local} views.
The former is an abstraction that yields a description of the system
from a \emph{holistic} point of view.
A global view indeed describes the coordination necessary among the
various components of the system \quo{altogether}.
In contrast, the local view of a choreography specifies the behaviour
of the single components in \quo{isolation}.
In the so called top-down approach to choreographies, the local view
can be derived by \emph{projecting} the global one on each single
component.
This yields results that typically guarantee that the execution of the
local components reflects the specification of the global one without
spoiling communication soundness (e.g., deadlock freedom, liveness, etc.).
These results do not hold in general and require to restrict to
\emph{well-formed} global views.

The composition of local views is simple.
The local components just execute in parallel on a specific
communication infrastructure.
The situation is more complex for global views and, despite some
attempts (cf. \cref{sec:related}), the problem of composing global views
of choreographies is still open.

This paper, after recalling the  choreographic model proposed in \cite{BLT20}, 
describes a proposal for the global views composition problem via a
 composition operation over \emph{interfaces}.
The modelling language of global views of \cite{BLT20} are \emph{choreography
  automata} (c-automata for short), which are basically finite-state
automata whose transitions are labelled with \emph{interactions}
expressing communications between some participants (also called
roles).
More precisely, an interaction $\gint[]$ represents the fact that
participant \p\ sends message $\msg$ to participant \q\ and
participant \q\ receives it.
For instance, consider the following c-automata
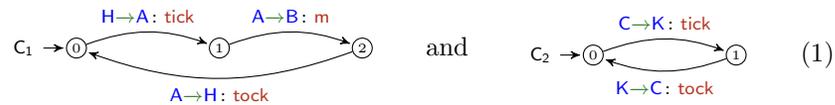
\begin{equation}
  \begin{array}{c@{\qqand}c}
    \begin{tikzpicture}[node distance=1.9cm]
      \tikzstyle{every state}=[cnode]
      \tikzstyle{every edge}=[carrow]
\node[state] (zero) {$0$};
      \node[state] (one) [right   of=zero]   {$1$};
      \node[draw=none,fill=none] (start) [left  = 0.3cm  of zero]{$\chora[_1]$};
      \node[state]  (two) [right of=one] {$2$};
\path  (start) edge node {} (zero) 
      (zero) edge[bend left=20] node[above] {$\gint[][h][tick][A]$} (one)
      (one) edge[bend left=20] node[above] {$\gint[]$} (two)
      (two) edge[bend left=20] node[below] {$\gint[][@][tock][H]$} (zero)
      ;
    \end{tikzpicture}
    &
    \begin{tikzpicture}[node distance=1.9cm]
      \tikzstyle{every state}=[cnode]
      \tikzstyle{every edge}=[carrow]
\node[state] (zero) {$0$};
      \node[state] (one) [right   of=zero]   {$1$};
      \node[draw=none,fill=none] (start) [left  = 0.3cm  of zero]{$\chora[_2]$};
\path  (start) edge node {} (zero) 
      (zero) edge[bend left = 20] node[above] {$\gint[][c][tick][K]$} (one)
      (one) edge[bend left = 20]  node[below]  {$\gint[][K][tock][c]$} (zero)
      ;
    \end{tikzpicture}
  \end{array}\label{eq:CAintro}
\end{equation}
where the left-most c-automaton $\chora[_1]$ represents a system where
\p\ sends \q\ message $\msg$ once it receives a $\msg[tick]$ from an
external \quo{cron} service \iH.
Participant \p\ acknowledges the completion of the exchange with \q\
by sending \iH message $\msg[tock]$.
The right-most c-automaton $\chora[_2]$ specifies that the cron
service \p[c] repeatedly sends message $\msg[tick]$ and receives
message $\msg[tock]$ from \iK.

Our model of \emph{local views} are a variant of communicating
systems~\cite{bz83} in which we adopt the synchronous semantics
defined in~\cite{BLT20}.

For us, an \emph{interface} is a designated role of a c-automaton
representing behaviour \quo{scattered in the enviroment}, namely
provided by other c-automata.
For instance, \iH is an interface of $\chora[_1]$ in
\eqref{eq:CAintro} whereby we \quo{delegate} the cron service; this
outsourced behaviour is provided by another system in the execution
environment of $\chora[_1]$ such as the c-automaton $\chora[_2]$.

The composition of c-automata, say $\chora$ and $\chora[B]$, over
their interfaces, say \iH for $\chora$ and \iK for $\chora[B]$, is the
result of the combination of two operations:
\begin{enumerate}[(i)]
\item the first operation is simply the product
  $\chora \times \chora[B]$ of the two automata and
\item the second operation is \emph{(interface) blending}, namely the
  removal of the interfaces transforming participants \iH and \iK into
  \quo{forwarders}.
\end{enumerate}
More precisely, if the first transition $t$ of two consecutive
transitions in the product $\chora \times \chora[B]$ has an interface
role as the receiver of a message $\msg$ and the second, say $t'$, the
other interface role as sender of $\msg$, then $t$ and $t'$ are
replaced with a transition where the sender of $t$ and the receiver of
$t'$ exchange $\msg$.
Any transition involving \iH or \iK that is not replaced is then
removed.

Let us illustrate how to compose the c-automata in \eqref{eq:CAintro}.
The product of $\chora[_1]$ and $\chora[_2]$ and their blending on \iH
and \iK are respectively the c-automata
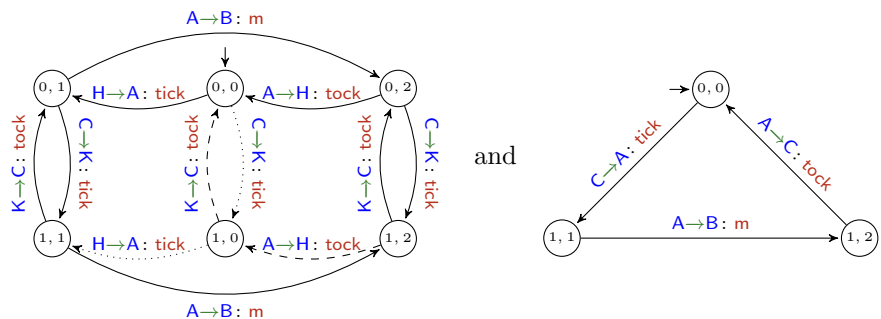
\[
  \begin{tikzpicture}[node distance=2.3cm]
    \tikzstyle{every state}=[cnode]
    \tikzstyle{every edge}=[carrow]
\node[draw=none,fill=none] (start) {};
    \node[state, below = .3cm of start] (00) {$0,0$};
    \node[state] (01) [left of=00]   {$0,1$};
    \node[state] (02) [right of=00] {$0,2$};
    \node[state, below = 1.5cm of 00] (10) {$1,0$};
    \node[state] (11) [left of=10]   {$1,1$};
    \node[state] (12) [right of=10] {$1,2$};
\path
    (start) edge node {} (00)
    (00) edge[bend left=20] node[above] {$\gint[][h][tick][A]$} (01)
    (01) edge[bend left=30] node[above] {$\gint[]$} (02)
    (02) edge[bend left=20] node[above] {$\gint[][@][tock][H]$} (00)
    (10) edge[bend left=20, dotted] node[above] {$\gint[][h][tick][A]$} (11)
    (11) edge[bend right=30] node[below] {$\gint[]$} (12)
    (12) edge[bend left=20, dashed] node[above] {$\gint[][@][tock][H]$} (10)
    ;
    \path
    (00) edge[bend left=20, dotted] node[above] {$\gint[][c][tick][k]$} (10)
    (10) edge[bend left=20, dashed] node[above] {$\gint[][k][tock][c]$} (00)
    (01) edge[bend left=20] node[above] {$\gint[][c][tick][k]$} (11)
    (11) edge[bend left=20] node[above] {$\gint[][k][tock][c]$} (01)
    (02) edge[bend left=20] node[above] {$\gint[][c][tick][k]$} (12)
    (12) edge[bend left=20] node[above] {$\gint[][k][tock][c]$} (02)
    ;
  \end{tikzpicture}
  \qand
  \begin{tikzpicture}[node distance=2.3cm]
    \tikzstyle{every state}=[cnode]
    \tikzstyle{every edge}=[carrow]
\node[draw=none,fill=none] (start) {};
    \node[state, right = .3cm of start] (00) {$0,0$};
    \node[state, below left = of 00] (11) {$1,1$};
    \node[state, below right = of 00] (12) {$1,2$};
\path
    (start) edge node {} (00)
    (00) edge node[above] {$\gint[][c][tick][A]$} (11)
    (11) edge node[above] {$\gint[]$} (12)
    (12) edge node[above] {$\gint[][@][tock][c]$} (00)
    ;
  \end{tikzpicture}
\]
where the dotted and the dashed transitions in the product automaton
on the left form the consecutive transitions to blend because one
interface participant receives a message and the other interface
participant sends it.

As shown in this simple example, blending enables modular design of
choreographies.
We note that modularity can be attained also with other mechanisms.
For instance, in the approach in~\cite{MontesiY13} one can specify
partial designs and then check that their combination enjoys
properties of interest, without explicitly building a composed
choreography.
Another example is the use of refinement of global views as suggested
in~\cite{dmt20}.
Choreography automata instead are equipped with an internal operation
of composition that given two c-automata yields another
one. Composition preserves the interesting behaviours of the
components and guarantees communication soundness (liveness and
deadlock freedom) of the resulting c-automaton under mild conditions.
We argue that blending-based composition is more flexible an approach
to modularity than other proposals; for instance, the result of the
composition of two c-automata is another c-automaton that provides a
global specification of the composition.
Also, the blending operation allows one to \quo{self-compose} a
c-automaton and this corresponds to a sort of refinement operation
where blending two roles \iH and \iK of a c-automaton amounts to
consider them as \quo{forwarders} that are then made implicit and removed.
We will further demonstrate the flexibility of blending-based
composition on a non-trivial working example where more systems can be
composed together.

\paragraph{Structure \& Contributions of the paper}
The basic notions used through the paper are surveyed in
\cref{sec:bkg}.
\cref{sec:c-automata} recalls c-automata and their projections; the former
model is essentially borrowed from~\cite{BLT20} and the latter are basically
the communicating systems of~\cite{bz83}.
Choreography automata yield a simple yet very expressive
language for global views.
\cref{sec:wf,sec:comp} report our main results.

\begin{description}
\item[Well-formedness] The notion of well-formedness and its relations
  with the projection operation \cite{BLT20} are recalled in
  \cref{sec:wf}.
The projection is correct (cf. \cref{th:projectionCorrectness}), and
  guarantees liveness as well as lock- and deadlock-freedom
  (cf. \cref{prop:live,prop:lock,prop:deadlock}).
\item[Composition] The composition of c-automata is defined in
  \cref{sec:comp} as a refinement of the product of c-automata through
  the blending operation.
This operation is inspired by one of the composition approaches
  in~\cite{francoICE,BDLT21}.
The composition of arbitrary c-automata may not preserve
  well-formedness (and hence the behaviour at the local level).
  
\item[Properties] We identify a class of c-automata that, imposing
  mild conditions (cf. \cref{def:univocity}) on interfaces, guarantees
  that our notion of composition preserves well-formedness
  (cf. \cref{prop:wf-presbycomp}).
Moreover, interface blending ensures that the composition does not
  add new behaviours to the roles not in the interface
  (cf. \cref{lemma:sim}) for all c-automata, while for
  \emph{reflective} c-automata we can also prove that behaviours are
  preserved (cf. \cref{lemma:bisim}).
\item[Compatibility] Finally, the notion of \emph{compatibility} of
  interfaces (cf. \cref{def:compatibility}, borrowed
  from~\cite{francoICE}) is sufficient to establish reflectiveness
  (\cref{thm:comprod}).
\end{description}

Besides being based on the well-known theory of formal languages,
c-automata are also technically convenient as they allow us to reuse
the rich theory of automata.
\cref{sec:conc} discusses these points more extensively together with
other concluding remarks, related work, and future lines of research.
\ifhideproofs
Additional material and complete proofs can be found in the technical report \cite{BLT-TechRep}.
\else
\fi

\section{Preliminaries}\label{sec:bkg}
We recall some basic notions of finite-state automata (FSAs).
A \emph{finite-state automaton} is a tuple
$A = \conf{\sset, \lset, \tset, s_0}$ where
\begin{itemize}
\item $\sset$ is a finite set of states (ranged over by $s,q,\ldots$),
\item $\lset$ is a finite set of labels (ranged over by
  $\al,\lambda,\ldots$) and $\epsilon \not\in \lset$,
\item
  $\tset \subseteq \sset \times (\lset \cup \Set \epsilon) \times
  \sset$ is a set of transitions (where $\epsilon$ represents the
  empty word), and
\item $s_0 \in \sset$ is the initial state.
\end{itemize}
We use the usual notation  $\arro{}^*$ for the reflexive and transitive
closure of $\arro{}$.
The set of \emph{reachable states} of $A$ is
$\RS[A] = \Set{s \mid s_0\arro{}^* s}$.
For $t = (s_1,\lambda,s_2)$ we write $s_1 \arro{\lambda} s_2$ and
occasionally $t \in A$ when $t \in \tset$.

\begin{remark}
  Our definition of FSA omits the set of \emph{accepting} states since
  we consider only FSAs where each state is accepting.
We discuss this point further at the end of the paper.
  \rmkend
\end{remark}

We define on FSAs traces, trace equivalence, and bisimilarity.
\begin{definition}[Trace equivalence]
  A \emph{run} of an FSA $A = \conf{\sset, \lset, \tset, s_0}$ is a
  (possibly empty or infinite) sequence of consecutive transitions
  starting at $s_0$.
The \emph{trace} (or \emph{word}) $w$ of a run
  $(s_{i-1} \arro{\lambda_{i-1}} s_i)_{1 \leq i \leq n}$
  is\footnote{Assume $n = \infty$ if the run is infinite.} obtained by
  concatenating the labels of the run, namely
  $w = \lambda_0 \cdot \lambda_1 \cdots \lambda_n$; if the run is empty
  then $w = \epsilon$.
The \emph{language} of $A$, written $\lang{A}$, is the set of traces
  of $A$.
Two FSAs $A$ and $B$ are trace equivalent iff $\lang{A} = \lang{B}$.
We say that \emph{$A$ accepts $w$ from $s$} if
  $w \in \lang{\conf{\sset, \lset, \tset, s}}$.
\end{definition}

Since we have to deal with possibly infinite traces, it is natural to
use {\em coinduction} as the main logical tool.
We recall the notion of bisimulation.
\begin{definition}[Bisimilarity]
  A \emph{bisimulation} on two FSAs
  $A = \conf{\sset,\lset,\tset,s_0}$ and
  $A' = \conf{\sset',\lset',\tset',s'_0}$ is a
  relation $R \subseteq \sset \times \sset'$ such that if
  $(s,s') \in R$ then
  \begin{itemize}
  \item if $s \arro{\lambda} s_1$ then $s' \arro{\lambda} s'_1$ and
    $(s_1,s'_1) \in R$,
  \item if $s' \arro{\lambda} s'_1$ then $s \arro{\lambda} s_1$ and
    $(s_1,s'_1) \in R$.
  \end{itemize}
  Bisimilarity is the largest bisimulation.
\end{definition}

Recall that trace equivalence and bisimulation equivalence coincide on
deterministic labelled transition systems (see e.g.,~\cite[Theorem
2.3.12]{groote-mousavi:2014}).

\medskip

In this paper we adopt communicating finite-state machines
(CFSMs)~\cite{bz83} to model the local behaviour of distributed
components.
The following definitions are borrowed from~\cite{bz83} and adapted to
our context.

Let $\ptpset$ be a set of \emph{participants} (or \emph{roles}, ranged
over by $\p$, $\p[B]$, etc.) and $\msgset$ a set of \emph{messages}
(ranged over by $\msg$, $\msg[x]$, etc.).
We take $\ptpset$ and $\msgset$ disjoint.
\begin{definition}[Communicating system]
  A \emph{communicating finite-state machine} (CFSM) is an FSA $\aCM$
  on the set
  \begin{align*}
    \lact & = \{\aout, \ain \mid \p[A], \p[B] \in \ptpset,  \msg \in \msgset\}
  \end{align*}
  of \emph{actions}.
The \emph{subject} of an output (resp. input) action $\aout$
  (resp. $\ain$) is $\p$ (resp. \q).
Machine $\aCM$ is \emph{$\p$-local} if all its transitions have
  subject $\p$.

  A \emph{(communicating) system} is a map
  $\aCS = (\aCM_{\p})_{\p \in \ptpset}$ assigning a $\p$-local CFSM
  $\aCM_{\p}$ to each participant $\p \in \ptpset$.
\end{definition}

We consider the synchronous semantics for communicating systems as a
transition system labelled in a set of \emph{interactions}\\
 \centerline{
    $\lint \ =\  \{\gint[] \mid \p \neq  \q \in \ptpset, \msg \in \msgset\}$
  }
defined as follows.

\begin{definition}[Synchronous semantics]
  Let $\aCS = (\aCM_{\p})_{\p \in \ptpset}$ be a communicating system
  where
  $\aCM_{\p} = \conf{\sset_{\p}, \lact, \tset_{\p}, q_{0\p}}$
  for each participant $\p \in \ptpset$.
A \emph{(synchronous) configuration} for $\aCS$ is a map
  $\vec q = (q_{\p})_{\p \in \ptpset}$ assigning a \emph{local state}
  $q_{\p}\in \sset_{\p}$ to each participant $\p \in \ptpset$.
We denote $q_{\p}$ with $\pstate{q}{\p}$.

  The \emph{(synchronous) semantics} of $\aCS$ is the transition
  system $\ssem{\aCS}=\conf{\sset,\lint,\tset,\vec q_0}$ defined
  as follows
  \begin{itemize}
  \item $\sset$ is the set of synchronous configurations and
    $\vec q_0=(q_{0\p})_{\p \in \ptpset} \in \sset$ is the
    \emph{initial} configuration
  \item $\vec q_1 \arro{\gint[]} \vec q_2 \in \tset$ if
    \begin{itemize}
    \item $\pstate{q_1}{\p} \trans[\p]{\aout} \pstate{q_2}{\p}$ and
      $\pstate{q_1}{\p[B]} \trans[{\p[B]}]{\ain} \pstate{q_2}{\p[B]}$,
      and
    \item for all $\p[C] \neq \p,\q$,
      $\pstate{q_1}{\p[C]} = \pstate{q_2}{\p[C]}$
    \end{itemize}
    In this case, we say that $\pstate{q_1}{\p} \trans[\p]{\aout} \pstate{q_2}{\p}$ and
      $\pstate{q_1}{\p[B]} \trans[{\p[B]}]{\ain} \pstate{q_2}{\p[B]}$ are component transitions of $\vec q_1 \arro{\gint[]} \vec q_2$.
  \item $\vec{q_1} \trans[\p]{\epsilon} \vec{q_2} \in \tset$\hspace{6pt} if\hspace{6pt}
    $\pstate{q_1}{\p} \trans[{\p}]{\epsilon} \pstate{q_2}{\p}$, and
    for all $\q \neq \p$,
    $\pstate{q_1}{\q} = \pstate{q_2}{\q}$
  \end{itemize}
  If $\vec q_1 \arro{\alpha} \vec q_2$, we say that 
$\vec{q_2}$ is \emph{(synchronously) reachable from the
    configuration $\vec{q_1}$ by firing the transition $\alpha$}.
\end{definition}

\section{Choreography Automata}\label{sec:c-automata}
We introduce \emph{choreography automata} (c-automata) as an
expressive and flexible model of global specifications, following the
styles of choreographies~\cite{BravettiZ07,WS-CDL,BPMN}, global
graphs~\cite{gt18} and multiparty session types~\cite{HYC08,survey}.
As customary in choreographic frameworks, we show how to project
c-automata on local specifications.
As anticipated, our projections yield CFSMs formalising the local
behaviour of the participants of a choreography.

C-automata (ranged over by $\chora$, $\chora[B]$, etc.) are FSAs with
labels in $\lint$.
\begin{definition}[Choreography automata]\label{def:chorautomata}
  A \emph{choreography automaton} (c-automaton) is an $\epsilon$-free
  FSAs on the alphabet $\lint$.
Elements of $\lint^*$ are \emph{choreography words}, subsets of
  $\lint^*$ are \emph{choreography languages}.
\end{definition}

Admitting $\epsilon$-transitions in c-automata would force us to
deal with specifications where participants may take inconsistent
decisions. For instance, consider the following patterns
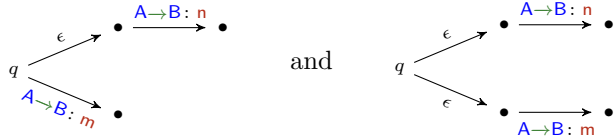
\[
  \begin{tikzpicture}[->,shorten >=1pt,auto,node distance=.2cm and 1cm]
    \tikzstyle{every edge}=[carrow]
\node (q) {$q$};
    \node[above right = of q] (q1) {$\bullet$};
    \node[below right = of q] (q2) {$\bullet$};
    \node[right = of q1] (q3) {$\bullet$};
\path (q) edge node[above] {$\epsilon$} (q1)
         (q) edge node[below] {$\gint[]$} (q2);
   \path (q1) edge node[above] {$\gint[][a][n][b]$} (q3);
 \end{tikzpicture}
 \qqand
  \begin{tikzpicture}[->,shorten >=1pt,auto,node distance=.2cm and 1cm]
    \tikzstyle{every edge}=[carrow]
\node (q) {$q$};
    \node[above right = of q] (q1) {$\bullet$};
    \node[below right = of q] (q2) {$\bullet$};
    \node[right = of q1] (q3) {$\bullet$};
    \node[right = of q2] (q4) {$\bullet$};
\path (q) edge node[above] {$\epsilon$} (q1)
         (q) edge node[below] {$\epsilon$} (q2);
   \path (q1) edge node[above] {$\gint[][a][n][b]$} (q3);
   \path (q2) edge node[below] {$\gint[]$} (q4);
 \end{tikzpicture}
\]
At state $q$, \p\ and \q\ should locally decide which message to
exchange; for instance, \p\ may decide for message $\msg$ while
\q\ may choose for message $\msg[n]$ leading to miscommunications.
Note that we still admit non-determinism in c-automata.

\newcommand{\weclock}{
  \begin{tikzpicture}[node distance=1.9cm]
    \tikzstyle{every state}=[cnode]
    \tikzstyle{every edge}=[carrow]
\node[state] (zero) {$0$};
    \node[state] (one) [right   of=zero]   {$1$};
    \node[draw=none,fill=none] (start) [left  = 0.3cm  of zero]{$\chora[C]$};
    \node[state] (two) [below of = one] {$2$};
\path
    (start) edge node {} (zero) 
    (zero) edge node[above] {$\gint[][r][tick][d]$} (one)
    (one) edge node[above] {$\gint[][d][count][S]$} (two)
    (two) edge  node[below]  {$\gint[][f][tock][r]$} (zero)
    ;
  \end{tikzpicture}
}
\newcommand{\wevalidator}{
  \begin{tikzpicture}[node distance=1.9cm]
    \tikzstyle{every state}=[cnode]
    \tikzstyle{every edge}=[carrow]
\node[state] (zero)     {$0$};
    \node[state] (one)  [right  of=zero]    {$1$};
    \node[draw=none,fill=none] (start) [above left = 0.3cm  of zero]{$\chora[V]$};
    \node[state] (two) [right  of=one] {$2$};
    \node[state] (three) [below right of=two,xshift=4mm,yshift=4mm] {$3$};
    \node[state] (four) [above right of=two,xshift=4mm,yshift=-4mm] {$4$};
    \node[state] (five) [right  of=four] {$5$};
\path (start) edge node {} (zero) 
    (zero) edge node [above] {$\gint[][C][tick][Q]$} (one)
    (one) edge node [above] {$\gint[][Q][text][H]$} (two)
    (two) edge node  [below]  {$\gint[][H][ack][Q]$} (three)
    edge node [above] {$\gint[][H][nack][Q]$} (four)
    (three) edge [bend left=25]       node [above] {$\gint[][Q][ack][I]$} (zero)
    (four) edge node [below] {$\gint[][Q][text][I]$} (five)
    (five) edge [bend right=33]     node [above] {$\gint[][I][text][Q]$} (zero)
    ;
  \end{tikzpicture}
}
\newcommand{\wepublisher}{
  \begin{tikzpicture}[node distance=1.9cm]
    \tikzstyle{every state}=[cnode]
    \tikzstyle{every edge}=[carrow]
\node[state] (one)  {$1$};
    \node[draw=none,fill=none] (start) [above left = 0.3cm  of one]{$\chora[P]$};
    \node[state] (two) [right of=one] {$2$};
    \node[state] (three) [below right of=two,xshift=4mm,yshift=4mm] {$3$};
    \node[state] (four) [above right of=two,xshift=4mm,yshift=-4mm] {$4$};
    \node[state] (five) [above right of=one]    {$5$}; 
    \node[state] (six) [above right of=three,yshift=-4mm,xshift=4mm] {$6$};
    \node[state] (seven) [right of=six] {$7$};
    \node[state] (eight) [right of=seven] {$8$};
    \node[state] (nine) [below right of=eight,xshift=4mm,yshift=4mm] {$9$};
    \node[state] (ten) [below of=seven] {$10$};
    \node[state] (eleven) [above right of=eight,xshift=4mm,yshift=-4mm] {$11$};
    \node[state] (twelve) [above right of=seven] {$12$};
\path  (start) edge node {} (one) 
    (one) edge node [above] {$\gint[][K][text][A]$} (two)
    (two) edge node [above] {$\gint[][A][ack][K]$} (three)
    edge node [above] {$\gint[][A][nack][K]$} (four)
    (three) edge node[above] {$\gint[][A][tock][E]$} (six)
    (seven) edge node[above] {$\gint[][K][text][B]$} (eight)
    (eight) edge node[above]  {$\gint[][B][ack][K]$} (nine)
    edge node [above]  {$\gint[][B][nack][K]$} (eleven)
    (ten) edge [bend left=20]        node [above,pos=0.15] {$\gint[][B][go][A]$} (one)
    (five) edge [bend right=30]     node [above] {$\gint[][A][wait][B]$} (one)
    (four) edge [bend right=30]     node [above] {$\gint[][A][tock][E]$} (five)
    (twelve) edge [bend right=30]     node [above] {$\gint[][B][wait][A]$} (seven)
    (six) edge node [above] {$\gint[][A][go][B]$} (seven)
    (eleven) edge [bend right=30]      node  [above] {$\gint[][B][tock][E]$} (twelve)
    (nine) edge [bend left=20]      node  [above] {$\gint[][B][tock][E]$} (ten)
    ;
  \end{tikzpicture}
}
\begin{figure}
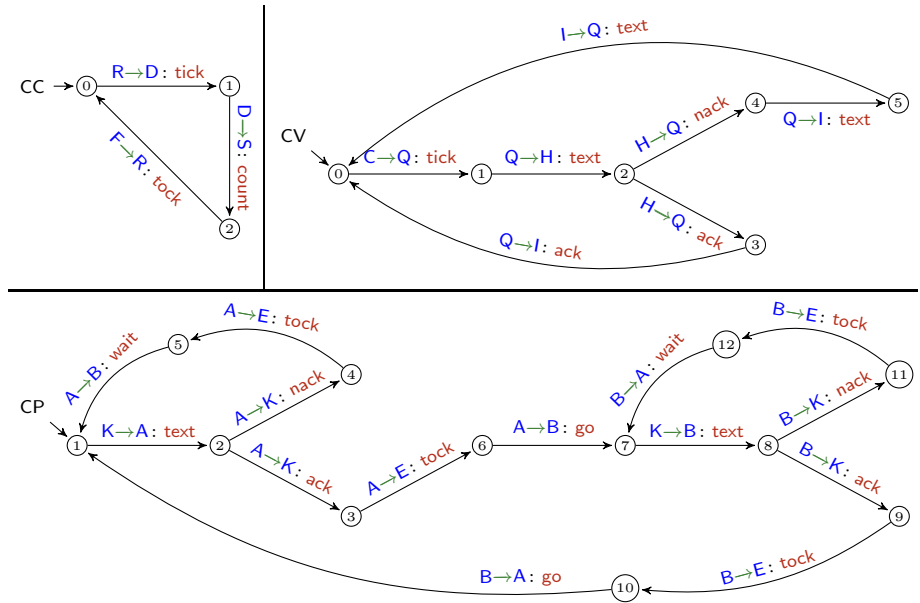
\centering
  $\begin{array}{c|c}
     \weclock
     &
     \wevalidator
     \\\hline\multicolumn 2 c {
     \wepublisher
     }
  \end{array}$
  \caption{\label{fig:automata}Three sample c-automata}
\end{figure}

\begin{example}[Working Example]\label{ex:we1}
We consider a publication system $S_{\p[p]}$ and a validation
  system $S_{\p[v]}$ in combination with a cron system $S_{\p[c]}$
  (alike the one in \cref{sec:intro}).
The c-automata $\chora[C]$, $\chora[P]$, and $\chora[V]$
  in~\cref{fig:automata} formalise the overall behaviour of
  $S_{\p[c]}$, $S_{\p[p]}$, and $S_{\p[v]}$, respectively.
\begin{description}
  \item[\textbf{System $S_{\p[c]}$}] is composed of participants \p[r],
    \p[d], \p[s], and \p[f]; it repeatedly ticks \p[d] to execute some
    tasks; before starting, \p[d] informs \p[s] on the number of
    tasks, and \p[f] informs \p[r] with message $\msg[tock]$ that the
    tasks are completed.
  \item[\textbf{System $S_{\p[v]}$}] is composed of participants \p[c], \p[q],
  \p[h], and \p[i] as follows.
    \begin{enumerate}
    \item Participant \p[h] repeatedly checks if texts received
      from \p[q] can be accepted or not, accordingly informing
      \p[q] by means of messages $\msg[ack]$ or $\msg[nack]$;
      \begin{itemize}
      \item refusal ($\msg[nack]$): \p[q] sends the
        text to participant \p[i] which modifies it and sends it
        back to \p[q] for resubmission to \p[h];
      \item acceptance ($\msg[ack]$): \p[q] informs
        \p[i] that her help is not needed.
      \end{itemize}
    \item Each cycle of the above protocol can start only when \p[q]
      does receive a $\msg[tick]$ message from participant \p[c].
    \end{enumerate}
  \item[\textbf{System $S_{\p[p]}$}] is composed of participants \p, \q,
    \p[k], and \p[e] as follows.
    \begin{enumerate}
    \item Participant \p[k] repeatedly sends text messages to,
      alternately, participants \p\ and \q\ (starting with \p), who
      can accept or refuse the texts, sending back a message
      $\msg[ack]$ or $\msg[nack]$, respectively;
      \begin{itemize}
      \item in case of acceptance ($\msg[ack]$): the participant who
        received the text sends the message $\msg[go]$ to the other
        receiver in order to inform her that it is her turn to get a
        text message from \p[k];
      \item in case of refusal ($\msg[nack]$): the participant who
        received the text, say \p, sends an $\msg[alt]$ to the other
        receiver \q\ in order to inform her that it is not her turn
        yet to get a text message from \p[k], since \p\ will keep on
        receiving texts from \p[k] until one is accepted;
      \end{itemize}
    \item Participants \p\ and \q\ inform participant \p[e], by means
      of a message $\msg[tock]$, that one of their cycles has been
      completed.\finex
    \end{enumerate}
  \end{description}
\end{example}

Given a c-automaton, our projection operation builds the corresponding
communicating system consisting of the set of projections of the
c-automaton on each participant, which yields a CFSM.

\begin{definition}[Projection]\label{def:projection}
  The \emph{projection on \p} of a transition $t = s \arro{\alpha} s'$
  of a c-automaton, written $\proj{t}{A}$, is defined by:
  \[
    \proj{t}{A} \ = \begin{cases}        
      s \arro{\aout[A][C]} s' & \text{ if } \alpha = \gint[][B][@][C] \land \q=\p
      \\
      s \arro{\ain[B][A]} s' &  \text{ if } \alpha = \gint[][B][@][C] \land \p[C]=\p
      \\
      s \arro{\epsilon} s' & \text{ if } \alpha = \gint[][B][@][C] \land \q, \p[C] \neq \p
      \\
      s \arro{\epsilon} s' & \text{ if } \alpha = \epsilon
    \end{cases}
  \]
The \emph{projection} of a c-automaton
  $\chora = \conf{\sset, \lint, \tset, s_0}$ on a participant
  $\p \in \ptpset$, denoted $\proj{\chora}{A}$, is obtained by
  minimising up-to-language equivalence the \emph{intermediate}
  automaton
  \[
    A_{\p} = \conf{\sset, \lact, \Set{s \arro{\proj{t}{A}} s' \,|\, s \arro{t} s' \in \tset}, s_0}
  \]
The \emph{projection of $\chora$}, written
  $\proj{\chora}{}$, is the communicating system
  $(\proj{\chora}{\p})_{\p \in \ptpset}$.
\end{definition}
\begin{remark}
  The projection with respect to \p\ of a c-automaton is
  \p-local.
  \rmkend
\end{remark}
Notably, minimisation also removes $\epsilon$-transitions.
For technical reasons, we assume the process of minimisation to be
performed through the partition refinement algorithm~\cite{pt87}.
In fact, to establish the correspondence between runs of a c-automaton
and the ones of its blending (e.g., c.f \cref{lemma:bisim}) we need to
map the states of the former on the states of the latter.
This correspondence is immediate in the partition refinement algorithm
(a state of the c-automaton corresponds to the equivalence class
containing it in the minimised automaton).
Our results do not depend on the use of the partition refinement
algorithm, however, adopting a different algorithm might require to
explicitly track this correspondence.
Therefore, given $\chora=\conf{\sset,\lint,\tset,s_0}$ and
$\p\in \ptpset$, the states of $\proj{\chora}{A}$, as well as the
components of $\ssem{\proj{\chora}{}}$, are subsets of $\sset$.

\ifhideproofs
{ }
\else
Besides, by construction, we have
 \begin{fact}\label{fact:uniquestate}
   Let $\chora=\conf{\sset,\lset,\tset,s_0}$ be a c-automaton.
 \begin{enumerate}[i)]
 \item
   \label{fact:uniquestate-i}
Given $s \in \RS[\chora]$, there is a unique configuration $\vec q
\in \RS[\ssem{\proj{\chora}{}}]$
   such that
   $s \in \pstate{q}{\p}$ for each $\p \in \ptpset$.
 \item
   \label{fact:uniquestate-ii}
   For any configuration $\vec{q} \in \RS[\ssem{\proj{\chora}{}}]$, there exists
   $s\in \RS[\chora]$  such that for each $\p \in \ptpset$ we have
   $s \in \pstate{q}{\p}$.
 \end{enumerate}
\end{fact}
\fi

\begin{example}[Working example: projections]\label{ex:projectionwe}
  The projections of \p[c] are
  
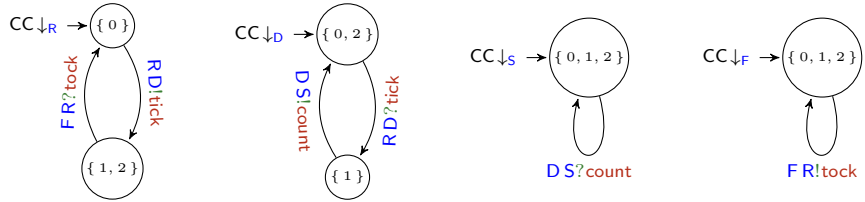
\[
    \begin{tikzpicture}[node distance=1.9cm]
      \tikzstyle{every state}=[cnode]
      \tikzstyle{every edge}=[carrow]
\node[state] (zero) {$\Set 0$};
      \node[state] (one) [below of=zero]   {$\Set{1,2}$};
      \node[draw=none,fill=none] (start) [left  = 0.3cm  of zero]{$\proj{\chora[C]}{r}$};
\path
      (start) edge node {} (zero) 
      (zero) edge[bend left] node[above] {$\aout[r][d][][tick]$} (one)
      (one) edge[bend left] node[above] {$\ain[f][r][][tock]$} (zero)
      ;
    \end{tikzpicture}
    \qquad
    \begin{tikzpicture}[node distance=1.9cm]
      \tikzstyle{every state}=[cnode]
      \tikzstyle{every edge}=[carrow]
\node[state] (zero) {$\Set{0,2}$};
      \node[state] (one) [below of=zero]   {$\Set 1$};
      \node[draw=none,fill=none] (start) [left  = 0.3cm  of zero]{$\proj{\chora[C]}{d}$};
\path
      (start) edge node {} (zero) 
      (zero) edge[bend left] node[below] {$\ain[r][d][][tick]$} (one)
      (one) edge[bend left] node[below] {$\aout[d][s][][count]$} (zero)
      ;
    \end{tikzpicture}
    \qquad
    \begin{tikzpicture}[node distance=1.9cm]
      \tikzstyle{every state}=[cnode]
      \tikzstyle{every edge}=[carrow]
\node[state] (zero) {$\Set{0,1,2}$};
      \node[draw=none,fill=none] (start) [left  = 0.3cm  of zero]{\proj{$\chora[C]}{S}$};
\path
      (start) edge node {} (zero) 
      (zero)[loop below,looseness=10] edge node[below] {$\ain[d][s][][count]$} (zero)
      ;
    \end{tikzpicture}
    \qquad
    \begin{tikzpicture}[node distance=1.9cm]
      \tikzstyle{every state}=[cnode]
      \tikzstyle{every edge}=[carrow]
\node[state] (zero) {$\Set{0,1,2}$};
      \node[draw=none,fill=none] (start) [left  = 0.3cm  of zero]{\proj{$\chora[C]}{f}$};
\path
      (start) edge node {} (zero) 
      (zero)[loop below,looseness=10] edge node[below] {$\aout[f][r][][tock]$} (zero)
      ;
    \end{tikzpicture}
  \]
Note that the states are subsets of states of $\chora[C]$ due to minimisation and that
  all projections do not have $\epsilon$-transitions.
\finex
\end{example}

The projection operation is well-behaved with respect to trace equivalence.
\begin{proposition}
  If $\chora$ and $\chora'$ are trace-equivalent c-automata then\linebreak
  $\proj{\chora}{A}=\proj{\chora'}{A}$ for each participant
  $\p \in \ptpset$.
\end{proposition}
\begin{proof}
  Let us consider the intermediate automata $A_{\p}$ and $A_{\p[A']}$
  in the projection of $\chora$ and $\chora'$, respectively. If
  $A_{\p}$ and $A_{\p[A']}$ are trace-equivalent then their minimal
  automaton up to trace equivalence is of course the same. We have
  to show that $A_{\p}$ and $A_{\p[A']}$ are trace equivalent. Take
  a word $w_{\p}$ in the language of $A_{\p}$. It is obtained from
  some computation that accepts a word $w_{\chora}$ in $\chora$. By
  trace equivalence, the same word is accepted in
  $\chora'$. Executing the computation generating it in $A_{\p[A']}$
  will produce $w_A$. This proves the thesis.
\qed
\end{proof}

Thanks to the result above, without loss of generality, we can assume
that c-automata are minimal up-to language equivalence (in particular,
they do not have $\epsilon$-transitions and they are deterministic).

The following proposition relates the language of the projection with
the language of the original automaton.

\begin{proposition}
  For all c-automata $\chora$ and $\p \in \ptpset$,
  $\lang{\proj{\chora}{\p}} = \proj{\lang{\chora}}{\p}$.
\end{proposition}
\begin{proof}
Let $A_{\p}$ be the intermediate automaton built in the first step in the projection of $\chora$ on $\p$.
We have
$$\lang{\proj{\chora}{\p}} = \lang{A_{\p}} =  \proj{\lang{\chora}}{\p}$$
where the first equality follows since $\proj{\chora}{\p}$ and
$A_{\p}$ are trace-equivalent by definition, and the second
equality follows since $A_{\p}$ is obtained transition per transition
by projecting the labels of $\chora$ on $\p$.
\qed
\end{proof}

\section{Well-formed Choreography Automata}\label{sec:wf}

In order to ensure that the projection of a c-automaton on a
communicating system is well-behaved we require some conditions on the
c-automaton.
We first define the notion of concurrent transitions.

\begin{definition}[Concurrent transitions]
  Let $\chora=\conf{\sset,\lset,\tset,s_0}$.  Two transitions $s
  \arro{\al_1} s_1$ and $s \arro{\al_2} s_2$ are
  concurrent iff there is a state $s_3 \in \sset$ and transitions $s_1
  \arro{\al_2} s_3$ and $s_2 \arro{\al_1} s_3$.
\end{definition}

We can now define the key notion of well-branchedness, which
intuitively states that each participant is aware of \quo{what it needs to
  do in the current state}.
In other words, each participant is aware of choices made by others
when its behaviour depends on those choices.
The awareness of choice is checked on \emph{spans}, namely runs that
may constitute alternative branches of choices.
Spans are formalised as follows.
\begin{definition}[$s$-span]
  Given a state $s$ of a c-automaton $\chora$, a pair of runs
  $(\pi_1,\pi_2)$ is an \emph{$s$-span} if $\pi_1$ and $\pi_2$ are
  coinitial at $s$, transition-disjoint, acyclic, and either $\pi_1$
  and $\pi_2$ are cofinal otherwise they are both maximal in $\chora$
  and share only the state $s$.
\end{definition}

\begin{example}[Working example: spans]\label{ex:spanswe}
  In the c-automaton $\chora[V]$ the only state with spans is state
  $2$.
The unique $2$-span is $(\pi,\pi')$ where
  \[\begin{tikzpicture}[node distance=1.5cm]
      \tikzstyle{every state}=[cnode]
      \tikzstyle{every edge}=[carrow]
\node (pi1) {$\pi =$};
      \node[state, right = -.05cm of pi1] (two) {$2$};
      \node[state, right = of two] (three) {$3$};
      \node[state, right = of three] (zero) {$0$};
\path
      (two) edge node [above]  {$\gint[][H][ack][Q]$} (three)
      (three) edge node [above] {$\gint[][Q][ack][I]$} (zero)
      ;
\node[right = of zero] (pi2) {$\pi' =$};
      \node[state, right = -.05cm of pi2] (2) {$2$};
      \node[state, right = of 2] (4) {$4$};
      \node[state, right = of 4] (5) {$5$};
      \node[state, right = of 5] (0) {$0$};
\path
      (2) edge node [above] {$\gint[][H][nack][Q]$} (4)
      (4) edge node [above] {$\gint[][Q][text][I]$} (5)
      (5) edge node [above] {$\gint[][I][text][Q]$} (0)
      ;
    \end{tikzpicture}
  \]
  Indeed, $\pi$ and $\pi'$ are coinitial, cofinal, acyclic, and
  transition disjoint.
Note that any other pair of paths from $2$ in not a $2$-span since
  the transition from $0$ to $1$ of $\chora[V]$ would make the paths
  non-cofinal and either make the paths non transition-disjoint or
  have the state $0$ on both paths.
\finex
\end{example}

Intuitively, a choice is well-branched when it has a determined
selector, and the other participants either behave uniformly in each
branch, or can ascertain which branch has been chosen from the
messages they receive.
\begin{definition}[Well-branchedness]\label{def:wb}
  A c-automaton $\chora$ is \emph{well-branched} if for each state $s$
  in $\chora$ and participant $\q \in \ptpset$ sender in a
  transition from $s$:
  \begin{enumerate}[(1)]
  \item \label{it:par1} all transitions from $s$ involving \q, have sender \q\ and
    pairwise distinct labels;
  \item \label{it:par2} for each transition $t$ from $s$ whose sender is not \q\ and
    each transition $t'$ from $s$ whose sender is \q, $t$ and $t'$
    are concurrent
  \item \label{it:inp} for each participant $\p \neq \q \in \ptpset$
    and $s$-span $(\pi_1,\pi_2)$,
the first pair of different actions in $\proj{\pi_1}{A}$ and
    $\proj{\pi_2}{A}$ (if any) is of the form
    $(\ain[C][A][@][m],\ain[D][A][@][n])$ with $\p[C] \neq \p[D]$ or
    $\msg \neq \msg[n]$.
\end{enumerate}
  We dub \q\ a \emph{selector at $s$}.
\end{definition}

\begin{example}[Working example: well-branchedness]\label{ex:wbwe}
  In~$\chora[V]$ all the states satisfy the conditions of
  \cref{def:wb}; the only non-trivial case is state $2$.
Condition~\eqref{it:par1} holds for \iH, which is the selector of
  the choice at $2$; condition~\eqref{it:par2} holds vacuously, and
  condition~\eqref{it:inp} holds for both \ptp[Q]\ and \p[i] in the unique
  $2$-span $(\pi,\pi')$, described in \cref{ex:spanswe}.
Indeed, the
  first actions of \ptp[q]\ on $\pi$ and $\pi'$ are the input from \iH
  which have different messages; likewise, the first actions of \p[i]
  on $\pi$ and $\pi'$ are the inputs from $\ptp[Q]$, which have different
  messages as well.
\finex
\end{example}

Condition~\eqref{it:par2}, vacuously true in the example, is needed
when multiple participants act as sender in the same state: this
ensures that the only possibility is that actions of different
participants are concurrent so that possible choices at a state are
not affected by independent behaviour.

We also rule out c-automata having consecutive transitions involving
disjoint participants and not independent of each other.
\begin{definition}[Well-sequencedness]
\label{def:wellseq}
  A c-automaton $\chora$ is \emph{well-sequenced} if  each pair
  of consecutive transitions $s \arro{\gint[][A][m][B]} s' \arro{\gint[][C][n][D]}
  s''$ either
  \begin{itemize}
  \item share a participant, that is
    $\{\p,\q\} \cap \{\p[C],\p[D]\} \neq \emptyset$, or
  \item are part of a diamond, i.e.  there is $s'''$ such that
    $s \arro{\gint[][C][n][D]} s''' \arro{\gint[][A][m][B]} s''$.
  \end{itemize}
\end{definition}

\noindent
\textbf{Notation.} For the sake of readability, a
well-sequenced c-automaton $\chora$ can be represented by omitting,
for each diamond, two of its consecutive transitions.
We call such representation \emph{compact}. Notice that, given a
compact representation, it is always possible to recover the original
c-automaton.
So far and hereafter we assume that all c-automata are compactly
represented.

\begin{example}[Working example]\label{ex:notationwe}
  Both $\chora[V]$ and $\chora[P]$ are well-sequenced because
  they enjoy the first condition of \cref{def:wellseq}.
Instead, the participants on the transitions from $1$ to $2$
  and from $2$ to $0$ in $\chora[C]$ are disjoint; in fact,
  we assume $\chora[C]$ to be compactly represented.
\finex
\end{example}

Well-sequencedness is necessary to establish a precise
correspondence between the language of a c-automaton and its
corresponding communicating system
(cf. \cref{th:projectionCorrectness}) as well as to ensure the
applicability of the blending operation (and hence systems
composition to work).

\begin{definition}[Well-formedness]
  A c-automaton is well-formed if it is both well-branched and
  well-sequenced.
\end{definition}

\ifhideproofs

\else
\begin{remark}
Not any c-automaton can be ``completed'' to a well-sequenced automaton,
as shown by the following example.\\
Let us consider the following c-automaton

$$
 
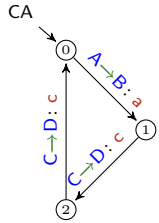
\begin{tikzpicture}[node distance=1.5cm]
      \tikzstyle{every state}=[cnode]
      \tikzstyle{every edge}=[carrow]
 
  \node[state]           (zero)                        {$0$};
   \node[draw=none,fill=none] (start) [above left = 0.3cm  of zero]{$\chora$};
     \node[state]           (one)  [below right  of =zero]                      {$1$};
       \node[state]           (two)  [below left  of =one]                      {$2$};

  \path  (start) edge node {} (zero) 
            (zero)    edge                  node [sloped,above] {$\gint[][A][a][B]$} (one)
            (one)    edge                  node [sloped,above] {$\gint[][C][c][D]$} (two)
            (two)    edge                  node [sloped,above] {$\gint[][C][c][D]$} (zero)
             ;

 \end{tikzpicture}
 $$      
 
 The above c-automaton is not well sequenced because of the transitions $0-1$ and $1-2$.
 So we complete the diamond.
 
 $$
 
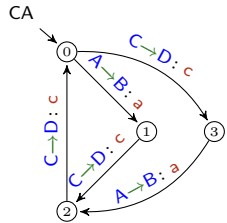
\begin{tikzpicture}[node distance=1.5cm]
      \tikzstyle{every state}=[cnode]
      \tikzstyle{every edge}=[carrow]
 
  \node[state]           (zero)                        {$0$};
   \node[draw=none,fill=none] (start) [above left = 0.3cm  of zero]{$\chora$};
     \node[state]           (one)  [below right  of =zero]                      {$1$};
     \node[state]           (two)  [below left  of =one]                      {$2$};
     \node[state]           (three)  [right  of =one,xshift=-6mm]                      {$3$};

  \path  (start) edge node {} (zero) 
            (zero)    edge                  node [sloped,above] {$\gint[][A][a][B]$} (one)
            (one)    edge                  node [sloped,above] {$\gint[][C][c][D]$} (two)
            (two)    edge                  node [sloped,above] {$\gint[][C][c][D]$} (zero)
            (zero)    edge  [bend left]          node [sloped,above] {$\gint[][C][c][D]$} (three)
          (three)    edge   [bend left]               node [sloped,above] {$\gint[][A][a][B]$} (two)  
            
             ;

 \end{tikzpicture}
 $$      
  
 The resulting automaton is still not well-sequenced, because of the transitions $3-2$ and $2-0$.
 So we complete the diamond.
 
 $$
 
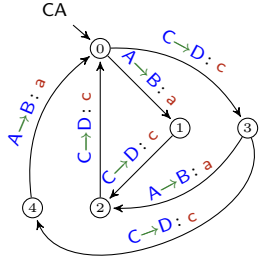
\begin{tikzpicture}[node distance=1.5cm]
      \tikzstyle{every state}=[cnode]
      \tikzstyle{every edge}=[carrow]
 
  \node[state]           (zero)                        {$0$};
   \node[draw=none,fill=none] (start) [above left = 0.3cm  of zero]{$\chora$};
     \node[state]           (one)  [below right  of =zero]                      {$1$};
     \node[state]           (two)  [below left  of =one]                      {$2$};
     \node[state]           (three)  [right  of =one,xshift=-6mm]                      {$3$};
     \node[state]           (four)  [ left of =two,xshift=--6mm]                      {$4$};

  \path  (start) edge node {} (zero) 
            (zero)    edge                  node [sloped,above] {$\gint[][A][a][B]$} (one)
            (one)    edge                  node [sloped,above] {$\gint[][C][c][D]$} (two)
            (two)    edge                  node [sloped,above] {$\gint[][C][c][D]$} (zero)
            (zero)    edge  [bend left]          node [sloped,above] {$\gint[][C][c][D]$} (three)
          (three)    edge   [bend left]               node [sloped,above] {$\gint[][A][a][B]$} (two)  
           (three)    edge   [bend left=90]               node [sloped,above] {$\gint[][C][c][D]$} (four)  
          (four)    edge   [bend left]               node [sloped,above] {$\gint[][A][a][B]$} (zero)  

             ;

 \end{tikzpicture}
 $$      
 
  The resulting automaton is still not well-sequenced, because of the transitions $4-0$ and $0-3$.
 So we complete the diamond.
 
 $$
 
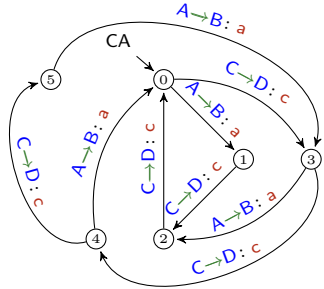
\begin{tikzpicture}[node distance=1.5cm]
      \tikzstyle{every state}=[cnode]
      \tikzstyle{every edge}=[carrow]
 
  \node[state]           (zero)                        {$0$};
   \node[draw=none,fill=none] (start) [above left = 0.3cm  of zero]{$\chora$};
     \node[state]           (one)  [below right  of =zero]                      {$1$};
     \node[state]           (two)  [below left  of =one]                      {$2$};
     \node[state]           (three)  [right  of =one,xshift=-6mm]                      {$3$};
     \node[state]           (four)  [ left of =two,xshift=--6mm]                      {$4$};
    \node[state]           (five)  [ left of =zero]                      {$5$};

  \path  (start) edge node {} (zero) 
            (zero)    edge                  node [sloped,above] {$\gint[][A][a][B]$} (one)
            (one)    edge                  node [sloped,above] {$\gint[][C][c][D]$} (two)
            (two)    edge                  node [sloped,above] {$\gint[][C][c][D]$} (zero)
            (zero)    edge  [bend left]          node [sloped,above] {$\gint[][C][c][D]$} (three)
          (three)    edge   [bend left]               node [sloped,above] {$\gint[][A][a][B]$} (two)  
           (three)    edge   [bend left=90]               node [sloped,above] {$\gint[][C][c][D]$} (four)  
          (four)    edge   [bend left]               node [sloped,above] {$\gint[][A][a][B]$} (zero)  
           (four)    edge   [bend left=90]               node [sloped,above] {$\gint[][C][c][D]$} (five)  
          (five)    edge   [bend left=90]               node [sloped,above] {$\gint[][A][a][B]$} (three)  

             ;

 \end{tikzpicture}
 $$      
 
  The resulting automaton is still not well-sequenced, because of the transitions $5-3$ and $3-4$.
 So we complete the diamond.... And we can go on forever.\\
 
 The impossibility to complete the initial c-automaton depends on the fact that
 the intended completed automaton should generate a non regular language (since we should 
 generate strings with a number of $\gint[][C][c][D]$ interaction which is, roughly, double of
 the number of $\gint[][A][a][B]$ interactions.
It would hence be interesting to know whether, in case the expected completed interaction language of
a c-automaton is regular and prefix-closed, it is possible to generate it also by means of a well-formed c- automaton.  It would be also interesting to know whether a condition can be given on cycles such that the completion of a c-automaton is possible.
\rmkend
\end{remark}
\fi
 \subsection{Languages and well-formedness }\label{sec:wflang}
We show a closure property of the language of well-sequenced
c-automata.
\begin{definition}[Commutation of independent interactions]
  Two  choreography words $w$ and $w'$ are
  \emph{equal up-to-commutation of independent interactions}, written
  $w \comm w'$, if one can be obtained from the other by repeatedly
  swapping consecutive interactions with disjoint sets of
  participants.

  Given a choreography language $L$
  \[
    \close{L} = \Set{w \in \lint \sst \exists w' \in L .\ w \comm w' }
  \]
  is the \emph{closure under commutation of independent interactions} of
  $L$.
\end{definition}
\begin{proposition}\label{prop:wsimpliesclosed}
  Let $\chora$ be a well-sequenced c-automaton.  
  Then $\lang{\chora}$ is closed under
    commutation of independent actions, i.e. $\lang{\chora} = \close{\lang{\chora}}$.
\end{proposition}
\begin{proof}
  Let us assume $\lang{\chora} \neq \close{\lang{\chora}}$.  Clearly,
  $\lang{\chora} \subseteq \close{\lang{\chora}}$.  Then there exist
  $w' \in \close{\lang{\chora}}$ such that $w' \not\in
  \lang{\chora}$. By definition of closure, there is
  $w \in \lang{\chora}$ such that $w \comm w'$.
By definition of $\comm$, there exist two choreographic words $w_1$
  and $w_2$ such that
  $w \comm w_1\cdot(\gint[][A][b][B])\cdot(\gint[][C][c][D])\cdot w_2
  \in \lang{\chora}$ and
  $w_1\cdot (\gint[][C][c][D])\cdot(\gint[][A][b][B])\cdot w_2 \not\in
  {\lang{\chora}}$, where
  $\Set{\ptp[A],\ptp[B]}\cap\Set{\ptp[C],\ptp[D]}=\emptyset$.  This
  implies that $\chora$ is not well-sequenced. In fact, if that were
  not the case, let $s$ and $s'$ be the states of $\chora$ from which
  the subwords $(\gint[][A][b][B])\cdot(\gint[][C][c][D])$ and $w_2$
  are, respectively, generated.  By well-sequencedness, there is also
  a path from $s$ to $s'$ generating
  $(\gint[][C][c][D])\cdot(\gint[][A][b][B])$, so implying that
  $w_1\cdot (\gint[][C][c][D])\cdot(\gint[][A][b][B])\cdot w_2 \in
  {\lang{\chora}}$, which is a contradiction.
\qed
\end{proof}

Notice that the converse of the above proposition does not hold
without the assumption of implicit diamonds.
In fact, consider the following c-automaton
\[
  \begin{tikzpicture}[node distance=1.9cm]
    \tikzstyle{every state}=[cnode]
    \tikzstyle{every edge}=[carrow]
\node[state] (zero)                        {$0$};
    \node[state] (one)   [above right   of=zero,yshift=-8mm,xshift=4mm]   {$1$};
    \node[draw=none,fill=none] (start) [above left  = 0.3cm  of zero]{$\chora[A]$};
    \node[state] (two) [right  of=one] {$2$};
    \node[state] (three)   [below right   of=zero,yshift=8mm,xshift=4mm]   {$3$};
    \node[state] (four) [right  of=three] {$4$};
\path (start) edge node {} (zero) 
   (zero) edge node [above] {$\gint[][A][a][B]$} (one)
   edge node [below] {$\gint[][C][c][D]$} (three)
   (one) edge node [above] {$\gint[][C][c][D]$} (two)
   (three) edge node  [below]  {$\gint[][A][a][B]$} (four)          
   ;
 \end{tikzpicture}
\]
we can check that $\lang{\chora} = \close{\lang{\chora}}$ but $\chora$
is not well-sequenced.

The next result establishes that the language of a well-formed c-automaton
coincides with the language of the communicating system made of its
projection.
This provides a correctness criterion for our projection operation.
\ifhideproofs
{}
\else
The proof requires to reason coinductively to account for possibly infinite runs.
In particular, we adopt the coinduction style advocated in
\cite{KozenS17} which, without any loss of formal rigour, promotes
readability and conciseness.
\fi

\begin{theorem}\label{th:projectionCorrectness}
  If $\chora$ is a well-formed
  c-automaton then
  $\lang{\ssem{\proj{\chora}{}}} = \lang{\chora}$.
\end{theorem}
\begin{proof}
  We first show that $\lang{\chora} \subseteq
  \lang{\ssem{\proj{\chora}{}}}$.  \\
  Let us take a trace $\sigma \in
  \lang{\chora}$. We need to show that $\sigma \in
  \lang{\ssem{\proj{\chora}{}}}$. We shall actually prove a stronger thesis:
  \begin{quote}
  if $\chora$ accepts $\sigma$ starting from a state $s$,
  then $\ssem{\proj{\chora}{}}$ accepts $\sigma$ starting from the unique
  (by \cref{fact:uniquestate}\eqref{fact:uniquestate-i})
  state $\vec q=(q_{\p})_{\p \in \ptpset}$ such that for each
  $\p \in \ptpset$ we have $s \in \pstate{q}{\p}$ .
  \end{quote}

\noindent
  The proof is by coinduction. If $\sigma$ is empty, there is nothing
  to prove. \\
  Otherwise, let $\sigma$ be $(\gint[][B][@][C])\cdot \sigma'$,
  where $\gint[][B][@][C]$ is accepted through a transition
  $s \arro{\gint[][B][@][C]} s'$ 
   and where $\sigma'$ is accepted by  $\chora$ starting from $s'$. \\
  Consider now the intermediate automata built in the first step of
  the projection of $\chora$ (\cref{def:projection}).
   By construction, 
    $s \arro{\aout[B][C]} s'$ in $A_{\q}$,
  $s \arro{\ain[B][C]} s'$ in $A_{\p[C]}$ and
  $s \arro{\epsilon} s'$ in  $A_{\p[D]}$ for each
  $\p[D] \neq \q,\p[C]$.
By coinduction,  $\sigma'$ is accepted by
  $\ssem{\proj{\chora}{}}$ starting from the unique
  state $\vec {q'}=(q'_{\p})_{\p \in \ptpset}$ such that for each
  $\p \in \ptpset$ we have $s' \in \pstate{q'}{\p}$.
  Hence, 
  by definition of minimization and by definition of 
   $\ssem{\proj{\chora}{}}$, we get that
   $\ssem{\proj{\chora}{}}$ accepts $\sigma$ starting from the 
  state $\vec q=(q_{\p})_{\p \in \ptpset}$ such that for each
  $\p \in \ptpset$ we have $s \in \pstate{q}{\p}$.\\

  We now  show that $\lang{\ssem{\proj{\chora}{}}} \subseteq
  \lang{\chora}$.  
As before, we consider a stronger thesis:
\begin{quote}
if $\ssem{\proj{\chora}{}}$ accepts $\sigma$ starting from a state $\vec
  q=(q_{\p})_{\p \in \ptpset}$, then $\chora$ accepts $\sigma$
  starting from any state $s$ such that for each $\p \in \ptpset$ we
  have $s \in \pstate{q}{\p}$. 
 \end{quote} 
 The thesis will follow since by definition of synchronous semantics
 $s_0 \in \pstate{q_{0}}{\p}$
 for each $\p \in \ptpset$.\\
 The proof is by coinduction. If $\sigma$ is empty, there is nothing
 to prove.  Otherwise, let $\sigma$ be
 $(\gint[][B][@][C])\cdot\sigma'$, where $\gint[][B][@][C]$ is
 accepted by a transition $\vec q \arro{\gint[][B][@][C]}
 \vec{q'}$. Hence, by definition of synchronous semantics, there are
 transitions $\pstate{q}{\q} \arro{\aout[B][C]} \pstate{q'}{\q}$ and
 $\pstate{q}{\p[C]} \arro{\ain[B][C]} \pstate{q'}{\p[C]}$.  Moreover,
 $\pstate{q}{\p[D]} = \pstate{q'}{\p[D]}$ for each
 $\p[D] \neq \q,\p[C]$.
  
  Let now $s$ be any state such that for each $\p \in \ptpset$ we have
  $s \in \pstate{q}{\p}$ (by \cref{fact:uniquestate}\eqref{fact:uniquestate-ii}).  
  Hence, in the intermediate automata
   built in the first step of
  the projection of $\chora$ (\cref{def:projection}),
  by definition of minimization up-to language equivalence, we have 
  \begin{equation}
  \label{eq:eps}
  s \arro{\epsilon} \cdots \arro{\epsilon} \hat s_{\q}
  \arro{\aout[B][C]} s'_{\q}
  \text{ in $A_{\q}$ and }
   s \arro{\epsilon} \cdots
  \arro{\epsilon} \hat s_{\p[C]} \arro{\ain[B][C]} s'_{\p[C]} \text{ in $A_{\p[C]}$}.
  \end{equation}

  The two transitions $\hat s_{\q} \arro{\aout[B][C]}
  s'_{\q}$ and $\hat s_{\p[C]} \arro{\ain[B][C]} s'_{\p[C]}$
  may be both obtained from the projection of the same transition of $\chora$ or not.
  Let us consider the two cases separately.
\begin{description}
\item
$\hat s_{\q} \arro{\aout[B][C]}
  s'_{\q}$ and $\hat s_{\p[C]} \arro{\ain[B][C]} s'_{\p[C]}$
  are both obtained from the projection of some transition $t: \hat s \arro{\gint[][B][@][C]} s'$ of $\chora$, 
  with $\hat s= \hat s _{\q}=\hat s_{\p[C]}$
  and $s'=s'_{\q}=s'_{\p[C]}$ .\\
  In a such a case we will show that $t$ is enabled in $s$, that is $t$ is of the
  form $s \arro{\gint[][B][@][C]} s'$, with
  $s'=s'_{\q}=s'_{\p[C]}$. 
  By contradiction, let us assume that not to be the
  case, and let $\pi_1$ be  one of the  runs with minimal length $n>0$ from
  $s$ to a state $\hat s$ such that $\hat s \arro{\gint[][B][@][C]}
  \hat s'$. 
   Let us call $t'$ the last transition in
  $\pi_1$.  By well-sequencedness (\cref{def:wellseq}) $t'$ involves either
  $\q$ or $\p[C]$, or it is concurrent to $t$.
  \begin{itemize}
  \item If $t'$ is
    concurrent with $t$,\\
    then there is a transition with label $\gint[][B][@][C]$ coinitial
    with $t'$. This implies the existence of a run shorter than
    $\pi_1$, which was assumed to be minimal, and hence a
    contradiction.
  \item
   If $t'$ is not concurrent with  $t$, 
   then $t'$ should
  involve either $\q$ or $\p[C]$. Let us assume it to involve
  $\q$ but not $\p[C]$, the other cases are analogous. 
   By~\eqref{eq:eps}, we  know that 
    there is a  run 
    $s \arro{\epsilon} \cdots \arro{\epsilon} \hat s_{\q}$
   in $A_{\q}$ 
   and a run $s \arro{\epsilon} \cdots \arro{\epsilon} \hat s_{\p[C]}$
   in $A_{\ptp[C]}$
   both with all labels $\epsilon$
   and such that $\hat s(=\hat s_{\q}=\hat s_{\p[C]})$  enables $t$ in $\chora$.
   Because of the projection on $\q$,
   there exists a run $\pi_2$, in $\chora$, not involving $\q$.
   We can assume
   that the last transition of $\pi_2$ is not concurrent to $t$, as above.
   Thus, thanks to well-sequencedness, 
   it necessarily ends
   with a transition involving $\p[C]$.
   We can also assume $\pi_2$ to have minimal length. \\
Note that $\pi_1$ and $\pi_2$ are coinitial, cofinal, and different
  since one contains transitions involving $\q$ but not $\p[C]$
  and the other one contains transitions involving $\p[C]$ but not
  $\q$. Since we assumed that they are minimal they are also
  acyclic. By removing common transitions and selecting any of the
  remaining run fragments $\pi'_1$ of $\pi_1$ involving $\q$ and
  the corresponding fragment $\pi'_2$ of $\pi_2$ (which does not
  involve $\q$) we obtain runs which are coinitial, cofinal,
  acyclic and transition disjoint. By construction, $\proj{{\pi'_1}}{B}$
  and $\proj{\pi'_2}{B}$ accept different words (since the second
  is empty while the first is not) and the first pair of different
  characters is not an input for $\q$ in both the cases (since in
  $\pi'_2$ no such character exists). However, this contradicts
  well-branchedness hence this case can never happen.
  \end{itemize}
  Thus, there is a transition $s \arro{\gint[][B][@][C]} s'$ in
  $\chora$.  The state $s'$ is such that $s' \in \pstate{q'}{\p}$
  for each $\p \in \ptpset$, thanks to the properties of $s$ and of
  the definition of synchronous semantics. By coinduction, this state
  accepts $\sigma'$, and the thesis follows.
\item $\hat s_{\q} \arro{\aout[B][C]} \hat s'_{\q}$ and
  $\hat s_{\p[C]} \arro{\ain[B][C]} \hat s'_{\p[C]}$ are obtained by
  projecting two different transitions $t_1$ and $t_2$ in $\chora$
  with the
  same label $\gint[][B][@][C]$. \\
  Since $\chora$s are deterministic (\cref{def:chorautomata}), there
  are two different runs $\pi_1$ and $\pi_2$ in $\chora$, the former
  ending with $t_1$ and the latter ending with $t_2$.  We consider now
  two possible subcases.
  \begin{itemize}
  \item
  If $\proj{\pi_1}{B}=\proj{\pi_2}{B}$ and
  $\proj{\pi_1}{C}=\proj{\pi_2}{C}$,\\
   then the two runs project to the
  same run in $\proj{\chora}{B}$ and $\proj{\chora}{C}$. As a result,
  any of the two transitions can be used to answer the challenge,
  and the thesis follows by coinduction 
  as above.
  \item
  If instead the projections on either $\q$ or $\p[C]$ are
  different,\\
   then we can remove common transitions and cycles and
  select any of the pairs of run fragments (which are coinitial,
  transition disjoint and acyclic) highlighting the difference in the
  projection. If they are not the final fragments, then they are also
  cofinal, otherwise we can extend them till they become cofinal or
  both maximal. By well-branchedness the first action involving the
  role on which the projections differ should be an input on both the
  runs, and the two inputs should be different. If the projection is
  different on $\q$, since we assumed that $\gint[][B][@][C]$ is
  the first action involving $\q$, then we have a contradiction
  (since it is an output). If the projection is different on
  $\p[C]$, then $\gint[][B][@][C]$ is the first action involving
  $\p[C]$, and it is the same on both runs, against the hypothesis
  that the two projections on $\p[C]$ were different.

  Anyway, this case cannot happen.
\qed
\end{itemize}
\end{description}
\end{proof}

 \subsection{Communication soundness and well-formedness }\label{sec:wfsound}
In this section we show that the projection of well-formed
choreography automata enjoys relevant correctness properties.

The projection of a well-formed c-automaton is live~\cite{ScalasY19}:
if there is a participant willing to take a transition, there is a
computation executing such transition.

\begin{proposition}[Liveness]\label{prop:live}
  Let $\chora$ be a well-formed c-automaton.
For each $\vec q \in \RS[{\ssem{\proj{\chora}{}}}]$, if there is $\p
  \in \ptpset$ with an outgoing transition $t$ from $\pstate{q}{\p}$
  in $\proj{\chora}{\p}$, then there exists a run of
  $\ssem{\proj{\chora}{}}$ including a transition which has $t$ as
  component transition.
\end{proposition}
\begin{proof}
  Since $\vec q$ is reachable, there is a run to $\vec q$ producing
  some word $w$. Consider the unique state $q_s$ of $\chora$ such that
  $q_s \in \pstate{q}{\ptp}$ for each $\ptp \in \ptpset$. The
  c-automaton $\chora_s$ obtained from $\chora$ by setting $q_s$ as
  initial state is well-formed. Now, assume there exists $\p$ such
  that $\pstate{q}{\p}$ has an outgoing transition $t$. Thanks to the
  definition of projection, there is a transition $t'$ in $\chora_s$
  involving $\p$ and another participant, say $\q$.  Such a transition
  is on a run with word $w'$.
A run with word $w'$ exists also in $\ssem{\proj{\chora}{}}$ thanks
  to \cref{th:projectionCorrectness}, it goes through $\vec q$, since
  the state reached is fully determined by the label (since
  participants are deterministic), and it executes $t$ as
  desired. This proves the thesis.
\qed
\end{proof}

\cref{prop:lock} below shows that the projection of a well-formed
c-automaton is lock-free~\cite{survey}, namely that enabled
transitions of a participant are sooner or later fired.
We formalise locks first.
(Recall that the transitions of projected CFSMs are minimal up-to
language equivalence and hence deterministic.)
\begin{definition}[Lock]
  Let $\chora$ be a c-automaton.
A state $\vec q \in \RS[{\ssem{\proj{\chora}{}}}]$ is a \emph{lock}
  if there is $\p \in \ptpset$ with an outgoing transition $t$ from
  $\pstate{q}{\p}$ in $\proj{\chora}{\p}$, yet in all runs of
  $\ssem{\proj{\chora}{}}$ starting from $\vec q$ there is no
  transition $t'$ involving $\p$.
\end{definition}

\begin{proposition}[Lock freedom]\label{prop:lock}
  Let $\chora$ be a well-formed c-automaton.
For each $\vec q \in \RS[{\ssem{\proj{\chora}{}}}]$, $\vec q$ is not a lock.
\end{proposition}
\begin{proof}
  Since $\vec q$ is reachable, there is a run to $\vec q$ producing
  some word $w$. Consider the unique state $q_s$ of $\chora$ such that
  $q_s \in \pstate{q}{\ptp}$ for each $\ptp \in \ptpset$. The
  c-automaton $\chora_s$ obtained from $\chora$ by setting $q_s$ as
  initial state is well-formed. Now, assume there exists $\p$ such
  that $\pstate{q}{\p}$ has an outgoing transition $t$. Thanks to the
  definition of projection, there is a transition $t'$ in $\chora_s$
  involving $\p$ and another participant, say $\q$.  Transition $t'$
  is on a run with word $w'$ in $\chora_s$, and thanks to
  \cref{th:projectionCorrectness} also in
  $\ssem{\proj{\chora_s}{}}$. We have to show that each run in
  $\ssem{\proj{\chora_s}{}}$ has a transition involving $\p$.  This
  will imply the thesis.

  Assume towards a contradiction that there is a run not
  involving $\p$. Such a run and the one including $t'$ are
  coinitial. By removing common transitions they can be made
  transition disjoint, and by removing cycles they can also be taken
  cycle free. Thanks to well-branchedness, since one of the two paths
  contains a transition with $\ptp$, the other one also need to
  include a transition with $\ptp$, against the hypothesis. This
  proves the thesis.
\qed
\end{proof}
Interestingly, \cref{prop:lock} does not require any notion of
fairness like in frameworks that syntactically constrain loops (eg., in
behavioural types loops do not have continuations).
Instead, look-freedom requires some notion of fairness in frameworks
admitting continuations after loops (eg., \cite{gt18}).
In our case, fairness is not required since
well-branchdness imposes that each participant involved in a
continuation after a loop should also be involved within the loop.

The next proposition shows that the projections of a well-formed
c-automaton form a deadlock-free communicating system.
We first define \emph{deadlock} configurations as those of
communicating systems with no outgoing transitions, yet with at least
one component willing to take a transition.
\cref{def:deadlock} adapts the ones in~\cite{LangeTY15,gt18} to a
synchronous setting.

\begin{definition}[Deadlock]\label{def:deadlock}
  A state $\vec q \in \RS[\aCS]$ reachable in a communicating system
  $\aCS = (\aCM_{\p})_{\p \in \ptpset}$ is a \emph{deadlock} if
  $\vec q$ has no outgoing transitions, yet there exists 
  $\p \in \ptpset$ such that $\pstate{q}{\p}$ has an outgoing
  transition in the CFSM of $\aCS(\p) = \aCM_{\p}$.
\end{definition}

\begin{proposition}[Deadlock freedom]\label{prop:deadlock}
  If $\chora$ is a well-formed c-automaton and
  $\vec q \in \RS[\ssem{\proj{\chora}{}}]$, $\vec q$ is not a
  deadlock.
\end{proposition}
\begin{proof}
  We prove the contrapositive.
If participant $\p \in \ptpset$ has an outgoing transition $t$ from
  $\pstate{q}{\p}$ in $\proj{\chora}{\p}$, then (\cref{prop:lock})
  there is a run involving $\vec q$ where a transition
  $\pstate{q}{\p}$ is fired.
\qed
\end{proof}

\section{Composition of Choreography Automata}\label{sec:comp}
In this section we study how to compose two or more (well-formed) c-automata into a
single c-automaton.
Composition is obtained by means of two operations:
\begin{enumerate}
\item
 the product of c-automata, building a
c-automaton corresponding to the concurrent execution of the two
original c-automata;
\item a \emph{blending} operation that, given two roles of a
  c-automaton, removes them and adjusts the c-automaton as if they
  became two \quo{coupled} forwarders.
\end{enumerate}
For instance, blending \iH and \iK removes participants \iH and \iK
and sends each message $\msg$ originally sent to \iH to whoever \iK
used to
send $\msg$, and vice versa.

\subsection{The blending algorithm}\label{subsec:blending}
We present and discuss the blending operation, which will then be used
for defining c-automata composition.
We start by giving an informal algorithmic presentation of the
blending operation on two roles \iH and \iK in a given c-automaton.
(A more formal presentation of the algorithm is in \cref{app:blending}.)
\\[2mm]
{\small{\sc The (informal) BLENDING algorithm}\\
  {\bf Input:} a c-automaton and two roles, \iH and \iK.\\
  {\bf Output:} a c-automaton.\\[1mm]
{\bf begin}
\vspace{-2mm}
\begin{enumerate}[I)]
\item 
\label{stepI}
each transition with $\cauttr[p][{\gint[][A][@][H]}][@][][]$
  (resp. $\cauttr[p][{\gint[][A][@][k]}][@][][]$) is removed, and for
  each transition $\cauttr[q][{\gint[][K][@][B]}][r][][]$
  (resp. $\cauttr[q][{\gint[][h][@][B]}][r][][]$) a transition
  $\cauttr[p][{\gint[][A][@][B]}][r][][]$ (resp.   $\cauttr[p][{\gint[][A][@][B]}][r][][]$) is added
  provided that $\p \neq \q$; if $\p = \q$ the blending is not
  defined;
\item 
\label{stepII}
transitions involving neither \iH nor \iK are
  preserved, whereas all other transitions are removed;
\item
\label{stepIII}
 states and transitions unreachable from the initial state are
  removed.
\end{enumerate}
\vspace{-2mm}
{\bf end}}
\\
Notice how the blending operation can be interpreted as a sort of
\quo{self-composition} of a single system.
It is worth remarking that self-composition is a relevant operation
per-se; it can be used, for instance, to get rid of two participants
\iH and \iK by transforming them into a \quo{hidden} bidirectional
forwarder, letting their partners in the interaction to communicate
directly.
Of course, blending \iH and \iK could cause the overall behaviour of
the system to drastically change, unless some conditions are satisfied
(see the notion of reflectiveness in \cref{def:reflective} and
\cref{lemma:bisim}).
Conditions essentially require that if a message is forwarded from
\iH to \iK then \iK expects it, and vice versa.

\begin{remark}
  The blending operation can be performed in time $O(n^2)$ where $n$
  is the number of transitions in the c-automaton, see \cref{prop:complexity}
  in \cref{app:blending}.
\rmkend
\end{remark}

Before proceeding with the formalisation of the blending operation, it
is convenient to introduce some notations.
Let $\widehat{\chora}$ denote the c-automaton obtained by removing the
states unreachable from the initial state of the c-automaton $\chora$.
The set of transitions of $\chora$ involving participants
in $P \subseteq \ptpset$ is
\[
  \chora_{@ P} = \{\cauttr[@][{\gint[]}][@][][] \in \chora
  \sst \{\p, \q\} \cap P \neq \emptyset\}
\]
Given two participants \iH and \iK,
\[
  \chora_{(\iH,\iK)} = \{\cauttr \sst \exists r \in \sset \qst
  \cauttr[@][{\gint[][a][m][h]}][r][][],
  \cauttr[r][{\gint[][k][m][B]}][q][][] \in \chora\}
\]
is the set of \emph{blended}\footnote{The decoration $(\iH,\iK)$ is
  pleonastic; it is a convenience for \cref{def:uptoproj} and some
  proofs\ifhideproofs{ reported in~\cite{BLT-TechRep}}\else{}\fi.}
transitions of $\chora$ with respect to \iH and \iK.
 
\begin{definition}[Blending]\label{def:sync}
  Given two participants \iH and \iK, the \emph{blending in a
    c-automaton $\chora = \conf{\sset,\lint,\tset,s_0}$ of \iH and
    \iK}, written $\sync{\chora}{H}{K}$, is the c-automaton
  \[\begin{array}{llcl}
    & \sync \chora h k & = &
      \begin{cases}
        \widehat{\conf{\sset,\lint,\tset_{\bowtie},s_0}},
        & \text{if } \forall \cauttr \in \tset_{\bowtie} \qst \p \neq \q
        \\
        \bot, & \text{otherwise}
      \end{cases}
    \end{array}
  \]
  where
  $\tset_{\bowtie} =  \big(\tset \cup \chora_{(\iH,\iK)} \cup \chora_{(\iK,\iH)}\big) \setminus \chora_{@ \{\iH,\iK\}}$.
\end{definition}

\begin{example}
  In \cref{sec:intro}, the construction in \cref{def:sync} is used to
  perform the blending on \iH and \iK of the product automaton on the
  left, producing the automaton on the right.
\end{example}

The blending operation enjoys two different forms of commutativity.

\begin{lemma}\label{lem:commutativity}
  Given a c-automaton $\chora$ and participants \iH, \iK, \iI, and \iJ, we have
  \begin{enumerate}
  \item\label{lem:commutativity1} $\sync{\chora}{H}{K}=\sync{\chora}{K}{H}$
  \item\label{lem:commutativity2} $\sync{(\sync{\chora}{H}{K})}{I}{J} = \sync{(\sync{\chora}{I}{J})}{H}{K}$
  \end{enumerate}
\end{lemma}
\begin{proof}
  The proof of \eqref{lem:commutativity1} straighfordwardly follows
  from \cref{def:sync} since
  \[
    \tset_{\bowtie} =  \big(\tset \cup \chora_{(\iH,\iK)} \cup \chora_{(\iK,\iH)}\big) \setminus \chora_{@ \{\iH,\iK\}}
    = \big(\tset \cup \chora_{(\iK,\iH)} \cup \chora_{(\iH,\iK)}\big) \setminus \chora_{@ \{\iK,\iH\}}
  \]
To prove \eqref{lem:commutativity2} we show that the c-automata
  on the two sides of the equality have the same set of transitions.
By \cref{def:sync}, the transitions of $\sync{\chora}{H}{K}$ are
  \[
    \tset_1 =  \big(\tset \cup \chora_{(\iH,\iK)} \cup \chora_{(\iK,\iH)}\big) \setminus \chora_{@ \{\iH,\iK\}}
  \]
  where $\tset$ are the transitions of $\chora$.
Hence, again by \cref{def:sync},
  \[
    \tset_2 =  \big(\tset_1 \ \cup \ \sync \chora h k _{(\iI,\iJ)} \ \cup \ \sync \chora h k_{(\iJ,\iI)}\big) \setminus \chora_{@ \{\iI,\iI\}}
  \]
  are the transitions of $\sync{(\sync{\chora}{H}{K})}{I}{J}$.
Similarly, we have that
  \[
    \tset_2' =  \big(\tset_1' \ \cup \ \sync \chora i j _{(\iH,\iK)} \ \cup \ \sync \chora i j_{(\iK,\iH)}\big) \setminus \chora_{@ \{\iI,\iI\}}
  \]
  are the transitions of $\sync{(\sync{\chora}{i}{j})}{h}{k}$ provided
  that $\tset_1'$ are the transitions of $\sync \chora i j$.

  We prove $\tset_2 = \tset_2'$.
Let  $t \in \tset_2$ be a transition with label $\gint[]$.
It is a simple observation that
  $\{\p,\q\} \cap \{\iI, \iJ\} = \emptyset$ (by construction any
  transition where participants are those in the interface are
  removed).
Hence we have three cases:
  \begin{itemize}
  \item If $t \in \tset_1$
  \item If $t \in \sync \chora h k _{(\iI,\iJ)}$ then we also have
    $\{\p,\q\} \cap \{\iH, \iK\} = \emptyset$.
And, assuming $t = \cauttr$, there is $r$ such that
    $\cauttr[@][{\gint[][a][m][i]}][r][][]
    \cauttr[][{\gint[][j][m][b]}][q][][]$ are in $\tset_1$.
We procede by case analysis:
    \begin{itemize}
    \item Both $\cauttr[@][{\gint[][a][m][i]}][r][][]$ and
      $\cauttr[r][{\gint[][j][m][b]}][q][][]$ are in $\chora$.
In this case, $\cauttr[p][{\gint[][a][m][i]}][q][i][j]$
      is in $\sync \chora i j$ and hence in $\tset_2'$ since
      $\{\p,\q\} \cap \{\iI,\iJ\} = \emptyset$.
This implies that $\cauttr[p][{\gint[]}][q][i][j]$ is
      in $\sync{(\sync \chora i j)} h k$ since
      $\{\p,\q\} \cap \{\iH,\iK\} = \emptyset$.
    \item Only one between $\cauttr[@][{\gint[][a][m][i]}][r][][]$ and
      $\cauttr[r][{\gint[][j][m][b]}][q][][]$ is in $\chora$.
With no loss of generality, assume that only
      $\cauttr[r][{\gint[][j][m][b]}][q][][]$ is in $\chora$.
Then $\cauttr[@][{\gint[][a][m][i]}][r][][]$ is in
      $\sync \chora h k$ and therefore there is $r'$ such that
      $\cauttr[@][{\gint[][a][m][h]}][r'][][]
      \cauttr[][{\gint[][k][m][i]}][r][][]$ in $\chora$ (the case
      $\cauttr[@][{\gint[][a][m][k]}][r'][][]
      \cauttr[][{\gint[][h][m][i]}][r][][]$ is $\chora$ is similar).
Therefore, 
      $\cauttr[r'][{\gint[][k][m][b]}][q][][]$ in $\sync \chora i j$
      which in turn implies that $t \in \sync{(\sync \chora i j)} h k$.

    \item None of     $\cauttr[@][{\gint[][a][m][i]}][r][][]$ and
      $\cauttr[r][{\gint[][j][m][b]}][q][][]$ is in $\chora$.
Applying twice the argument in the previous item we get the
      thesis.
    \end{itemize}
  \item If $t \in \sync \chora h k_{(\iJ,\iI)}$ we procede as in the previous case.
  \end{itemize}

  The inclusion $\tset_2' \subseteq \tset_2$ is proved similarly.
\qed
\end{proof}
Whereas it is quite natural to expect property
\eqref{lem:commutativity1} to hold -- due to the perfect symmetry
between \iH and \iK in the definition of blending -- property
\eqref{lem:commutativity2} sounds less natural.
In fact, one could imagine that blending on \iH and \iK could affect
the behaviour of the other participants of $\chora$, including $\II$
and $\JJ$.
Remarkably, blending is such that property \eqref{lem:commutativity2}
holds.
Such result, besides being interesting on its own, will be used also
to prove associativity of the composition.
This is quite relevant if one takes into account that the development
of systems in a modular way is one of the main goals of our framework
for choreographies composition.
In fact modularity naturally requires the composition operation to be
insensitive on the order of application.

We provide a condition ensuring that blending does not essentially
affect the behaviour of participants, namely that the only difference
between $\chora$ and $\sync \chora H K$ is that some messages
sent/received by \iH or \iK in $\chora$ are sent/received by other
participants in $\sync \chora H K$.
The following definition is instrumental to formally establish the
informal property above; it casts the projection operation on
c-automata where some interfaces have been blended.
\begin{definition}[Up-to-interface projection]\label{def:uptoproj}
  The \emph{projection on participant \p\ up-to-interface $(\iH,
    \iK)$ of a c-automaton $\chora =
    \conf{\sset,\lint,\tset,s_0}$} is the automaton
  $\proji{\chora}{A}{H}{K} = \conf{\sset,\lact,\tset',s_0}$ where
  \[
    \tset' = \{\proj{t}{A}\}_{t \in \tset}
    \cup
    \{\cauttr[@][{\aout[a][x]}][@][][] \,|\, {\cauttr[@][{\gint[][a][m][b]}][@][x][y] \in \tset'}\}
    \cup
    \{\cauttr[@][{\ain[y][a]}][@][][] \,|\, {\cauttr[@][{\gint[][b][m][a]}][@][x][y] \in \tset'}\}
\]
\end{definition}

\ifhideproofs
\else
\begin{lemma}\label{lem:states}
  If $\sync \chora h k$ is defined, its states are included in the
  states of $\chora$.
\end{lemma}
\begin{proof}
  By construction, since the first two steps do not change the set of
  states, and the last one may only remove states.
\qed
\end{proof}
\fi

\begin{proposition}[Simulation]\label{lemma:sim}
  If $\sync{\chora}{H}{K}$ is defined then for each participant
  $\p \neq \iH,\iK$ we have that $\proj{\chora}{A}$ simulates
  $\proji{\sync{\chora}{H}{K}}{A}{H}{K}$.
\end{proposition}
\begin{proof}
  Recall that states of projection are equivalence classes of states of
  the projected automata up to trace equivalence.
Let us consider the relation
  \[
    \rel = \{(S',S) \sst S \cap S' \neq \emptyset, S \in \proj{\chora}{A}, S' \in \proji{\sync{\chora}{H}{K}}{A}{H}{K}\}
  \]
  The proof is by coinduction.
Take any transition $\cauttr[S'][\alpha][S'_1][][]$ in
  $\proji{\sync{\chora}{H}{K}}{A}{H}{K}$.
By definition of projection there is a transition
  $t' = \cauttr[s'][\alpha'][s'_1][x][y]$ in $\sync{\chora}{H}{K}$
  with $s'_1 \in S'_1$ whose projection up-to-interface on \p\ yields
  label $\alpha$.
By definition of blending, transition $t'$ is derived by transitions
  of $\chora$.
There are a few cases:
  \begin{itemize}
  \item if $t'$ involves neither \iH nor \iK then
    $\cauttr[s'][\alpha][s'_1][][]$ is in $\chora$, and
    $\cauttr[{[s']}][\alpha'][{[s'_1]}][][]$ is in $\proj{\chora}{A}$ as
    desired since the two target states are $\rel$-related;
  \item if $t' = \cauttr[s'][{\gint[][A][@][B]}][s'_1][h][k]$ then
    $\chora$ has transitions
    \[
      \cauttr[s'][{\gint[][A][@][H]}][s'_2][][]
      \qquad \text{and} \qquad
      \cauttr[s'_2][{\gint[][K][@][B]}][s'_1][][]
    \]
    hence in $\proj{\chora}{A}$ we have
    $\cauttr[{[s']}][{\aout[A][H]}][{[s'_2]}][][] = [s'_1]$ as
    desired.
  \end{itemize}
  The case where \p\ is the receiver is analogous to the last case
  above.
\qed
\end{proof}

One would hope the operation of blending to affect \emph{only} the
participants that are blended; however, this is not the case in
general as shown by the following example.
\begin{example}\label{ex:nonrefl}
  Consider the following well-formed c-automaton and its (empty) blending
  on \iH and \iK:
\[
    \begin{tikzpicture}[node distance=1.7cm,]
      \tikzstyle{every state}=[cnode]
      \tikzstyle{every edge}=[carrow]
\node[state] (one)                        {$1$};
      \node[draw=none,fill=none] (start) [left = 0.3cm  of one]{$\chora[A]$};
      \node[state] (two)  [ right  of =one]                      {$2$};
      \node[state] (three) [  right  of =two]                      {$3$};
\path  (start) edge node {} (one) 
      (one)edge                    node [below] {$\gint[][K][m][B]$} (two)
      (two)edge                    node [below] {$\gint[][B][n][D]$} (three)
;
    \end{tikzpicture}
    \qquad
    \begin{tikzpicture}[node distance=1.7cm,]
      \tikzstyle{every state}=[cnode]
      \tikzstyle{every edge}=[carrow]
\node[state] (one)                        {$1$};
      \node[draw=none,fill=none] (start) [left = 0.3cm  of one]{$\sync{\chora}{H}{K}$};
\path  (start) edge node {} (one)
;
    \end{tikzpicture}
  \]
Trivially, \p[D] behaves differently in $\chora$ than in
  $\sync{\chora}{H}{K}$.
\finex
\end{example}

\begin{remark}
  The above example highlights an undesirable feature of blending.
We therefore identify a condition on c-automata that avoids that
  blending modifies the behaviour on non-interface participants.
\rmkend
\end{remark}

\begin{definition}[Reflective c-automata]\label{def:reflective}
  Let $\iH,\iK$ be two participants.
A c-automaton $\chora$ is \emph{\iH-to-\iK reflective} if
  \begin{enumerate}[(1)]
  \item \label{it:out} for each input transition
    $\cauttr[S][{\ain[A][H][][m]}][S'][][]$ in $\proj{\chora}{H}$ we
    have:
    \begin{enumerate}
    \item\label{item:existsout} there is
      $\cauttr[S''][{\aout[K][B][][m]}][S'''][][]$ in
      $\proj{\chora}{K}$ with $S' \cap S'' \neq \emptyset$;
    \item\label{item:forallout} for each
      $\cauttr[S''][{\aout[K][X][][m]}][S'''][][]$ in
      $\proj{\chora}{K}$, there are states
      $s \in S, s' \in S' \cap S'', s''' \in S'''$ such that
      $\cauttr[s][{\gint[][A][@][H]}][s'][][]
      \cauttr[][{\gint[][K][@][X]}][s'''][][]$ in $\chora$.
    \end{enumerate}
  \item \label{it:in} for each output transition
    $\cauttr[S][{\aout[H][a][][m]}][S'][][]$ in $\proj{\chora}{H}$ we
    have:
  \begin{enumerate}
  \item\label{item:existsin} there is
    $\cauttr[S''][{\ain[b][K][][m]}][S'''][][]$ in $\proj{\chora}{K}$
    with $S''' \cap S \neq \emptyset$;
  \item\label{item:forallin} for each
    $\cauttr[S''][{\ain[x][K][][m]}][S'''][][]$ in
    $\proj{\chora}{K}$, there are states
    $s'' \in S'', s''' \in S''' \cap S, s' \in S'$ such that
    $\cauttr[s''][{\gint[][x][@][k]}][s'''][][]
    \cauttr[][{\gint[][h][@][a]}][s'][][]$ in $\chora$.
  \end{enumerate}
 \end{enumerate}
  If $\chora$ is \emph{\iH-to-\iK reflective} and \emph{\iK-to-\iH
    reflective} then we say that $\chora$ is \emph{reflective on \iH
    and \iK}.
\end{definition}

A non-trivial example of reflective c-automaton is given below.
\begin{example}\label{ex:reflective}
  The c-automaton below is reflective on \iH and \iK.
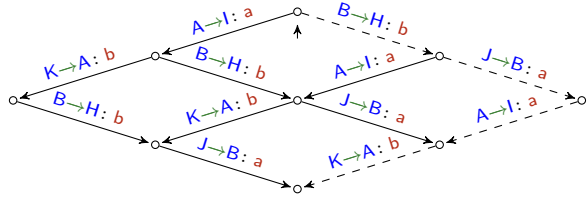
  \[
    \begin{tikzpicture}[node distance=1.4cm,]
      \tikzstyle{every state}=[cnode]
      \tikzstyle{every edge}=[carrow]
\node[state]           (00)  {};\node[state]           (10)  [below  left of=00, xshift=-9mm,yshift=4mm] {};\node[draw=none,fill=none] (start) [below = 0.3cm of 00]{};
      \node[state]            (01) [below  right of=00, xshift=9mm,yshift=4mm] {};\node[state]           (11) [below left of=01, xshift=-9mm,yshift=4mm] {};\node[state]            (02) [below right of=01, xshift=9mm,yshift=4mm] {};\node[state]            (20) [below left of=10, xshift=-9mm,yshift=4mm] {};\node[state]            (21) [below right of=20, xshift=9mm,yshift=4mm] {};\node[state]            (12) [below left of=02, xshift=-9mm,yshift=4mm] {};\node[state]            (22) [below left of=12,  xshift=-9mm,yshift=4mm] {};\path  (start) edge node {} (zero) 
      (00)    edge                                  node [above] {$\gint[][A][a][I]$} (10)
      edge[dashed]                                  node [above] {$\gint[][B][b][H]$} (01)
      (01)    edge                                   node [above] {$\gint[][A][a][I]$} (11)
      edge[dashed]                                   node [above] {$\gint[][J][a][B]$} (02)
      (10)    edge                                   node [above] {$\gint[][K][b][A]$} (20)
      edge                                   node [above] {$\gint[][B][b][H]$} (11)
      (20)    edge                                   node [above] {$\gint[][B][b][H]$} (21)
      (02)    edge[dashed]                                   node [above] {$\gint[][A][a][I]$} (12)
      (11)    edge                                   node [above] {$\gint[][K][b][A]$} (21)
      edge                                   node [above] {$\gint[][J][a][B]$} (12)
      (21)    edge                                   node [above] {$\gint[][J][a][B]$} (22)
      (12)    edge[dashed]                                   node [above] {$\gint[][K][b][A]$} (22)
      ;
    \end{tikzpicture}
  \]
  For instance, the three inner states on the dashed path would
  collapse in the same equivalence class when projecting on \iH.

  Instead, the c-automaton $\chora$ in \cref{ex:nonrefl} is not
  reflective on \iH and \iK (in particular it is not \iK-to-\iH
  reflective).
In fact, on the projections
  \[
    \begin{array}{c@{\hspace{18mm}}c}
      \begin{tikzpicture}[node distance=2.4cm,]
        \tikzstyle{every state}=[cnode]
        \tikzstyle{every edge}=[carrow]
\node[state] (A)                        {$\Set{1,2,3}$};
        \node[draw=none,fill=none] (start) [left = 0.3cm  of A]{$\proj{\chora[A]}{H}$};
\path  (start) edge node {} (A) 
;
      \end{tikzpicture}
      &
        \begin{tikzpicture}[node distance=2.4cm,]
          \tikzstyle{every state}=[cnode]
          \tikzstyle{every edge}=[carrow]
\node[state] (A)                        {$\Set{1}$};
          \node[draw=none,fill=none] (start) [left = 0.3cm  of one]{$\proj{\chora[A]}{K}$};
          \node[state] (B)  [ right  of =A]                      {$\Set{2,3}$};
\path  (start) edge node {} (A) 
          (A)    edge                                 node [above] {${\ain[K][B][][m]}$} (B)
          ;
        \end{tikzpicture}
    \end{array}
  \]
  condition \eqref{item:existsout} of \cref{def:reflective} does not
  hold. \finex
\end{example}

We now formally show that reflectiveness on \iH and \iK does guarantee
that the behaviour of any participant $\ptp[A]\neq\iH,\iK$ is not
\quo{affected} by the blending \iH and \iK.

\begin{theorem}[Bisimulation]\label{lemma:bisim}
  If $\chora$ is reflective on \iH and \iK then $\proj{\chora}{A}$ and
  $\proji{\sync{\chora}{H}{K}}{A}{H}{K}$ are bisimilar, for each
  participant $\p \neq \iH,\iK$.
\end{theorem}
\begin{proof}
  Fix the participant $\p \in \ptpset$.
Let us consider the relation $\rel$ defined as follows:
  $\rel = \{(S',S) | S \cap S' \neq \emptyset, S \in \proj{\chora}{A},
  S' \in \proji{\sync{\chora}{H}{K}}{A}{H}{K}\}$
  
  The proof is by coinduction.
One direction follows from \cref{lemma:sim}.
Let us consider the other direction, namely that
  $\proji{\sync{\chora}{H}{K}}{A}{H}{K}$ simulates $\proj{\chora}{A}$.
  Take any transition $\cauttr[S][\alpha][S'][][]$ in
  $\proj{\chora}{A}$.
There are a few cases depending on the transition:
  \begin{itemize}
  \item If the transition does not involve \iH nor \iK
    then by definition of projection in $\chora$ there is $s \in S$
    such that $\cauttr[s][\alpha'][s'][][]$ for some $s' \in S'$ such
    that the projection of $\alpha'$ on $\ptp[A]$ is $\alpha$.
Also, in $\sync{\chora}{H}{K}$ $\cauttr[s][\alpha'][s'][][]$, and
    in $\proji{\sync{\chora}{H}{K}}{A}{H}{K}$
    $\cauttr[{[s]}][\alpha'][{[s']}][][]$ as desired.
The two target states are in the relation hence we are done.
  \item If the transition is $\cauttr[S][{\ain[A][H][][m]}][S'][][]$
    then by \eqref{item:existsin} in \cref{def:reflective} there is a
    transition $\cauttr[S''][{\aout[K][B][][m]}][S'''][][]$ in
    $\proj{\chora}{K}$ with $S' \cap S'' \neq \emptyset$.
By the \eqref{item:forallin} in \cref{def:reflective}, there are
    states $s \in S, s' \in S' \cap S'', s''' \in S'''$ such that
    $\cauttr[s][{\gint[][A][@][H]}][s'][][]
    \cauttr[][{\gint[][K][@][B]}][s''][][]$ in $\chora$.
As a consequence in $\sync{\chora}{H}{K}$ we have
    $\cauttr[s][{\gint[][A][@][B]}][s''']$.
Thus, in $\proji{\chora}{A}{H}{K}$ we have
    $\cauttr[{[s]}][{\ain[A][H][][m]}][{[s''']}][][]$.
The thesis follows since $[s''']=[s'']$ given that in the
    projection on $\ptp[A]$ they are connected by an
    $\epsilon$-transition and $[s''] \cap [s'] \neq \emptyset$.
  \item If the transition is $\cauttr[S][{\aout[H][A][][m]}][S'][][]$
    then by \eqref{item:existsout} in \cref{def:reflective} there is
    $\cauttr[S''][{\ain[B][K][][m]}][S'''][][]$ in $\proj{\chora}{K}$
    with $S''' \cap S \neq \emptyset$.
Also, by \eqref{item:forallout}in \cref{def:reflective} there are states
    $s'' \in S'', s''' \in S''' \cap S, s' \in S'$ such that
    $\cauttr[s''][{\gint[][B][@][K]}][s'''][][]
    \cauttr[][{\gint[][H][@][A]}][s'][][]$ in $\chora$.
As a consequence in $\sync{\chora}{H}{K}$ we have
    $\cauttr[s''][{\gint[][B][@][A]}][s']$.
Thus, in $\proji{\chora}{A}{H}{K}$ we have
    $\cauttr[{[s'']}][{\aout[H][A][][M]}][{[s']}][][]$ as desired.
\qed
  \end{itemize}
\end{proof}

\begin{remark}\label{rem:wellseqsnecessity}
  The blending operation is sensitive with respect to the
  representation of c-automata, as shown in the following
  example.
If we apply the blending operation on participants \iH and \iK
  directly on the canonical automaton below
  \[
    \begin{tikzpicture}[node distance=2.0cm and 0cm]
      \tikzstyle{every state}=[cnode]
      \tikzstyle{every edge}=[carrow]
      \node[state]           (zero)                        {$0$};
      \node[state]           (one)  [right of=zero]  {$1$};
      \node[draw=none,fill=none] (start) [left = .3cm of zero]{$\chora$};
      \node[state]           (two)  [right of=one]  {$2$};
      \node[state]           (three)  [right   of=two]  {$3$};
      
      \path  (start) edge node {} (zero) 
      (zero)    edge                                  node [above] {$\gint[][A][a][H]$} (one)
      (one)         edge                                  node [above] {$\gint[][C][b][D]$} (two)
      (two)    edge                                   node [above] {$\gint[][K][a][B]$} (three)
      ;
    \end{tikzpicture}
  \]
  then we get the empty automaton, whereas a \quo{correct} blending
  should give
  
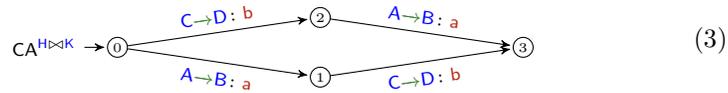
\begin{equation}
    \label{eq:a}
    \begin{tikzpicture}[node distance=.2cm and 2.5cm]
      \tikzstyle{every state}=[cnode]
      \tikzstyle{every edge}=[carrow]
      \node[draw=none,fill=none] (start) {$\sync \chora h k$};
      \node[state, right = .3cm of start]           (zero)                        {$0$};
      \node[state]           (one)  [below right = of zero]  {$1$};
      \node[state]           (two)  [above right = of zero]  {$2$};
      \node[state]           (three)  [above right = of one]  {$3$};
\path  (start) edge node {} (zero) 
      (zero)  edge[sloped] node [below] {$\gint[][A][a][B]$} (one)
      edge[sloped] node [above] {$\gint[][C][b][D]$} (two)
      (one)   edge[sloped] node [below] {$\gint[][C][b][D]$} (three)
      (two)   edge[sloped] node [above] {$\gint[][A][a][B]$} (three)
      ;
   \end{tikzpicture}
 \end{equation}
 since the interaction $\gint[][C][b][D]$ is independent from
 both $\gint[][A][a][H]$ and $\gint[][K][a][B]$.
 
 Let us consider now a well-sequenced c-automaton
 producing the language $\close{\lang{\chora}}$, i.e. where the independent interactions were
 made \quo{visible} by the presence of \quo{diamonds}, and precisely
 
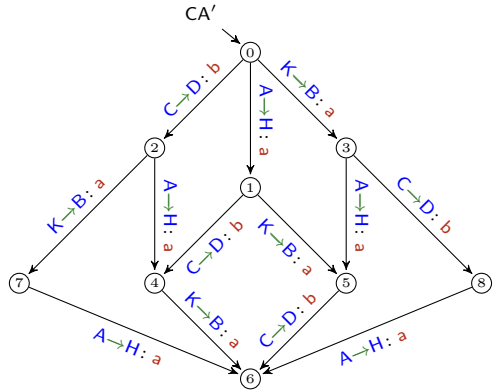
\[
   \begin{tikzpicture}[node distance=1.8cm and 7cm]
       \tikzstyle{every state}=[cnode]
       \tikzstyle{every edge}=[carrow]
\node[state] (zero)                        {$0$};
       \node[state] (one)  [below   of=zero]  {$1$};
       \node[draw=none,fill=none] (start) [above left = 0.3cm  of zero]{${\chora[A']}$};
       \node[state] (two)  [below left   of=zero]  {$2$};
       \node[state] (three)  [below right  of=zero]  {$3$};
       \node[state] (four)  [below left of=one]  {$4$};
       \node[state] (five)  [below right  of=one]  {$5$};
       \node[state] (six)  [below  right  of=four]  {$6$};
       \node[state] (seven)  [left  of=four]  {$7$};
       \node[state] (eight)  [right   of=five]  {$8$};
\path  (start) edge node {} (zero) 
       (zero)  edge node [above] {$\gint[][C][b][D]$} (two)
       edge node [above] {$\gint[][A][a][H]$} (one)
       edge node [above] {$\gint[][K][a][B]$} (three)
       (one)   edge node [below] {$\gint[][C][b][D]$} (four)
       edge node [below] {$\gint[][K][a][B]$} (five)
       (four)  edge node [above] {$\gint[][K][a][B]$} (six)
       (five)  edge node [above] {$\gint[][C][b][D]$} (six)
       (two)   edge node [above] {$\gint[][A][a][H]$} (four)
       edge node [above] {$\gint[][K][a][B]$} (seven)
       (three) edge node [above] {$\gint[][A][a][H]$} (five)
       edge node [above] {$\gint[][C][b][D]$} (eight)
       (seven) edge node [below] {$\gint[][A][a][H]$} (six)
       (eight) edge node [below] {$\gint[][A][a][H]$} (six)            
       ;
     \end{tikzpicture}
   \]
The blending of \iH and \iK in $\chora'$ yields exactly the expected
   c-automaton~\eqref{eq:a}.
\rmkend
\end{remark}

We show that blending preserves well-formedness. This means that
c-automata obtained by blending can be correctly projected. We begin
with well-sequencedness.

\begin{proposition}[Blending preserves well-sequencedness]
\label{prop:blendpresws}
  If $\chora$ is a well-sequenced c-automaton then so is $\sync{\chora}H K$ (if defined).
\end{proposition}
\begin{proof}
By contradiction, let us assume $\sync \chora h k$ not to be well-sequenced.
So there exist in it a pair of transitions of the following form
 \[
   \begin{tikzpicture}[node distance=1.8cm and 7cm]
       \tikzstyle{every state}=[cnode]
       \tikzstyle{every edge}=[carrow]
\node[state] (zero)                        {$p$};
       \node[state] (one)  [right   of=zero]  {$q$};
       \node[state] (three)  [right   of=one]  {$r$};
\path  
       (zero)  edge node [above] {$\gint[][C][c][D]$} (one)
       (one)   edge node [above] {$\gint[][A][a][B]$} (three)
       ;
     \end{tikzpicture}
   \] 
   such that 
   \begin{itemize}
   \item $\Set{\ptp[C],\ptp[D]}\cap\Set{\ptp[A],\ptp[B]}=\emptyset$;
     and there is no state $q'$ such that
   \[
   \begin{tikzpicture}[node distance=1.8cm and 7cm]
       \tikzstyle{every state}=[cnode]
       \tikzstyle{every edge}=[carrow]
\node[state] (zero)                        {$p$};
       \node[state] (one)  [right   of=zero]  {$t$};
       \node[state] (three)  [right   of=one]  {$r$};
\path  
       (zero)  edge node [above] {$\gint[][A][a][B]$} (one)
       (one)   edge node [above] {$\gint[][C][c][D]$} (three)
       ;
     \end{tikzpicture}
   \] 
\end{itemize} 
Moreover by definition of Blending,
\begin{itemize}   
  \item 
  the transitions correspond, in $\chora$, to 
   \[
   \begin{tikzpicture}[node distance=1.8cm and 7cm]
       \tikzstyle{every state}=[cnode]
       \tikzstyle{every edge}=[carrow]
\node[state] (zero)                        {$p$};
       \node[state] (one)  [right   of=zero]  {$q$};
       \node[state] (two)  [right   of=one]  {$s$};
       \node[state] (three)  [right   of=two]  {$r$};

\path  
       (zero)  edge node [above] {$\gint[][C][c][D]$} (one)
       (one)   edge node [above] {$\gint[][A][a][H]$} (two)
       (two)   edge node [above] {$\gint[][K][a][B]$} (three)
       ;
     \end{tikzpicture}
   \]
  \item
  $\iH\not\in\Set{\ptp[C],\ptp[D]}$ (otherwise the $p$-to-$q$ transition would have been dropped by the blending procedure)
\end{itemize}  
By well-sequencedness of $\chora$, it follows that there exist diamonds in it enabling
to go from $p$ to $q$ following all possible combinations of transitions with labels
in $\Set{\gint[][C][c][D], \gint[][A][a][H], \gint[][K][a][B]}$.
This implies that, by the blending procedure, in $\sync{\chora}H K$ we have actually
 
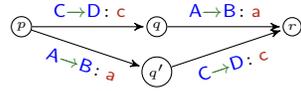
\[
   \begin{tikzpicture}[node distance=1.8cm and 7cm]
       \tikzstyle{every state}=[cnode]
       \tikzstyle{every edge}=[carrow]
\node[state] (zero)                        {$p$};
       \node[state] (one)  [right   of=zero]  {$q$};
       \node[state] (two)  [below   of=one,yshift=12mm]  {$q'$};
       \node[state] (three)  [right   of=one]  {$r$};
\path  
       (zero)  edge node [above] {$\gint[][C][c][D]$} (one)
                  edge node [below] {$\gint[][A][a][B]$} (two)
       (two)   edge node [below] {$\gint[][C][c][D]$} (three)          
       (one)   edge node [above] {$\gint[][A][a][B]$} (three)
       ;
     \end{tikzpicture}
   \] 
   Contradiction.
\qed
\end{proof}

Well-branchedness, unfortunately, is not generally preserved by blending,
as shown by the following counterexample.
Consider
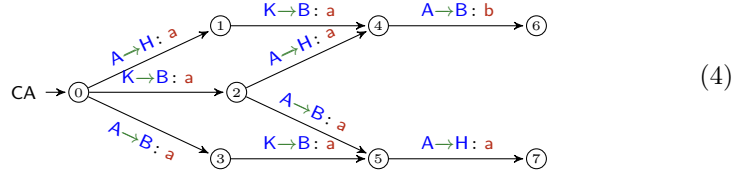
\begin{equation}
  \begin{tikzpicture}[node distance = 2.1cm]
    \tikzstyle{every state}=[cnode]
    \tikzstyle{every edge}=[carrow]
\node[draw=none,fill=none] (start) {$\chora$};
    \node[state, right = .3cm of start] (zero) {$0$};
    \node[state] (one)   [above right  of=zero, yshift=-6mm,xshift=4mm]   {$1$};
    \node[state] (two) [ right of=zero] {$2$};
    \node[state] (three) [below right  of=zero, yshift=--6mm,xshift=4mm] {$3$};
    \node[state] (four) [right  of=one] {$4$};
    \node[state] (five) [ right of=three] {$5$};
    \node[state] (six) [right  of=four] {$6$};
    \node[state] (seven) [right  of=five] {$7$};

\path  (start) edge node {} (zero) 
    (zero)  edge node [above] {$\gint[][A][a][H]$} (one)
                edge node [above] {$\gint[][K][a][B]$} (two)
                edge node [below] {$\gint[][A][a][B]$} (three)
    (two)  edge node [above] {$\gint[][A][a][H]$} (four)
                edge node [above] {$\gint[][A][a][B]$} (five)
    (one) edge node [above] {$\gint[][K][a][B]$} (four)
    (three)   edge node [above]  {$\gint[][K][a][B]$} (five)
    (four)   edge node [above]  {$\gint[][A][b][B]$} (six)
    (five)   edge node [above]  {$\gint[][A][a][H]$} (seven)

    ;
  \end{tikzpicture}\label{eq:syncvswb}
\end{equation}
which can be checked to be well-branched.
By blending \iH and \iK in $\chora$ we get
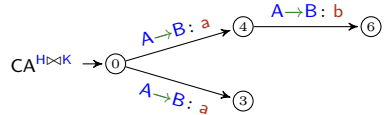
\[
  \begin{tikzpicture}[node distance = .70cm]
    \tikzstyle{every state}=[cnode]
    \tikzstyle{every edge}=[carrow]
\node[state] (zero)                        {0};
    \node[draw=none,fill=none] (start) [left   = 0.3cm  of zero]{$\sync{\chora}H K$};
    \node[state] (three) [below right  of=zero,xshift = 1.2cm] {3};
    \node[state] (four) [above right of=zero,xshift = 1.2cm] {4};
    \node[state] (six) [right  of=four,xshift = 1cm] {$6$};
\path  (start) edge node {} (zero) 
    (zero)    edge                                     node [sloped,above] {$\gint[][A][a][B]$} (four)
                 edge                                     node [sloped,below] {$\gint[][A][a][B]$} (three)
    (four)   edge node [above]  {$\gint[][A][b][B]$} (six)

    ;
  \end{tikzpicture}
\]
which is not well-branched.
We therefore restrict to the class of \emph{\p-univocal} c-automata.
\begin{definition}[\p-univocity]\label{def:univocity}
  A c-automaton $\chora$ is \emph{\p-univocal} if for any transitions
  $\cauttr[p][{\gint[][x][@][a]}][@][][]$ and
  $\cauttr[p][{\gint[][y][@][c]}][q'][][]$ in $\chora$, $\p[c] = \p$
  and for any transitions $\cauttr[p][{\gint[][a][@][x]}][@][][]$ and
  $\cauttr[p][{\gint[][c][@][y]}][q'][][]$ in $\chora$, $\p[c] = \p$
\end{definition}
It is immediate to check that $\chora$ in \eqref{eq:syncvswb} above is
not \iH-univocal.

\ifhideproofs
{}
\else
\begin{lemma}\label{lemma:incl}
  If $\sync \chora h k$ is defined then for each run $\pi$ in
  $\sync \chora h k$ there is a run $\hat \pi$ in $\chora$ such that
  if $t$ precedes $t'$ in $\pi$ and both occur in $\chora$ then $t$
  precedes $t'$ in $\hat \pi$.
\end{lemma}
\begin{proof}
  It suffices to observe that for any transition $t'' = \cauttr$ in
  $\pi$ which is not in $\chora$ there is a state $r$ such that
  transitions
  $\cauttr[@][{\gint[][@][@][h]}][r][][]
  \cauttr[][{\gint[][k]}][@][][]$ are in $\chora$ (and likewise for
  any $t'' = \cauttr[@][@][@][k][h]$ in $\pi$ which is not in $\chora$).
Hence, $\pi'$ is obtained by replacing $t''$ with such transitions.
\qed
\end{proof}

\begin{corollary}\label{cor:incl}
  If $\sync \chora h k$ is defined then for each acyclic run $\pi$ in
  $\sync \chora h k$ there is an acyclic run $\hat \pi$ in $\chora$
  such that if $t$ precedes $t'$ in $\pi$ and $t$ and $t'$ occur in
  $\chora$ then $t$ precedes $t'$ in $\hat \pi$.
\end{corollary}
\begin{proof}
  Observe that the construction in the proof of \cref{lemma:incl}
  never uses a same transition to replace different transitions on
  $\pi$.
\qed
\end{proof}
\fi

The next results establish that blending preserves well-branchedness
on univocal and reflective c-automata.
\begin{proposition}\label{prop:wb-vs-sync}
  Fix two participants \iH and \iK and a well-branched c-automaton
  $\chora$ such that $\sync \chora h k$ is defined.
If $\chora$ is \iH- and \iK-univocal then conditions \eqref{it:par1}
  and \eqref{it:par2} of \cref{def:wb} hold on $\sync{\chora}{H}{K}$.
If moreover $\chora$ is \iH- and \iK-reflective then condition
  \eqref{it:inp} of \cref{def:wb} holds on $\sync{\chora}{H}{K}$.
\end{proposition}
\begin{proof}
  By \emph{reductio ad absurdum}.
Let $p$ be a state in $\sync \chora h k$ with no selector.
Then one of the following must hold for each participant $\q$ of
  $\sync \chora h k$ (note that $\q \neq \iH$ and $\q \neq \iK$):
  \begin{enumerate}
  \item\label{it:output} there are transitions
    $t = \cauttr[@][{\gint[][b][@][a]}][@][][]$ and
    $t' = \cauttr[@][{\gint[][b][@][a]}][q'][][]$ with $q \neq q'$, or
  \item\label{it:diamond} there are transitions
    $t = \cauttr[@][{\gint[][b][@][a]}][@][][]$ and
    $t' = \cauttr[@][{\gint[][x][n][y]}][q'][][]$ such that
    $\p[x] \neq \q$ and $t$ and $t'$ are not concurrent, or else
  \item\label{it:unware} there are a participant $\p[x] \neq \q$ and a
    $p$-span $(\pi_1,\pi_2)$ in $\sync \chora h k$ such that ($i$)
    $\proj{\pi_1}{x} \neq \proj{\pi_2}{x}$ and ($ii$) the 
    first pair of characters where $\proj{\pi_1}{x}$ and
    $\proj{\pi_2}{x}$ differ are not both inputs.
\end{enumerate}
  We derive a contradiction in each of the cases above.
\paragraph{Case \eqref{it:output}} Not both $t$ and $t'$ can be
  transitions of $\chora$ otherwise $\chora$ would not be
  well-branched.
With no loss of generality, suppose that $t$ is not in $\chora$.
There must be $r$ such that
  \begin{equation}\label{eq:thereisr}
    \cauttr[@][{\gint[][b][@][h]}][r][][]
    \cauttr[][{\gint[][k][@][a]}][q][][]
    \quad\text{or}\quad
    \cauttr[@][{\gint[][b][@][k]}][r][][]
    \cauttr[][{\gint[][h][@][a]}][q][][]
    \qquad \text{ in } \chora;
  \end{equation}
  Then $t' \not\in \chora$ otherwise $\chora$ would not be \iH- or \iK-univocal.
Hence, there is $r'$ s.t.
  \[
    \cauttr[@][{\gint[][b][@][h]}][r'][][]
    \cauttr[][{\gint[][k][@][a]}][q'][][]
    \quad\text{or}\quad
    \cauttr[@][{\gint[][b][@][k]}][r'][][]
    \cauttr[][{\gint[][h][@][a]}][q'][][]
    \qquad \text{ in } \chora;
  \]
  If
  $\cauttr[@][{\gint[][b][@][h]}][r][][]
  \cauttr[][{\gint[][k][@][a]}][q][][]$ and
  $\cauttr[@][{\gint[][b][@][h]}][r'][][]
  \cauttr[][{\gint[][k][@][a]}][q'][][]$ are in $\chora$ then $p$
  would have two transitions with the same label violating the
  well-branchedness of $\chora$.
If
  $\cauttr[@][{\gint[][b][@][h]}][r][][]
  \cauttr[][{\gint[][k][@][a]}][q][][]$ and
  $\cauttr[@][{\gint[][b][@][k]}][r'][][]
  \cauttr[][{\gint[][h][@][a]}][q'][][]$ are in $\chora$ then,
  contrary to our hypothesis, $\chora$ would not be \iH- and
  \iK-univocal.
Swapping \iH and \iK would yield similar contradictions
  in the other cases.
  \paragraph{Case \eqref{it:diamond}} Not both $t$ and $t'$ can be
  transitions of $\chora$ otherwise $\chora$ would not be
  well-branched.

  If $t \not\in \chora$, there is $r$ such that \eqref{eq:thereisr}
  holds.
By the well-branchedness of $\chora$, there is $r'$ such that
  
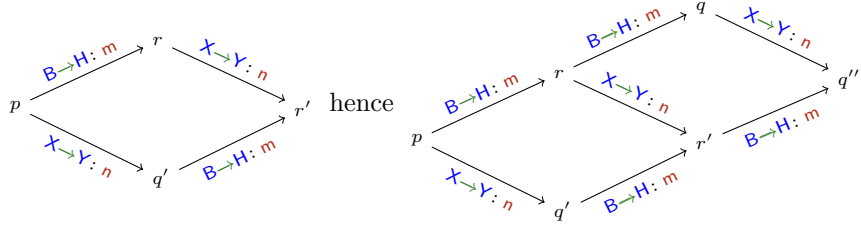
\[
    \begin{tikzpicture}[node distance = .5cm and 1.5cm]
      \node (p) {$p$};
      \node[above right = of p] (r) {$r$};
      \node[below right = of p] (q') {$q'$};
      \node[below right = of r] (r') {$r'$};
      \path[->] (p) edge node[sloped,above] {$\gint[][b][@][h]$} (r);
      \path[->] (p) edge node[sloped,below] {$\gint[][x][n][y]$} (q');
      \path[->] (r) edge node[sloped,above] {$\gint[][x][n][y]$} (r');
      \path[->] (q') edge node[sloped,below] {$\gint[][b][@][h]$} (r');
    \end{tikzpicture}
    \text{ hence }
    \begin{tikzpicture}[node distance = .5cm and 1.5cm]
      \node (p) {$p$};
      \node[above right = of p] (r) {$r$};
      \node[below right = of p] (q') {$q'$};
      \node[below right = of r] (r') {$r'$};
      \node[above right = of r] (q) {$q$};
      \node[below right = of q] (q'') {$q''$};
      \path[->] (p) edge node[sloped,above] {$\gint[][b][@][h]$} (r);
      \path[->] (p) edge node[sloped,below] {$\gint[][x][n][y]$} (q');
      \path[->] (r) edge node[sloped,above] {$\gint[][x][n][y]$} (r');
      \path[->] (q') edge node[sloped,below] {$\gint[][b][@][h]$} (r');
      \path[->] (r) edge node[sloped,above] {$\gint[][b][@][h]$} (q);
      \path[->] (q) edge node[sloped,above] {$\gint[][x][n][y]$} (q'');
      \path[->] (r') edge node[sloped,below] {$\gint[][b][@][h]$} (q'');
    \end{tikzpicture}
  \]
  for a $q''$ where the diamond $r$-$q$-$r'$-$q''$ exists by the
  well-branchedness of $\chora$.
Therefore, in $\sync \chora h k$ there is a diamond
  $p$-$q$-$q'$-$q''$ with labels $\gint[][b][@][h]$ and
  $\gint[][x][n][y]$, contrary to our hypothesis.
  \\
  If
  $\cauttr[@][{\gint[][b][@][k]}][r][][]
  \cauttr[][{\gint[][h][@][a]}][q][][]$
  the proof is similar.

  If $t' \not\in \chora$, there is $r$ such that
  
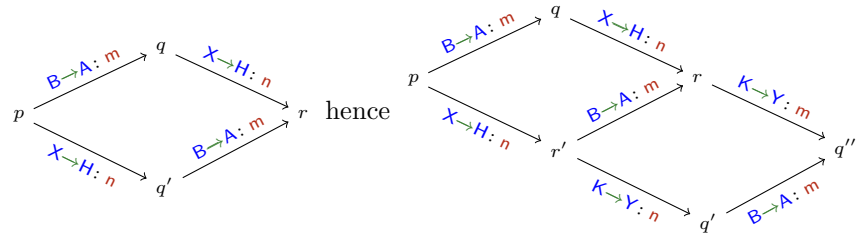
\[
    \begin{tikzpicture}[node distance = .5cm and 1.5cm]
      \node (p) {$p$};
      \node[above right = of p] (q) {$q$};
      \node[below right = of q] (r) {$r$};
      \node[below right = of p] (q') {$q'$};
      \path[->] (p) edge node[sloped,above] {$\gint[][b][@][a]$} (q);
      \path[->] (p) edge node[sloped,below] {$\gint[][x][n][h]$} (q');
      \path[->] (q) edge node[sloped,above] {$\gint[][x][n][h]$} (r);
      \path[->] (q') edge node[sloped,above] {$\gint[][b][@][a]$} (r);
    \end{tikzpicture}
    \text{ hence }
    \begin{tikzpicture}[node distance = .5cm and 1.5cm]
      \node (p) {$p$};
      \node[above right = of p] (q) {$q$};
      \node[below right = of q] (r) {$r$};
      \node[below right = of p] (r') {$r'$};
      \node[below right = of r] (q'') {$q''$};
      \node[below right = of r'] (q') {$q'$};
      \path[->] (p) edge node[sloped,above] {$\gint[][b][@][a]$} (q);
      \path[->] (p) edge node[sloped,below] {$\gint[][x][n][h]$} (r');
      \path[->] (r') edge node[sloped,below] {$\gint[][k][n][y]$} (q');
      \path[->] (q) edge node[sloped,above] {$\gint[][x][n][h]$} (r);
      \path[->] (r') edge node[sloped,above] {$\gint[][b][@][a]$} (r);
      \path[->] (r) edge node[sloped,above] {$\gint[][k][@][y]$} (q'');
      \path[->] (q') edge node[sloped,below] {$\gint[][b][@][a]$} (q'');
    \end{tikzpicture}
  \]
  where the diamond $r'$-$r$-$q'$-$q''$ exists by the
  well-branchedness of $\chora$ and the thesis follows as before.
  
  \paragraph{Case \eqref{it:unware}}
  The assumption $\proj{\pi_1}{x} \neq \proj{\pi_2}{x}$ implies that
  \begin{enumerate}[(a)]
  \item \label{it:sender} \p[x] is the sender in a transition of
    $\pi_1$ or of $\pi_2$ (otherwise both projections would be the
    automaton accepting the empty word), or
  \item \label{it:receiver} \p[x] is the receiver in a transition of
    $\pi_1$ or of $\pi_2$, but not both (otherwise either the inputs
    should be different by well-branchedness of $\chora$ contrary to
    our hypothesis).
  \end{enumerate}
With no loss of generality, suppose that \p[x] is the sender or the
  receiver of a transition in $\pi_1$.
\begin{description}
  \item[Case \eqref{it:sender}] 
Let $\pi'_1$ be the longest prefix of $\pi_1$ whose projection on
    \p[x] is also the projection of a prefix of $\pi_2$, and $t$ the
    first transition in $\pi_1$ after $\pi'_1$.
Then, for some $\pi_1''$, $\pi_2'$, and $\pi_2''$ we have
    \[
      \pi_1 = \pi_1', t, \pi_1'',
      \qquad
      \pi_2 = \pi_2', \pi_2'', \qquad
      \proj{\pi_1'}{x} = \proj{\pi_2'}{x}, \qquad\text{and}\qquad
      \proj{\big(t,\ \pi_1''\big)}{x} \neq \proj{\pi_2''}{x}
    \]
Using the construction of the proof of \cref{lemma:incl}, we have
    two co-initial runs $\hat \pi_1$ and $\hat \pi_2$ in $\chora$ such
    that, for $i \in \{1,2\}$, $\hat \pi_i$ has the transitions of
    $\chora$ occurring in $\pi_i$ preserving the order they have in
    $\pi_i$.
Note that the construction guarantees that $\hat \pi_1$ and
    $\hat \pi_2$ are transition disjoint and \cref{cor:incl} ensures
    that they are acyclic.
Moreover, if $\hat \pi_1$ and $\hat \pi_2$ are not co-final, then
    they can be extended to form a span adding transition from the
    maximal states of $\pi_1$ and $\pi_2$.
Let $\hat \pi_1'$ and $\hat \pi_2'$ be the part of $\hat \pi_1$
    and $\hat \pi_2$ which include the transitions of $\pi_1'$ and
    $\pi_2'$; then \p[x] cannot behave uniformly $\hat \pi_1'$ and
    $\hat \pi_2'$ (otherwise the first action of \p[x] distinguishing
    its behaviour on the two runs would be the output at $t$ and this
    is not possible by the well-branchedness of $\chora$).
This implies that the first actions distinguishing the behaviour
    of \p[x] on $\hat \pi_1'$ and $\hat \pi_2'$ should be labelled
    with $\gint[][h][@][x]$ and $\gint[][K][@][x]$ (obtained through
    the construction of \cref{lemma:incl} which should have replaced a
    transition in $\pi_1'$ and in $\pi_2'$ with receiver \p[x]).
Then the univocity of $\chora$ is violated.
  \item[Case \eqref{it:receiver}] As in the previous case, we can use
    the construction of \cref{lemma:incl} to extend runs
    $\pi_1$ and $\pi_2$ to two coinitial runs
    $\pi_1'$ and $\pi_2'$ in $\chora$.
Notice that this construction cannot yield a transition with
    receiver \p[x] in $\pi_2'$ since it at most adds transitions where
    receivers are either \iH or \iK.
Then, the span $(\pi_1,\pi_2)$ cannot be cofinal, otherwise we
    would violate the well-branchedness of $\chora$.
Let $q$ and $q'$ the states reached by $\pi_1$ and $\pi_2$ (and
    hence the states reached by $\pi_1'$ and $\pi_2'$).
By the well-brancheness of $\chora$, $\pi_2'$ cannot be maximal,
    hence there should be a transition from $q'$ involving an
    interface which the blending operation cuts away (otherwise $q'$
    would not be the last state on $\pi_2$) because it cannot be
    blended with a forwarding transition involving the other
    interface.
Call such transition $t$ and assume that its arrival state is
    $q''$ and, with no loss of generality, that it involves \iH.
    
    If \iH is the receiver of $t$ then, by reflectiveness
    (condition \eqref{it:out} in \cref{def:reflective}), we can find a
    transition $\cauttr[q''][{\gint[][h][@][Y]}][q'''][][]$ which
    violates the maximality of $\pi_2$ in $\sync \chora h k$.
Likewise, if \iH is the receiver of $t$ then, by reflectiveness
    (condition \eqref{it:in} in \cref{def:reflective}), we can find a
    transition $\cauttr[q''][{\gint[][h][@][Y]}][q'''][][]$ which
    violates the maximality of $\pi_2$ in $\sync \chora h k$.
  \end{description}
  In all the cases we obtain a contradiction, which gives the thesis.
\qed
\end{proof}

The following result is immediate from \cref{prop:wb-vs-sync}.
\begin{corollary}\label{cor:wb-vs-sync}
  Given a well-branched c-automaton $\chora$ and participants \iH and
  \iK such that $\chora$ is \iH- and \iK-univocal and reflective on
  \iH and \iK, then $\sync \chora h k$, if defined, is well-branched.
\end{corollary}

\subsection{Modular development of choreographies}
In the previous section we studied blending in isolation. Now, as
informally described at the beginning of \cref{sec:comp}, we show how
to combine blending with product to provide a composition operator,
which is the basis of our modular development of choreographies.

We first define the product of c-automata. The product will be defined on c-automata having disjoint sets of participants.
We wish in fact the systems we compose to interact only through the interfaces.
\begin{definition}[Product of c-automata]
 The \emph{product} of two c-automata is the standard product of FSA if the
 automata have disjoint sets of participants and it is undefined
 otherwise.
\end{definition}

We can now define the composition operation.

\begin{definition}[Composition]\label{def:comp}
  Let $\chora$ and $\chora[B]$ be two c-automata, \iH and
  \iK be participants respectively in $\chora$ and in $\chora[B]$.
The \emph{composition of $\chora$ and $\chora[B]$ via \iH and \iK}
  is
\[
    \compose \chora {\chora[B]}{H}{K} \ = \
    \sync{({\chora[A]}{\times}{\chora[B]})} H K
  \]
  provided that $\chora$ and $\chora[B]$ have disjoint participants.
\end{definition}

A reasonable modular development of systems must rely on composition
operations that do not \quo{break} the relevant properties that
modules separately possess.
The relevant properties are those considered in
\cref{sec:wfsound}, that are guaranteed by projecting well-formed c-automata.
Thus, we will show below that composition preserves well-formedness.

\ifhideproofs
\begin{proposition}[Composition preserves well-formedness]\label{prop:wf-presbycomp}
  Given two participants $\HH\in\chora[A]$ and $\KK\in\chora[B]$,
  where $\chora[A]$ and $\chora[B]$ are well-formed c-automata,
  $\compose{\chora[A]}{\chora[B]}H K$ is well-formed.
\end{proposition}
\else
We begin by showing that composition does preserve well-branchedness
and in order to do that the following definition will be handy.
\\
\textbf{Notation.} We shall often denote by $\p\in\chora$ the fact
that $p$ is a participant in the c-automaton $\chora$ and denote the
set of its participants by $\ptpset(\chora)$

\begin{definition}
Let $\pi$ be a run in $\sync{\chora{\times}\chora[B]} h k$. We define the {\em  restriction of $\pi$
to $\chora[A]$}, dubbed $\restr{\pi}\chora$, as follows
\[
\begin{array}{rcl}
\restr{\varepsilon}{\chora} & = & \varepsilon \hspace{4mm}\text{(where $\varepsilon$ is the empty string)}\\
\restr{(t\,\pi')}{\chora} &= &\left\{\begin{array}{l@{\hspace{4mm}}l}
                                               t\, \restr{\pi'}{\chora} & \text{if $t\in\chora$}\\
                                               \restr{\pi'}{\chora} &  \text{if $t\in\chora[B]$}\\
                                               {\cauttr[@][{\gint[][@][@][h]}][r][][]}\;\restr{\pi'}{\chora} & \text{if $t=\cauttr$}\\
                                                                              & \text{where (*)}
                                               \end{array}\right.
\end{array}
\]
(*)  $r$ is such that $\cauttr$ has been obtained by blending
$\cauttr[@][{\gint[][@][@][h]}][r][][]$ and $\cauttr[r][{\gint[][K][@][B]}][q][][]$ in
$\chora{\times}\chora[B]$.\\
Similarly for the definition of $\restr{\pi}{\chora[B]}$.
\end{definition}

\begin{lemma}
\label{lem:restr}
Let $\pi$ be a run in $\sync{\chora{\times}\chora[B]} h k$. Then  $\restr{\pi}{\chora}$
(resp. $\restr{\pi}{\chora[B]}$) is a run in
$\chora$ (resp. $\chora[B]$).
\end{lemma}
\begin{proof}
Easy, by definition of product, blending and restriction $\restr{(\cdot)}{(\cdot)}$.
\qed
\end{proof}

\begin{proposition}[Composition preserves well-branchedness]\label{prop:wb-vs-comp}
  Given two participants $\HH\in\chora[A]$ and $\KK\in\chora[B]$,
  where $\chora[A]$ and $\chora[B]$ are well-branched,
  $\sync{(\chora[A]{\times}\chora[B])}{H}{K}$ is well-branched.
\end{proposition}
\begin{proof}
  By \emph{reductio ad absurdum}.
Let $p$ be a state in $\sync{(\chora[A]{\times}\chora[B])}{H}{K}$ with no selector.
Then one of the following must hold for each participant $\q$ of
  $\sync{(\chora[A]{\times}\chora[B])}{H}{K}$ (note that $\q \neq \iH$ and $\q \neq \iK$):
  \begin{enumerate}
  \item\label{it:output1} there are transitions
    $t = \cauttr[@][{\gint[][b][@][a]}][@][][]$ and
    $t' = \cauttr[@][{\gint[][b][@][a]}][q'][][]$ with $q \neq q'$, or
  \item\label{it:diamond1} there are transitions
    $t = \cauttr[@][{\gint[][b][@][a]}][@][][]$ and
    $t' = \cauttr[@][{\gint[][x][n][y]}][q'][][]$ such that
    $\p[x] \neq \q$ and $t$ and $t'$ are not concurrent, or else
  \item\label{it:unware1} there are a participant $\p[x] \neq \q$ and a
    $p$-span $(\pi_1,\pi_2)$ in $\sync{(\chora \times \chora[B])} h k$ such that ($i$)
    $\proj{\pi_1}{x} \neq \proj{\pi_2}{x}$ and ($ii$) the 
    first pair of labels where $\proj{\pi_1}{x}$ and
    $\proj{\pi_2}{x}$ differ are not 
    of the form
    $(\ain[C][X][@][m],\ain[D][X][@][n])$ with $\p[C] \neq \p[D]$ or
    $\msg \neq \msg[n]$
  \end{enumerate}
  We derive a contradiction in each of the cases above.
  In the following we can assume, without any loss of generality, that $\ptp[B]\in\chora[A]$.
\paragraph{Case \eqref{it:output1}} It is impossible that  $\ptp[A]\in\chora[A]$,
  otherwise $\chora$ would not be well-branched. So, necessarily 
  there must be $r$ and $r'$ such that  $r\neq r'$ and
  \begin{equation}\label{eq:thereisr1}
    \cauttr[@][{\gint[][b][@][h]}][r][][]
    \cauttr[][{\gint[][k][@][a]}][q][][]
    \hspace{4mm} \text{ and }  \hspace{4mm}
    \cauttr[@][{\gint[][b][@][h]}][r'][][]
    \cauttr[][{\gint[][k][@][a]}][q'][][]
  \end{equation}
  then $p$
  would have two transitions in $\chora$ with the same label violating the
  well-branchedness of $\chora$.
\paragraph{Case \eqref{it:diamond1}} Not both $t$ and $t'$ can be
  transitions of $\chora$, otherwise $\chora$ would not be
  well-branched.

  If $t \not\in \chora[A]$ and $t' \in \chora[B]$ , there is $r$ such that the first conjunct of  \eqref{eq:thereisr1}
  holds. Moreover, by definition of product we can infer that
  there is $r'$ and $q''$ such that
  
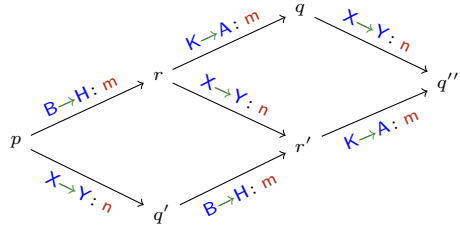
\[
    \begin{tikzpicture}[node distance = .5cm and 1.5cm]
      \node (p) {$p$};
      \node[above right = of p] (r) {$r$};
      \node[below right = of p] (q') {$q'$};
      \node[below right = of r] (r') {$r'$};
      \node[above right = of r] (q) {$q$};
      \node[below right = of q] (q'') {$q''$};
      \path[->] (p) edge node[sloped,above] {$\gint[][b][@][h]$} (r);
      \path[->] (p) edge node[sloped,below] {$\gint[][x][n][y]$} (q');
      \path[->] (r) edge node[sloped,above] {$\gint[][x][n][y]$} (r');
      \path[->] (q') edge node[sloped,below] {$\gint[][b][@][h]$} (r');
      \path[->] (r) edge node[sloped,above] {$\gint[][K][@][A]$} (q);
      \path[->] (q) edge node[sloped,above] {$\gint[][x][n][y]$} (q'');
      \path[->] (r') edge node[sloped,below] {$\gint[][k][@][a]$} (q'');
    \end{tikzpicture}
  \]
are in $\chora[A]{\times}\chora[B]$, and  hence

  \begin{equation}
    \label{eq:cgg}
    \begin{tikzpicture}[node distance = .5cm and 1.5cm]
      \node (p) {$p$};
      \node[above right = of p] (r) {$q$};
      \node[below right = of p] (q') {$q'$};
      \node[below right = of r] (r') {$q''$};
      \path[->] (p) edge node[sloped,above] {$\gint[][b][@][a]$} (r);
      \path[->] (p) edge node[sloped,below] {$\gint[][x][n][y]$} (q');
      \path[->] (r) edge node[sloped,above] {$\gint[][x][n][y]$} (r');
      \path[->] (q') edge node[sloped,below] {$\gint[][b][@][a]$} (r');
    \end{tikzpicture}
  \end{equation}
  is in $\sync{(\chora[A]{\times}\chora[B])}{H}{K}$.
  Contradiction.

  If $t \not\in \chora$ and $t' \in \chora$, then
  we would have 
  \[
    \begin{tikzpicture}[node distance = .5cm and 1.5cm]
      \node (p) {$p$};
      \node[above right = of p] (q) {$r$};
      \node[below right = of p] (r) {$q'$};
      \path[->] (p) edge node[sloped,above] {$\gint[][b][@][h]$} (q);
      \path[->] (p) edge node[sloped,above] {$\gint[][x][n][y]$} (r);
    \end{tikzpicture}
  \]
  both in $\chora$. Now, since we assumed that $\ptp[X]\neq\ptp[B]$, by well-branchedness of 
  $\chora$ we would have 
  \[
    \begin{tikzpicture}[node distance = .5cm and 1.5cm]
      \node (p) {$p$};
      \node[above right = of p] (r) {$r$};
      \node[below right = of p] (q') {$q'$};
      \node[below right = of r] (r') {$q''$};
      \path[->] (p) edge node[sloped,above] {$\gint[][b][@][H]$} (r);
      \path[->] (p) edge node[sloped,below] {$\gint[][x][n][y]$} (q');
      \path[->] (r) edge node[sloped,above] {$\gint[][x][n][y]$} (r');
      \path[->] (q') edge node[sloped,below] {$\gint[][b][@][h]$} (r');
    \end{tikzpicture}
  \]
  and hence
  \[
    \begin{tikzpicture}[node distance = .5cm and 1.5cm]
      \node (p) {$p$};
      \node[above right = of p] (r) {$r$};
      \node[below right = of p] (q') {$q'$};
      \node[below right = of r] (r') {$r'$};
      \node[above right = of r] (q) {$q$};
      \node[below right = of q] (q'') {$q''$};
      \path[->] (p) edge node[sloped,above] {$\gint[][b][@][h]$} (r);
      \path[->] (p) edge node[sloped,below] {$\gint[][x][n][y]$} (q');
      \path[->] (r) edge node[sloped,above] {$\gint[][x][n][y]$} (r');
      \path[->] (q') edge node[sloped,below] {$\gint[][b][@][h]$} (r');
      \path[->] (r) edge node[sloped,above] {$\gint[][K][@][A]$} (q);
      \path[->] (q) edge node[sloped,above] {$\gint[][x][n][y]$} (q'');
      \path[->] (r') edge node[sloped,below] {$\gint[][k][@][a]$} (q'');
    \end{tikzpicture}
  \]
  are in $\chora[A]{\times}\chora[B]$, where $r$-$q$ is in $\chora[B]$ and hence $q$-$q''$ and $r'$-$q''$
  exist by definition of product.
  So, we can obtain that (\ref{eq:cgg}) is in $\sync{(\chora[A]{\times}\chora[B])}{H}{K}$
  and we get a contradiction since we assumed $t$ and $t'$ not to be concurrent.
  
  If $t \not\in \chora$ and $\ptp[X]\in\chora[B]$ and $t' \not\in \chora[B]$, then,
  by reasoning similarly to the case ($t \not\in \chora[A]$ and $t' \in \chora[B]$)
  we would arrive to conclude (\ref{eq:cgg}) to be in $\sync{(\chora[A]{\times}\chora[B])}{H}{K}$.
  Contradiction.

  If $t \in \chora$ and $\ptp[X]\in\chora[A]$ and $t' \not\in \chora[A]$, then,
  we would have 
  \[
    \begin{tikzpicture}[node distance = .5cm and 1.5cm]
      \node (p) {$p$};
      \node[above right = of p] (q) {$q$};
      \node[below right = of p] (r) {$r$};
      \path[->] (p) edge node[sloped,above] {$\gint[][b][@][a]$} (q);
      \path[->] (p) edge node[sloped,above] {$\gint[][x][n][H]$} (r);
    \end{tikzpicture}
  \]
  both in $\chora$. Now, since we assumed that $\ptp[X]\neq\ptp[B]$, by the well-branchedness of $\chora$ there exists also the rest of the diamond and we can reason similarly to what done for the 
  case ($t \not\in \chora$ and $t' \in \chora$), getting to a contradiction with 
  $t$ and $t'$ not concurrent.

  If $t \in \chora$  and $t' \in \chora[B]$, then,
  we would infer the existence of   (\ref{eq:cgg}) by definition of product,
  getting a contradiction with $t$ and $t'$ not concurrent.

  \paragraph{Case \eqref{it:unware1}}
  Without loss of generality, we can assume $\ptp[X]\in\chora[A]$ (recall that we have also that $\ptp[X]\neq\ptp[H]$).\\
  The assumptions for the present case implies that the $p$-span ($\pi_1,\pi_2$) is  such that  
  there exist $t_1,t_2\not\in\chora[B]$ such that 
  \begin{enumerate}[a)]
  \item 
    either $t_1,t_2\in\chora[A]$ or [$t_1\in\chora[A]$ and $t_2\not\in\chora[A]$] or [$t_2\in\chora[A]$ and $t_1\not\in\chora[A]$];
  \item
    $\pi_1 = \pi_1'\, t_1\, \pi_1''$ and
    $\pi_2 = \pi_2'\, t_2\, \pi_2''$, for some $\pi_1''$, $\pi_2'$, and $\pi_2''$;
  \item
    \label{lab:lproj}
    $ \proj{\pi_1'}{x} = \proj{\pi_2'}{x}$
  \item
    \label{lab:eqlproj}
    $ (\proj{t_1}{x},\proj{t_2}{x}) \neq (\ain[C][X][@][m],\ain[D][X][@][n])$ 
    with $\p[C] \neq \p[D]$ or $\msg \neq \msg[n]$;
  \item
    $\pi_1'$ and $\pi_2'$ with the above properties have maximal lenght.
  \end{enumerate}
  Without loss of generality we can assume $t_1\in\chora[A]$.
  Let now $\hat \pi_1$ and $\hat \pi_2$ be the runs of $\chora[A]$ 
  defined out of $ \pi_1$ and $ \pi_2$ as in \cref{lem:restr}.
  It is not difficult to check that  $\hat \pi_1$ and $\hat \pi_2$ are 
  co-initial. 
Note that the construction guarantees that $\hat \pi_1$ and
  $\hat \pi_2$ are transition disjoint and acyclic.
Moreover, if $\hat \pi_1$ and $\hat \pi_2$ are not co-final, then
  $\pi_1$ and $\pi_2$
  can be extended (by adding transition from the
  maximal states of $\pi_1$ and $\pi_2$) in such a way\\
  \centerline{
    $\hat \pi_1 = \hat{\pi_1'}\, (\proj{(\restr{t_1}{\chora})}{X})\, \hat{\pi_1''}\, \hat{\pi_1'''}$
    \hspace{4mm}and\hspace{4mm}
    $\hat \pi_2 = \hat{\pi_2'}\, (\proj{(\restr{t_2}{\chora})}{X})\, \hat{\pi_2''}\, \hat{\pi_2'''}$}
  So  that $(\hat \pi'_1,\hat \pi'_2)$ forms a $p$-span of $\chora$.
  Besides, by (\ref{lab:lproj}) above, it is possible to infer that
  \begin{equation}
    \label{eq:pipiprojeq}
    {\hat{\pi_1'}} = {\hat{\pi_2'}} 
  \end{equation}
  
  We proceed now by considering the two possible cases:
  \begin{description}
  \item $t_1,t_2\in\chora[A]$\\
    By definition of restriction, we get  $\restr{t_i}{\chora} = t_i$ for $i=1,2$.
    Hence the above properties (\ref{eq:pipiprojeq})
    and (\ref{lab:eqlproj}) of 
    the $p$-span $(\hat \pi_1,\hat \pi_2)$ immediately
    contradict the  well-branchedness of $\chora$.
  \item $t_1\in\chora[A]$ and $t_2\not\in\chora[A]$

    We consider two further possibilities
    \begin{description}
    \item
      $t_2$  has been obtained by blending
      $\cauttr[p'][{\gint[][X][@][h]}][r][][]$ and $\cauttr[r][{\gint[][K][@][Y]}][q][][]$ in
      $\chora{\times}\chora[B]$.\\
      Since $t_1\in\chora[A]$, by definition of restriction we have $\restr{t_1}{\chora} = t_1$.
      Moreover, by definition, $\proj{(\restr{t_2}{\chora})}{X} = \aout[X][H][][m]$.
      Then the $p$-span $(\hat \pi_1,\hat \pi_2)$ 
      contradicts the  well-branchedness of $\chora$ both in case $\proj{t_1}{X}$ is an input
      and in case it is an output.
      
    \item
      $t_2$  has been obtained by blending
      $\cauttr[p'][{\gint[][Y][@][K]}][r][][]$ and $\cauttr[r][{\gint[][H][@][X]}][q][][]$ in
      $\chora{\times}\chora[B]$.\\
      In case $\proj{t_1}{X}$ is an output, we get a contradiction with the  well-branchedness of $\chora$.\\
      In case, instead, $\proj{t_1}{X}$ is an input,  we have necessarily 
      that $t_1$ is of the form $\cauttr[q'][{\gint[][C][n][X]}][q''][][]$ with $\ptp[C]\in\chora[A]$.
      Since we have also  that, necessarily, $\ptp[Y]\in\chora[B]$, it follows that
      $(\proj{t_1}{X},\proj{t_2}{X}) = (\ain[C][X][][n] , \ain[Y][X][][m])$ with  $\ptp[Y]\neq\ptp[C]$.
      This, however, is impossible. In fact in the present main case we are assuming condition
      (\ref{it:unware}) -- stated at the beginning of the proof -- to hold, according to which the
      first pair of labels where $\proj{\pi_1}{x}$ and
      $\proj{\pi_2}{x}$ differ are not 
      of the form
      $(\ain[C][X][@][m],\ain[D][X][@][n])$ with $\p[C] \neq \p[D]$ or
      $\msg \neq \msg[n]$.
\qed
    \end{description}
  \end{description}
\end{proof}

It is worth noticing that composition can be proven to preserve
well-branchedness without any univocity assumptions, which has been
instead necessary to prove well-branchedness preservation by blending
in \cref{prop:wb-vs-sync}.
This is due to the fact that \cref{prop:wb-vs-sync} deals with blending
on general c-automata, not those obtained from product.

For what concerns well-sequencedness, since we already know that
blending preserve well-sequencedness (see \cref{prop:blendpresws}), it
is enough to show well-sequencedness to be preserved by product.

\begin{lemma}[Product preserves well-sequencedness]\label{lem:productPreservesWS}
Let $\chora[A]$ and $\chora[B]$ be two well-sequenced
  c-automata. Then $\chora[A]{\times}\chora[B]$ is well-sequenced (if
  defined).
\end{lemma}
\begin{proof}
  Let us have two consecutive transitions in
  $\chora[A]{\times}\chora[B]$ with disjoint sets of participants.  If
  the two transitions belong both to $\chora[A]$ or both to
  $\chora[B]$, we have a diamond by well-sequencedness of $\chora[A]$
  and $\chora[B]$. Otherwise we have a diamond by definition of
  product.\qed
\end{proof}

Well-formedness preservation by composition can hence be obtained as a
corollary.
\begin{corollary}[Composition preserves well-formedness]\label{prop:wf-presbycomp}
  Given two participants $\HH\in\chora[A]$ and $\KK\in\chora[B]$,
  where $\chora[A]$ and $\chora[B]$ are well-formed c-automata,
  $\compose{\chora[A]}{\chora[B]}H K$ is well-formed.
\end{corollary}
\begin{proof}
  Well sequencedness of $\compose{\chora[A]}{\chora[B]}H K$ is a
  consequence of \cref{lem:productPreservesWS} and
  \cref{prop:blendpresws}, while its well branchedness is proved
  in \cref{prop:wb-vs-comp}.
\qed
\end{proof}
\fi

In order to show that composition does not lose behaviours (that is,
that \cref{lemma:bisim} holds) we also need to show that
$\chora[A]{\times}\chora[B]$ is reflective on \iH and \iK. We will
show below that in the special case of product automata,
reflectiveness can be proved using a simpler criterion, called
compatibility, which is checked on the behaviour of the projections on
\iH and \iK.

Informally, two CFSMs $\aCM_1$ and $\aCM_2$ are {\em compatible} if
$\aCM_1$ is bisimilar to the dual of $\aCM_2$ if the communicating
partners are abstracted away.

In order to define compatibility, a few simple definitions are handy.
Let
$\lio = \Set{\ain[][][][m], \aout[][][][m] \mid \msg[m]\in\msgset}$
and define the functions
\[
  (\cdot)^{\setminus \ptpset} : \lact\cup\Set{\epsilon}\rightarrow
  \lio\cup\Set{\epsilon}
  \qqand
  \Dual{(\cdot)} : \lio\cup\Set{\epsilon} \rightarrow \lio\cup\Set{\epsilon}
\]
by the following clauses
\[
  \begin{array}[c]{c@{\qquad}c@{\qquad}c}
    {(\ain)^{\setminus \ptpset}} = \ain[][][][m]
    &
      {(\aout)^{\setminus \ptpset}}= \aout[][][][m]
    &
      {\epsilon^{\setminus \ptpset}}= \epsilon
    \\[1em]
    \Dual{\ain[][][][m]} = \aout[][][][m]
    &
      \Dual{\aout[][][][m]} = \ain[][][][m]
    &
      \Dual{\epsilon}=\epsilon
  \end{array}
\]
which extend to CFSMs in the obvious way: given a CFSM
$\aCM = \conf{\sset,\lact,\tset,q_0}$, we define
$\aCM^{\setminus \ptpset} = \conf{\sset,\lio,\tset^{\setminus
    \ptpset},q_0}$ where
$\tset^{\setminus \ptpset} = \Set{p\trans[{}]{\alpha^{\setminus
      \ptpset}}q \mid p\trans[{}]{\alpha}q \in \tset}$; and likewise
for $\Dual{\aCM}$.

\begin{definition}[Compatibility]\label{def:compatibility}
  Two CFSMs $\aCM_1$ and $\aCM_2$ are \emph{compatible} if
  $\aCM_1^{\setminus \ptpset}$ is bisimilar to
  $\Dual{\aCM_2^{\setminus \ptpset}}$.
  \\
  Given $\chora$ and $\chora[B]$, participants \iH of $\chora$ and \iK of
  $\chora[B]$ are \emph{compatible} if $\proj{\chora}H$ and
  $\proj{\chora[B]}K$ are compatible.
\end{definition}

In general, reflectiveness implies compatibility, but, in the special
case of product automaton, compatibility of the components implies
reflectiveness of the product.

\begin{proposition}[Reflectiveness implies compatibility]\label{prop:refl2com}
  If a c-automaton $\chora$ is reflective on \iH and \iK, then
  $\proj{\chora}{H}$ and
  $\proj{\chora}{K}$ are compatible.
\end{proposition}
\begin{proof}
  We have to show that $\proj{\chora}{H}$ and $\proj{\chora}{K}$
  are compatible, namely that $\proj{\chora}{H}^{\setminus
    \ptpset}$ is bisimilar to $\Dual{\proj{\chora}{K}^{\setminus \ptpset}}$.

  Remember that the states of the projections are sets of states of the
  original c-automata.
  
  Let us consider the relation $\mathcal{R} \subseteq \RS[\proj{\chora}{H}] \times \RS[\proj{\chora}{K}]$ defined as:

  $$\mathcal{R} = \{(S_{\p[H]},S_{\p[K]}) \, | \, S_{\p[H]} \in \RS[\proj{\chora}{H}], S_{\p[K]} \in \RS[\proj{\chora}{K}], S_{\p[H]} \cap S_{\p[K]} \neq \emptyset\}$$

  By definition it relates the two initial states. Also, consider a
  challenge from $\proj{\chora}{H}$ (the other case is analogous),
  that is a transition $S_{\p[H]} \trans[{}]{\alpha} S'_{\p[H]}$.  Let
  us consider the case $\alpha = \ain[][][][m]$ (the case $\alpha =
  \aout[][][][m]$ is analogous), that is we have a transition $S_{\p[H]} \trans[{}]{\ain[A][H][][m]} S'_{\p[H]}$ in $\proj{\chora}{H}$.
  
  By reflectiveness we have a transition
  $S''_{\p[K]} \trans[{}]{\aout[K][B][][m]} S'''_{\p[K]}$ in $\proj{\chora}{K}$
  with $S'_{\p[H]} \cap S''_{\p[K]} \neq \emptyset$.

  Also, in $\chora$ we have transitions
  $\cauttr[s][{\gint[][a][@][h]}][s'][][]$ and
  $\cauttr[s'][{\gint[][k][@][b]}][s''][][]$. By projecting to $\p[H]$
  we get $s'' \in S'_{\p[H]}$ (since the second transition projects to
  $\epsilon$) as well as $s'' \in S'''_{\p[K]}$ and $S_{\p[K]}
  = S''_{\p[K]}$ (since the first transition projects to $\epsilon$).

  As a result, $\p[K]$ can answer the challenge with $S_{\p[K]} \trans[{}]{\aout[][][][m]} S'''_{\p[K]}$ with $S'_{\p[H]} \cap S'''_{\p[K]} \supseteq \{s''\} \neq \emptyset$ as desired.
\qed
\end{proof}

\begin{theorem}[Compatibility implies reflectiveness of products]\label{thm:comprod}
  If $\chora = \chora[B] \times \chora[C]$ and
  $\proj{\chora[B]}{H}$ and
  $\proj{\chora[C]}{K}$ are compatible then $\chora$ is reflective on \iH and \iK.
\end{theorem}
\begin{proof}
  For each $\cauttr[S][{\ain[A][H][][m]}][S'][][]$ in
  $\proj{\chora}{H}$ we have that $S = X \times Q_C$ and
  $S' = X' \times Q_c$ with $X,X' \subseteq Q_B$.
Hence, $\cauttr[X][{\ain[A][H][][m]}][X'][][]$ in
  $\proj{\chora[B]}{H}$.
Thanks to the bisimilarity condition there is
  $\cauttr[S_1][{\aout[K][B][][m]}][S'_1][][]$ in
  $\proj{\chora[C]}{K}$.
Hence,
  $\cauttr[Q_B \times S_1][{\aout[K][B][][m]}][Q_B \times S'_1][][]$
  in $\proj{\chora}{K}$.
Since $X' \times Q_c \cap Q_B \times S_1 \neq \emptyset$ the first
  condition follows.
The proofs of the other conditions are similar.
\qed
\end{proof}

As a corollary, if \iH in $\chora$ and \iK in $\chora[B]$ are
compatible, then roles different from \iH and \iK behave the same in
the original c-automata and in the composition.

\begin{corollary}
  If \iH in $\chora$ and \iK in $\chora[B]$ are compatible then
  $\proj{\chora}{A}$ and $\proji{\compose{\chora[A]}{\chora[B]}H K}{A}{H}{K}$ are
  bisimilar, for each participant $\p \neq \iH,\iK$.
\end{corollary}
\begin{proof}
  From \cref{lemma:bisim}, where the condition of reflectiveness on
  $\chora \times \chora[B]$ holds thanks to \cref{thm:comprod}.
\qed
\end{proof}

We show now, step by step, how composition works in a simple example.
\begin{example}[Composition at work]\label{ex:comp@work}
  Let us compose the c-automata below on the interfaces $\HH$ and $\KK$.
\[
  \begin{tikzpicture}[node distance=1.9cm]
    \tikzstyle{every state}=[cnode]
    \tikzstyle{every edge}=[carrow]
\node[state] (zero) {$0$};
    \node[state] (one) [left of=zero,xshift=-.3cm] {$1$};
    \node[draw=none,fill=none] (start) [above = 0.3cm  of zero]{$\chora$};
    \node[state] (two) [right of=zero,xshift=.3cm]  {$2$};
\path (start) edge node {} (zero)
    (zero) edge[bend right = 20] node[above] {$\gint[][H][ack][Q]$} (one)
    edge[bend left = 20] node[above] {$\gint[][H][nack][Q]$} (two)
    (one) edge[bend right=20] node[below] {$\gint[][Q][alt][I]$} (zero)
    (two)    edge    [bend left=20]        node [below] {$\gint[][Q][text][I]$} (zero)
    ;
  \end{tikzpicture}
  \qqand
  \begin{tikzpicture}[node distance=1.9cm]
    \tikzstyle{every state}=[cnode]
    \tikzstyle{every edge}=[carrow]
\node[state] (zero) {$0$};
    \node[state] (one) [left  of=zero, xshift=-.3cm]  {$1$};
    \node[draw=none,fill=none] (start) [above = 0.3cm  of zero]{$\chora[B]$};
    \node[state] (two) [right  of=zero, xshift=.3cm]  {$2$};
\path (start) edge node {} (zero) 
    (zero) edge[bend right=20] node[above] {$\gint[][A][ack][K]$} (one)
    edge[bend left=20] node[above] {$\gint[][A][nack][K]$} (two)
    (one) edge[bend right=20] node[below] {$\gint[][B][go][A]$} (zero)
    (two) edge[bend left=20] node[below] {$\gint[][B][go][A]$} (zero)
    ;
  \end{tikzpicture}
\]
Note that $\chora$ and $\chora[B]$ are well-formed, $\chora$ is \iH-univocal and $\chora[B]$ is \iK-univocal.
As discussed, we shall first compute $\chora \times \chora[B]$ and
then blending the two roles $\HH$ and $\KK$ in $\chora \times
\chora[B]$. The resulting c-automaton is well-formed thanks to
\cref{prop:wf-presbycomp}.

To save space, we represent these steps in the following decorated
c-automaton.
\vspace{-1.5cm}
\[
  \begin{tikzpicture}[node distance=1.9cm]
    \tikzstyle{every state}=[cnode]
    \tikzstyle{every edge}=[carrow]
\node[state] (00)                                         {$0,0$};
    \node[draw=none,fill=none] (start) [above left = .3cm of 00, xshift=10pt]{};
\node[state] (10)  [below right of=00, xshift=10mm, yshift=-10mm]  {$1,0$};
    \node[state] (01) [below  left of=00, xshift=-10mm, yshift=-10mm ] {$0,1$};
    \node[state] (20) [left of=01, xshift=-10mm] {$2,0$};
\node[state] (11) [below  of=00, yshift=-22mm] {$1,1$};
    \node[state] (02) [ right of=10, xshift=10mm] {$0,2$};
    \node[state, fill = black!20] (12) [below left of=02, yshift=-8mm ] {$1,2$};
\node[state, fill = black!20] (21) [below  right of=20, yshift=-8mm ] {$2,1$};
\node[state] (22) [above  of=00, yshift=2mm] {$2,2$};
\path  (start) edge node {} (zero) 
    (00)    edge[black!20] node [above] {$\gint[][H][ack][Q]$} (10)
    edge[black!20] node [above] {$\gint[][H][nack][Q]$} (20)
    edge[black!20, dashed] node [above] {$\gint[][A][ack][K]$} (01)
    edge[black!20, double] node [above] {$\gint[][A][nack][K]$} (02)
    edge[dotted] node [above] {$\gint[][A][nack][Q]$} (22)
    edge[dotted] node [above] {$\gint[][A][ack][Q]$} (11)
    (20)    edge    [bend left=40]        node [above] {$\gint[][Q][text][I]$} (00)
    edge    [black!20, bend right=60] node [above] {$\gint[][A][ack][K]$} (21)
    edge    [black!20, bend left=80] node [above] {$\gint[][A][nack][K]$} (22)
    (02)    edge    [bend right=40] node [above] {$\gint[][B][go][A]$} (00)
    edge    [black!20, double, bend right=80] node [above] {$\gint[][H][nack][Q]$} (22)
    edge    [black!20, bend left=60] node [above] {$\gint[][H][ack][Q]$} (12)
    (01)    edge[black!20, bend left=40] node [above] {$\gint[][H][nack][Q]$} (21)
    edge    [black!20, dashed, bend right=40] node [above] {$\gint[][H][ack][Q]$} (11)
    edge    [bend right=40] node [above] {$\gint[][B][go][A]$} (00)
    (10)    edge[black!20, bend left=40] node [above] {$\gint[][A][ack][K]$} (11)
    edge    [black!20, bend right=40] node [above] {$\gint[][A][nack][K]$} (12)
    edge    [bend left=40] node [above] {$\gint[][Q][alt][I]$} (00)
    (21)    edge[black!20] node [above] {$\gint[][B][go][A]$} (20)
    edge[black!20] node [above] {$\gint[][Q][text][I]$} (01)
    (11)    edge    [bend left=40] node [below] {$\gint[][B][go][A]$} (10)
    edge    [bend right=40] node [below] {$\gint[][Q][alt][I]$} (01)
    (22)    edge    [bend right=40] node [above] {$\gint[][B][go][A]$} (20)
    edge    [bend left=40] node [above] {$\gint[][Q][text][I]$} (02)
    (12)     edge[black!20] node [above] {$\gint[][B][go][A]$} (10)
    edge[black!20] node [above] {$\gint[][Q][alt][I]$} (02)
;
  \end{tikzpicture}
\]
where the decorations have the following interpretation:
\begin{itemize}
\item the transitions of $\chora \times \chora[B]$ are all the
  transitions except the dotted ones, that from state $(0,0)$ go to state
  $(1,1)$ and to state $(2,2)$, which are added by blending of the
  dashed transitions and the doubly-lined ones, respectively;
\item the grey transitions are the ones removed either because of
  blending (like the dashed and doubly-lined ones) or removal of
  unreachable states (like those involving greyed states $(2,1)$ and
  $(1,2)$);
\item the non-grey transitions (regardless if they are solid or
  dotted) form the final result of the composition.
\end{itemize}
We now apply step by step the blending algorithm on
${{\chora}{\times}{\chora[B]}}$ and $\HH$ and $\KK$.
\begin{quote}\em
  {\bf step \ref{stepI}:} Each transition
  $\cauttr[@][{\gint[][@][@][H]}][@][][]$ is removed, and for each
  transition $\cauttr[q][{\gint[][K][@][B]}][r][][]$ a transition
  $\cauttr[@][{\gint[]}][r][][]$ is added provided that $\p \neq \q$;
  and similarly, by swapping $\p[H]$ and $\p[K]$.
\end{quote}
In this step, the dashed and the doubly-lined transitions are replaced
 by the blended transitions, represented by
the dotted arrows.
\begin{quote}\em
  {\bf step \ref{stepII}:} Transitions involving neither \iH nor \iK
  are preserved, whereas all other transitions are removed.
\end{quote}
This step gets rid of grey transitions, but for the ones outgoing from the greyed states, which are removed in the next step.
\begin{quote}\em
  {\bf step \ref{stepIII}:} States and transitions unreachable from
  state $(0,0)$ are removed.
\end{quote}
This last step hence removes the greyed states and their outgoing
transitions, obtaining the overall composition of
$\chora$ and $\chora[B]$ through $\HH$ and $\KK$.
 \finex
\end{example}

It is worth pointing out that the composition operation can be
performed on any pair of roles, that is any role can be chosen as an
interface. However, if the conditions discussed above are not
verified, then the composition may not be well-behaved.
In particular, it may happen that the behaviour of some 
participants in the original systems (different from \iH and \iK) will
change, in particular some transitions may be disabled.

This can be seen on the simple example below.

$
\begin{array}{c@{\hspace{4mm}}c@{\hspace{4mm}}c}
\begin{tikzpicture}[node distance=1.9cm,]
     \tikzstyle{every state}=[cnode]
     \tikzstyle{every edge}=[carrow]

  \node[state] (zero)                        {$0$};
   \node[draw=none,fill=none] (start) [ left = 0.3cm  of zero]{$\chora[A]$};
    \node[state] (one)  [ right  of =zero]                      {$1$};

  \path  (start) edge node {} (zero) 
            (zero)    edge             node [above] {$\gint[][A][m][H]$} (one)
             ;

 \end{tikzpicture}
 &
 \text{and}
 &
\begin{tikzpicture}[node distance=1.9cm,]
     \tikzstyle{every state}=[cnode]
     \tikzstyle{every edge}=[carrow]

  \node[state] (zero)                        {$0$};
   \node[draw=none,fill=none] (start) [ left = 0.3cm  of zero]{$\chora[B]$};
    \node[state] (one)  [ right  of =zero]                      {$1$};

  \path  (start) edge node {} (zero) 
            (zero)    edge             node [above] {$\gint[][D][m][K]$} (one)
             ;

 \end{tikzpicture}
 \end{array}
 $
are well-formed  and, respectively \iH- and \iK- univocal.

Their composition $\compose{\chora}{\chora[B]}H K$ yields the empty
automaton, which is well-formed (thanks to \cref{prop:wf-presbycomp}).
However, participants \ptp[A] and \ptp[D] lose their transitions.
This is due to the fact that \iH in $\chora$ and \iK in
$\chora[B]$ are not compatible, hence ${\chora[A]}{\times}{\chora[B]}$
is not reflective.

\begin{remark}
  Compatibility does not ensure reflectiveness when taking into
  account c-automata which are
  not products, as the following example shows.
  \[
    \begin{tikzpicture}[node distance=2cm,]
      \tikzstyle{every state}=[cnode]
      \tikzstyle{every edge}=[carrow]
\node[state] (zero)                        {$0$};
      \node[draw=none,fill=none] (start) [left = 0.3cm  of zero]{};
      \node[state] (one)   [right   of=zero]   {$1$};
      \node[state] (two)   [right   of=one]   {$2$};
      \node[state] (three)   [right   of=two]   {$3$};

\path  (start) edge node {} (zero) 
      (zero)    edge               node [above] {$\gint[][A][a][H]$} (one)
      (one)    edge                node [above] {$\gint[][A][b][X]$} (two)
      (two)    edge                node [above] {$\gint[][K][a][X]$} (three)
      ;
    \end{tikzpicture}
  \]
It is easy to check that the c-automata $\chora$ is well-formed and
  that $\proj{\chora}{H}$ and $\proj{\chora}{K}$ are compatible.
Nonetheless $\chora$ is not \iH-\iK-reflective; in particular
  requirement (\ref{item:forallout}) of \cref{def:reflective} does not
  hold.
\rmkend
\end{remark}

Our composition operator is associative and commutative up-to isomorphism.
We first show that our composition operation commutes with the product
of c-automata.
\begin{lemma}\label{lem:abchk}
  If $\chora$, $\chora[B]$ and $\chora[C]$ are c-automata with
  pairwise disjoint sets of participants then
  $(\compose{\chora}{\chora[B]}{H}{K}) \times \chora[C] =
  \sync{((\chora \times \chora[B]) \times \chora[C])}{H}{K}$.
\end{lemma}
\begin{proof}
  We show that each run in
  $(\sync{(\chora \times \chora[B])}{H}{K}) \times \chora[C]$ is also
  a run in $\sync{((\chora \times \chora[B]) \times \chora[C])}{H}{K}$
  and viceversa.
  \\
  For each blended transition
  $t = \cauttr[((p,q),r)][{\gint[]}][((p',q'),r')]$ on a run $\pi$ in
  $(\sync{(\chora \times \chora[B])}{H}{K}) \times \chora[C]$ there is
  a state $((p'',q''),r'')$ in
  $(\chora \times \chora[B]) \times \chora[C]$ such that
  \begin{equation}\label{eq:unblend}
    \cauttr[((p,q),r)][{\gint[]}][]
    \cauttr[((p'',q''),r'')][{\gint[]}][((p',q'),r')][][]
    \text{ in } (\chora \times \chora) \times \chora[C]
  \end{equation}
Note that any of transitions above may be in $\chora[C]$ by
  hypothesis.
By replacing each $t$ in $\pi$ with the corresponding transitions in
  \eqref{eq:unblend} we obtain a run in
  $(\chora \times \chora[B]) \times \chora[C]$ and therefore $\pi$ is
  also a run in
  $\sync{((\chora \times \chora[B]) \times \chora[C])}{H}{K}$.
The other direction is similar.
\qed
\end{proof}

The following result establishes that our operation of composition
commutative and associative, modulo the structure of the states of the
c-automaton resulting from composition.
\begin{corollary}\label{cor:monoid}
  If $\chora$, $\chora[B]$ and $\chora[C]$ are c-automata with
  pairwise disjoint sets of participants then
  \begin{enumerate}
  \item $\compose{\chora}{\chora[B]}H K$ is isomorphic to
    $\compose{\chora[B]}{\chora}K H$ (commutativity up-to isomorphism)
  \item $\compose{(\compose{\chora}{\chora[B]}H K)}{\chora[C]}I J$ is
    isomorphic to
    $\compose{\chora}{(\compose{\chora[B]}{\chora[C]}I J)} H K$
    (associativity up-to isomorphism).
  \end{enumerate}
\end{corollary}
\begin{proof}
  For commutativity, it is a trivial observation that there is a
  label-preserving bijection between the runs in
  $\compose{\chora[A]}{\chora[B]}H K$ and those in
  $\compose{\chora[B]}{\chora[A]}K H$.
Since the blending operation does not depend on the structure of
  the states of a c-automaton then we have thesis.
  \\
  For associativity we have 
  \[\begin{array}{lcl@{\qquad}r}
    \compose{(\compose{\chora}{\chora[B]}H K)}{\chora[C]}I J
    & = & \sync{((\compose{\chora}{\chora[B]} H K) \times \chora[C])}I J
    & \text{by def. \ref{def:comp}} \\
    & = & \sync{(\sync{((\chora \times \chora[B]) \times \chora[C])} H K)}I J
    & \text{by cor. \ref{lem:abchk}} \\
    & \equiv & \sync{(\sync{(\chora \times (\chora[B] \times \chora[C])} H K)}I J
    & \text{by assoc. of } \_ \times \_ \\
    & = & \sync{(\compose{\chora}{(\chora[B] \times \chora[C])} H K)}I J
    & \text{by def. \ref{def:comp}} \\
    & = & \compose{\chora}{(\compose{\chora[B]}{\chora[C]} I J)} H K
    & \text{by def. \ref{def:comp}} \\
    \end{array}
  \]
  where $\equiv$ is the up-to isomorphism equality on FSAs.
\qed
\end{proof}
By \cref{cor:monoid}, the operation of composition basically induces a
sort of commutative monoidal algebraic structure on the class of
c-automata.
The term-algebra of this structure are trees that represent compositions
of automata.
We can picture those terms as a binary tree, dubbed \emph{composition
  tree}, where leaves are c-automata and edges are labelled with
participants; a node $n$ with children $n_1$ and $n_2$ such that the
edge from $n$ to $n_1$ is labelled \iH and the edge from $n$ to $n_1$
is labelled \iK represents $\compose{n_1}{n_2} H K$.
\cref{cor:monoid} is relevant for the use of our
choreographic framework for a modular development of systems
as it guarantees that the order of composition is immaterial.

Composition trees composition cannot produce \quo{circular}
compositions.
However, combining composition and blending allows us to produce
arbitrary graph structures.
This can be attained through the following alternative representation.
A \emph{composition graph} is a labelled undirected graph where nodes
are c-automata and edges between $\chora$ and $\chora[B]$ labelled
with $(\ptp[H],\ptp[K])$ iff there is a blending between role
$\ptp[H]$ of $\chora$ and role $\ptp[K]$ of $\chora[B]$.
For instance, \cref{ex:cntd} shows how we can produce working example
to produce a \quo{triangle-shaped} circular composition.
\begin{example}[Working example cont'd.]\label{ex:cntd}
  A compact representation of the composition of $\chora[P]$ and
  $\chora[V]$ via \iH and \iK is in \cref{pvfig}.
  \begin{figure}[t!]
  \[
    \begin{tikzpicture}[node distance=1.8cm]
      \tikzstyle{every state}=[cnode]
      \tikzstyle{every edge}=[carrow]
\node[state] (zero)     {$0$};
      \node[state] (one)  [ right  of=zero]   {$1$};
      \node[draw=none,fill=none] (start) [below left = 0.3cm  of zero]{};
      \node[state] (two) [ right  of=one] {$2$};
      \node[state] (three) [ below  left of=two] {$3$};
      \node[state] (four) [above right of=two,yshift=-6mm,xshift=6mm] {$4$};
      \node[state] (five)  [above left of=four,yshift=-6mm,xshift=-6mm]    {$5$}; 
      \node[state] (six) [ below  left of=three] {$6$};
      \node[state] (thirteen) [left of = five] {$13$};
      \node[state] (fourteen) [left of = thirteen] {$14$};  
      \node[state] (nineteen) [right of = six,xshift=6mm] {$19$};  
      \node[state] (seven) [right  of=nineteen] {$7$};
      \node[state] (eight) [right   of=seven] {$8$};
      \node[state] (nine) [below  right of=eight, yshift=6mm,xshift=6mm] {$9$};
      \node[state] (ten) [below left  of=nine, yshift=6mm,xshift=-6mm] {$10$};
      \node[state] (eleven) [above  right of=eight, yshift=-6mm,xshift=6mm] {$11$};
      \node[state] (twelve) [above left of= eleven, yshift=-6mm,xshift=-6mm] {$12$};
      \node[state] (fifteen) [left of = twelve] {$15$};
      \node[state] (sixteen) [left of = fifteen] {$16$};  
      \node[state] (seventeen) [left of = ten] {$17$};
      \node[state] (eighteen) [left of = seventeen] {$18$};  
\path  (start) edge node {} (zero) 
      (zero) edge node [above] {$\gint[][C][tick][Q]$} (one)
      (one) edge node [above] {$\gint[][Q][text][A]$} (two)
      (two) edge node [above] {$\gint[][A][ack][Q]$} (three)
      edge node [above] {$\gint[][A][nack][Q]$} (four)
      (three)   edge                                node  [above] {$\gint[][A][tock][E]$} (six)
      (seven)   edge                                 node [above] {$\gint[][Q][text][B]$} (eight)
      (eight)   edge                                  node [below]  {$\gint[][B][ack][Q]$} (nine)
      edge                                  node [above]  {$\gint[][B][nack][Q]$} (eleven)
      (ten)    edge                                 node [above] {$\gint[][B][go][A]$} (seventeen)
      (five)    edge                                  node [above] {$\gint[][A][wait][B]$} (thirteen)
      (four)    edge     [bend right=30]     node [above] {$\gint[][A][tock][E]$} (five)
      (twelve)    edge                              node [above] {$\gint[][B][wait][A]$} (fifteen)
      (six)    edge                                  node [below] {$\gint[][A][go][B]$} (nineteen)
      (eleven)   edge    [bend right=30]      node  [above] {$\gint[][B][tock][E]$} (twelve)
      (nine)   edge    [bend left=30]      node  [below] {$\gint[][B][tock][E]$} (ten)
      (thirteen)    edge                          node [above] {$\gint[][Q][text][I]$} (fourteen)
      (fourteen)    edge   [bend right=30]     node [below] {$\gint[][I][text][Q]$} (zero)
      (fifteen)    edge                          node [above] {$\gint[][Q][text][I]$} (sixteen)
      (sixteen)    edge   [bend right=30]     node [below] {$\gint[][I][text][Q]$} (nineteen)
      (seventeen)    edge                          node [above] {$\gint[][Q][text][I]$} (eighteen)
      (eighteen)    edge   [bend left=60]     node [below] {$\gint[][I][text][Q]$} (zero)
      (nineteen)    edge                           node [above] {$\gint[][C][tick][Q]$} (seven)
      ;
    \end{tikzpicture}
    \]
    \caption{A compact representation of the composition of $\chora[P]$ and
  $\chora[V]$ via \iH and \iK.}\label{pvfig}
  \end{figure}
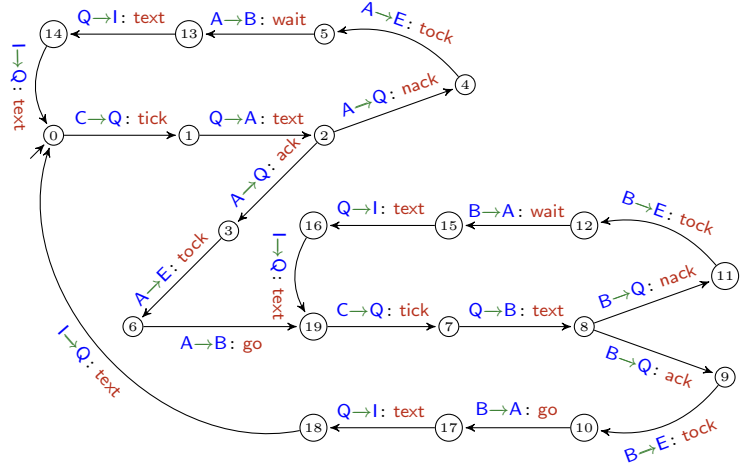
  (The full automaton $\compose{\chora[P]}{\chora[V]}{H}{K}$ is in
  \cref{app:we}.)
  
  The above c-automaton can be composed with $\chora[C]$ via $\ptp[C]$
  and $\ptp[D]$.
Then, in the resulting automaton, $\ptp[E]$ and $\ptp[F]$, originally
  belonging to different c-automata, can be blended.
A compact representation of
  $\sync{(\compose{(\compose{\chora[P]}{\chora[V]}{H}{K})}{\chora[C]}{C}{D})}E
  F$ is in \cref{pvcfig}.
  \begin{figure}[t!]
  \[
    \begin{tikzpicture}[node distance=2cm]
      \tikzstyle{every state}=[cnode]
      \tikzstyle{every edge}=[carrow]
\node[state] (zero)     {$0$};
      \node[state] (one)  [ right  of=zero]   {$1$};
      \node[draw=none,fill=none] (start) [left = 0.3cm  of zero]{};
      \node[state] (two) [ right  of=one] {$2$};
      \node[state] (three) [ below  left of=two,xshift=-4mm] {$3$};
      \node[state] (four) [above right of=two,xshift=35mm] {$4$};
\node[state] (six) [below left  of=three,xshift=-4mm] {$6$};
      \node[state] (twentytwo) [right of= six] {$22$};
      \node[state] (seven) [right  of=twentytwo] {$7$};
      \node[state] (eight) [right   of=seven] {$8$};
      \node[state] (nine) [below  right of=eight] {$9$};
      \node[state] (eleven) [ above right of=eight,xshift=4mm] {$11$};
      \node[state] (nineteen) [left of = four] {$19$};  
      \node[state] (five)  [ left  of=nineteen]    {$5$}; 
      \node[state] (thirteen) [left of = five] {$13$};
      \node[state] (fourteen) [left of = thirteen] {$14$};  
      \node[state] (twenty) [above right of = eleven] {$20$};      
      \node[state] (twentyone) [left of = nine] {$21$};      
      \node[state] (ten) [left   of=twentyone] {$10$};
      \node[state] (seventeen) [left of = ten] {$17$};
      \node[state] (twelve) [left  of= twenty] {$12$};
      \node[state] (eighteen) [left of = seventeen] {$18$};  
      \node[state] (fifteen) [left of = twelve,xshift=2mm] {$15$};   
      \node[state] (sixteen) [above right of = twentytwo,xshift=4mm] {$16$};  
\path  (start) edge node {} (zero) 
      (zero)    edge node [above] {$\gint[][R][tick][Q]$} (one)
      (one)    edge node [above] {$\gint[][Q][text][A]$} (two)
      (two)    edge node [above] {$\gint[][A][ack][Q]$} (three)
      edge node [above,xshift=-1cm] {$\gint[][A][nack][Q]$} (four)
      (three)   edge                                node  [above] {$\gint[][A][tock][R]$} (six)
      (seven)    edge                                 node [above] {$\gint[][Q][text][B]$} (eight)
      (eight)   edge                                  node [above]  {$\gint[][B][ack][Q]$} (nine)
      edge                                  node [above]  {$\gint[][B][nack][Q]$} (eleven)
      (ten)    edge                                 node [above] {$\gint[][B][go][A]$} (seventeen)
      (five)    edge                                  node [above] {$\gint[][A][wait][B]$} (thirteen)
      (nineteen)    edge                          node [above] {$\gint[][A][tock][R]$} (five)
      (twelve)    edge                              node [above] {$\gint[][B][wait][A]$} (fifteen)
      (six)    edge                                  node [above] {$\gint[][A][go][B]$} (twentytwo)
      (twenty)   edge                            node  [above] {$\gint[][B][tock][R]$} (twelve)
      (twentyone)   edge          node  [above] {$\gint[][B][tock][R]$} (ten)
      (thirteen)    edge                          node [above] {$\gint[][Q][text][I]$} (fourteen)
      (fourteen)    edge   [bend right=30]     node [above] {$\gint[][I][text][Q]$} (zero)
      (fifteen)    edge                          node [above] {$\gint[][Q][text][I]$} (sixteen)
      (sixteen)    edge                         node [above] {$\gint[][I][text][Q]$} (twentytwo)
      (seventeen)    edge                          node [above] {$\gint[][Q][text][I]$} (eighteen)
      (eighteen)    edge   [bend left=23]     node [below] {$\gint[][I][text][Q]$} (zero)
      (four)    edge node [above] {$\gint[][R][count][S]$} (nineteen)             
      (eleven)    edge node [above] {$\gint[][R][count][S]$} (twenty)                           
      (nine)    edge node [above] {$\gint[][R][count][S]$} (twentyone)                                         
      (twentytwo)   edge          node  [above] {$\gint[][R][tick][Q]$} (seven)             
      ;
    \end{tikzpicture}
    \]
    \caption{A compact representation of
  $\sync{(\compose{(\compose{\chora[P]}{\chora[V]}{H}{K})}{\chora[C]}{C}{D})}E
  F$.}\label{pvcfig}
  \end{figure}
In this way we connect systems $S_{\p[p]}$, $S_{\p[v]}$ and
  $S_{\p[c]}$ in a triangle shaped structure.
\finex
\end{example}

\ifhideproofs \else
In case we intended to compose a number of c-automata, it would be
helpful that the compatibility checks of the various interfaces we
would like connect could actually were all made at the very beginning.
This fact, relevant for the modular composition of system is
guaranteed by the following result.
\\
\textbf{Notation.} Below we denote by $\p\in\chora$ the fact that $p$
is a participant in the c-automaton $\chora$.
\begin{proposition}\label{prop:compafterblend}
  Let $\chora[A]$, $\chora[B]$ and $\chora[C]$ be c-automata with
  pairwise disjoint sets of participants and such that
  $\HH\in\chora[A]$, $\KK,\II\in\chora[B]$
  and $\JJ\in\chora[C]$.\\
  If $\proj{\chora[A]}{H}$ is compatible with $\proj{\chora[B]}{K}$,
  and $\proj{\chora[B]}{I}$ is compatible with $\proj{\chora[C]}{J}$,
  then $\proj{(\compose \chora {\chora[B]}{H}{K})}{I}$ is compatible
  with $\proj{\chora[C]}{J}$.
\end{proposition}
\begin{proof}[Sketch]
  By \cref{lemma:bisim}
  $\proj{(\compose \chora {\chora[B]}{H}{K}}{I})$ is bisimilar with
  $\proj{\chora[C]}{J}$.  The thesis follows by checking that
  compatibility is a subrelation of bisimilarity.
\qed
\end{proof}

\begin{remark}
  In general, compatibility is not a sufficient condition to guarantee
  that several choreographies can be composed in a circular stucture
  as we managed to do in \cref{ex:cntd}.
In fact, the possibility of obtaining \quo{sound} systems out of a
  circular composition is equivalent to the following property
\begin{quote}\em
    If $\chora[A]$ is a c-automaton with participants containing
    $\Set{\ptp[I],\ptp[J],\ptp[H],\ptp[K]}$, then
    compatibility of $\ptp[I]$ and $\ptp[J]$ is preserved by blending
    $\ptp[H]$ and $\ptp[K]$
  \end{quote}
This however, does not hold.
The intuitive reason is that a \quo{circular} sequence of
  compositions is \quo{sound} if they do not introduce deadlocks;
  the absence of these particular form of deadlocks cannot
  be guaranteed simply by compliance.
The following counterexample make this intuition more evident.
Let us consider the following c-automata with disjoint sets of
  participants.
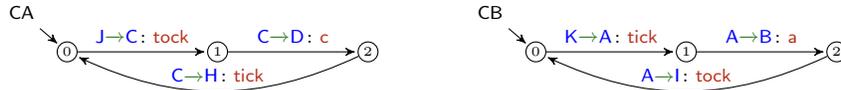
\[
    \begin{array}{c@{\hspace{12mm}}c}
      \begin{tikzpicture}[node distance=2cm,]
        \tikzstyle{every state}=[cnode]
        \tikzstyle{every edge}=[carrow]
        \node[state] (zero)                        {$0$};
        \node[state] (one)   [right   of=zero]   {$1$};
        \node[draw=none,fill=none] (start) [above  left = 0.3cm  of zero]{$\chora[A]$};
        \node[state] (two) [right  of=one] {$2$};
\path  (start) edge node {} (zero) 
        (zero)    edge                                     node [above] {$\gint[][J][tock][C]$} (one)
        (one)    edge                                      node [above] {$\gint[][C][c][D]$} (two)
        (two)   edge    [bend left=25]             node  [above]  {$\gint[][C][tick][H]$} (zero)
        ;
      \end{tikzpicture}
      &
      \begin{tikzpicture}[node distance=2cm,]
        \tikzstyle{every state}=[cnode]
        \tikzstyle{every edge}=[carrow]
        \node[state] (zero)                        {$0$};
        \node[state] (one)   [right   of=zero]   {$1$};
        \node[draw=none,fill=none] (start) [above  left = 0.3cm  of zero]{$\chora[B]$};
        \node[state] (two) [right  of=one] {$2$};
\path  (start) edge node {} (zero) 
        (zero)    edge                                     node [above] {$\gint[][K][tick][A]$} (one)
        (one)    edge                                      node [above] {$\gint[][A][a][B]$} (two)
        (two)   edge    [bend left=25]             node  [above]  {$\gint[][A][tock][I]$} (zero)
        ;
      \end{tikzpicture}
    \end{array}
  \]
We can immediately check that $\proj{\chora}{J}$ and
  $\proj{\chora[B]}{I}$ are compatible
  as well as $\proj{\chora}{H}$ and $\proj{\chora[B]}{K}$.
By \cref{thm:comprod}, compatibility of $\proj{\chora}{H}$ and
  $\proj{\chora[B]}{K}$
guarantees that the composition of ${\chora}$ and ${\chora[B]}$
  through \iH and \iK preserves the behaviour outside the
  interfaces.
Namely, the construction of the blended automaton $\sync{(\chora{\times}\chora[B])}{H}{K}$
  does not affect the behaviour of the participants other than
  \iH and \iK.
In fact,
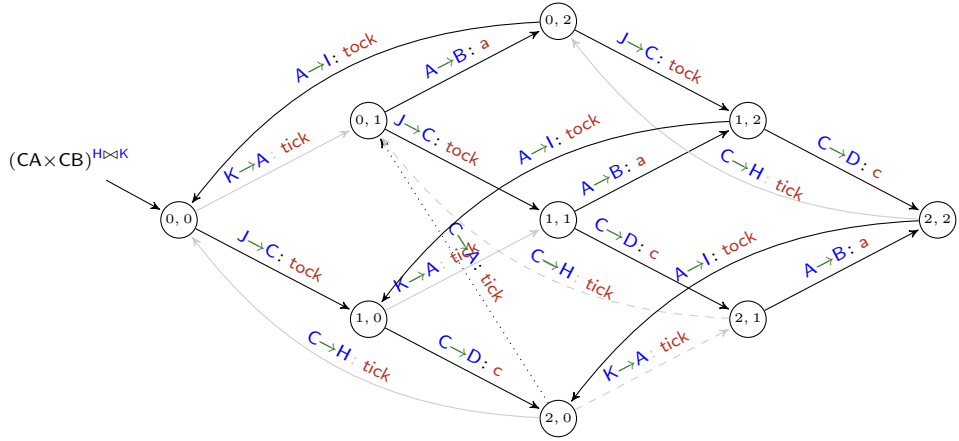
\[
    \begin{array}{c}
      \begin{tikzpicture}[node distance=3cm]
        \tikzstyle{every state}=[cnode]
        \tikzstyle{every edge}=[carrow]
\node[state] (00)                                         {$0,0$};
        \node[draw=none,fill=none] (start) [above left = 5mm of  00]{$\sync{(\chora{\times}\chora[B])}{H}{K}$};
        \node[state] (10)  [below right of=00 , xshift=4mm, yshift=8mm]  {$1,0$};
        \node[state] (01) [above  right of=00, xshift=4mm, yshift=-8mm] {$0,1$};
        \node[state] (20) [below right of=10, xshift=4mm, yshift=8mm] {$2,0$};
        \node[state] (11) [above right  of=10, xshift=4mm, yshift=-8mm] {$1,1$};
        \node[state] (02) [above right of=01, xshift=4mm, yshift=-8mm] {$0,2$};
        \node[state] (12) [above right of  =11, xshift=4mm, yshift=-8mm] {$1,2$};
        \node[state] (21) [above  right of=20, xshift=4mm, yshift=-8mm] {$2,1$};
        \node[state] (22) [above right of=21, xshift=4mm, yshift=-8mm] {$2,2$};
\path  (start) edge node {} (00) 
        (00) edge                                  node [above] {$\gint[][J][tock][C]$} (10)
        edge[black!20] node[above] {$\gint[][K][tick][A]$  } (01)
        (20) edge[black!20, dashed] node[above] { $\gint[][K][tick][A]$ } (21)
        edge     [black!20, bend left=25]     node [above] {$\gint[][C][tick][H]$} (00)
        edge [dotted]    node [above] {$\gint[][C][tick][A]$} (01)
        (02)    edge[bend right=25]        node [above] {$\gint[][A][tock][I]$} (00)
        edge                                     node [above] {$\gint[][J][tock][C]$  } (12)
        (01)    edge node [above] {$\gint[][A][a][B]$} (02)
        edge                                    node [above,xshift=-.5cm] {$\gint[][J][tock][C]$ } (11)
        (10)     edge                                  node [above] {$\gint[][C][c][D]$} (20)
        edge[black!20] node [above,xshift=-.3cm] {$\gint[][K][tick][A]$} (11)
        (21)    edge                                  node [above] {$\gint[][A][a][B]$ } (22)
        edge     [black!20, dashed, bend left=25]        node [above,xshift=.5cm] {$\gint[][C][tick][H]$ } (01)
        (11)    edge                                  node [above,xshift=-.5cm] {$\gint[][C][c][D]$ } (21)
        edge                                    node [above,xshift=-.5cm] {$\gint[][A][a][B]$ } (12)
        (22)    edge    [bend right=25]        node [above] {$\gint[][A][tock][I]$ } (20)
        edge[black!20, bend left=25] node[above,xshift=.5cm] {$\gint[][C][tick][H]$ } (02)
        (12)     edge                                  node [above] { $\gint[][C][c][D]$} (22)
        edge     [bend right=25]       node [above,xshift=.5cm] { $\gint[][A][tock][I]$} (10)
        ;
      \end{tikzpicture}
    \end{array}
  \]
where the product $\chora \times \chora[B]$ consists of all the
  transitions above but the dotted ones and the result of the
  composition is made of the solid and dotted transitions.
  
  Now we notice that the following c-automata
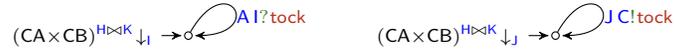
\[
    \begin{tikzpicture}
      \tikzstyle{every state}=[cnode]
      \tikzstyle{every edge}=[carrow]
\node[state] (zero)                        { };
      \node[draw=none,fill=none] (start) [left = 0.3cm  of zero] {$\proj{\sync{(\chora{\times}\chora[B])}{H}{K}}{I}$};
\path  (start) edge node {} (zero) 
      (zero)    edge   [loop right,min distance=10mm,in=0,out=60,looseness=10]                node[above,rotate=60,yshift=-.3cm,xshift=.5cm] {${\ain[A][I][][tock]}$} (zero)          
      ;
    \end{tikzpicture}
    \qquad
    \begin{tikzpicture}
      \tikzstyle{every state}=[cnode]
      \tikzstyle{every edge}=[carrow]
\node[state] (zero)                        {};
      \node[draw=none,fill=none] (start) [left = 0.3cm  of zero]{$\proj{\sync{(\chora{\times}\chora[B])}{H}{K}}{J}$};
\path  (start) edge node {} (zero) 
      (zero)    edge      [loop right,min distance=10mm,in=0,out=60,looseness=10]           node [above,rotate=60,yshift=-.3cm,xshift=.5cm] {${\aout[J][C][][tock]}$} (zero)
      ;
    \end{tikzpicture}
  \]
  are trivially compatible (as they were before the
  composition).
Nonetheless $\sync{(\chora{\times}\chora[B])}{H}{K}$
  is not reflective on $\II$ and $\JJ$.  We have in fact that 
  $\sync{(\sync{(\chora{\times}\chora[B])}{H}{K})}{I}{J}$ is the
  empty c-automaton.
    
  So, in order to safely compose choreography in a circular way,
  reflectiveness must be checked on the participants whose blending
  \quo{completes the circle}, namely $\II$ and $\JJ$.
  
  It is worth remarking that providing a condition for sound circular
  connections is a complex problem.  This has been already recognised
  in the literature.  Tree-like composition is the most adopted safe
  form of composition. For instance, \cite{LM13} discusses the problem
  (on page 2, paragraph 3) and provides a generalisation of acyclic
  architectures by considering so-called \quo{disjoint circular wait
    free component systems}.  In the context of session types, a
  similar problem arises when processes work on more than one session
  at the same time as in \cite{cdyp16}, where an \quo{interaction type
    system} is used to control the depencendies among roles.
\end{remark}

\section{Related and Future Work}\label{sec:conc}
We introduced a compositional model of choreographies and demonstrated
its suitability to cope with modular descriptions of global
specifications of message-passing applications.
The operation that we devise is basically the composition of classical
automata product and of the blending operation introduced here.
Our notion of composition preserves well-formedness under
mild conditions that, in practice, do not hinder its
applicability.

The adoption of an automata-based model brings in two main benefits.
Firstly, the constructions that we provided are based on basic
notions and mainly \emph{syntax-independent}.
In fact, it is often the case that syntax-driven models (such as
behavioural type systems) introduce complex constructions and
constraints to define well-formedness that
restrict~\cite{hlvlcmmprttz16} their applicability.
Secondly, we can re-use well-known results of the theory of automata
(e.g., we used notions of bisimulation, product and minimization) and
related tools.

\subsection{Related work}\label{sec:related}
The present paper includes two main contributions: ($i$) the
definition of a choreography model based on FSAs and ($ii$) the
definition of a notion of composition of choreographies.
We comment related work according to these contributions.

\subsubsection{Composing choreographies}
While choreographies have been deeply explored in the literature, the
problem of their composition has not. Indeed, only a few works
consider the issue of composition.

The closest to our approach are~\cite{francoICEprev} and
\cite{francoICE}, which inspired the present work. The former
introduces the idea of connecting choreographies via forwarders, but
did not provide a way to compute the corresponding choreography, that
is composition was performed only at the level of systems. The latter
provides a composition at the choreography level, but has stronger
well-formedness conditions than ours (e.g., in choices all
interactions should have the same pair of participants), and did not
allow for self-connections. As a result, only tree structures could be
created.

Modularity of choreographies has been considered
in~\cite{MontesiY13}.
There, one can specify partial choreographies where some roles are not
supposed to be fully specified.
Partial choreographies give some support to modular development,
although with some limitations.
For instance, complete choreographies cannot be composed, while in our
case any choreography can be composed by selecting one of its roles as
an interface.
In fact, a type system can check if partial choreographies are
consistent with respect to global types (using a \quo{dually
  incomplete} typing principle), but no composition of partial
choreographies is actually defined in~\cite{MontesiY13}.

Approaches to choreography composition in the setting of adaptive
systems has been discussed in~\cite{PredaGGLM16} and in~\cite{cdv15}.
In~\cite{PredaGGLM16}, choreographies form a programming language and
are executable, and composition is obtained by replacing a
choreography component $\chora[C_1]$ inside a choreography context
$\chora$ with a new choreography component $\chora[C_2]$ from outside
the system.
Notably, the only allowed interactions between the component and the
context are auxiliary interactions introduced by the projection and
needed to ensure well-formedness conditions at the border between the
component and the context.
In~\cite{cdv15} instead, adaptiveness is handled at the type level
rather than in a choreographic programming language. There the main
limitation is that adaptiveness only allows one to switch inside a
number of pre-defined behaviours (in a kind of context-driven choice),
and new behaviours cannot be introduced.

\subsubsection{Expressiveness of c-automata}
First, we note that the use of automata-based models for specifying
the local behaviour of distributed compontents is commonplace in the
literature~\cite{bz83,dah01}, less so for the global specifications
of choreographies.
To the best of our knowledge, the automata model used in the line of
work in~\cite{bbo12} (and references therein) is the only one
in the literature.
The main difference between models like in~\cite{bbo12} and c-automata
(besides some constraints on the usage of messages in interactions
that c-automata do not have), is that in this context the
realisability of choreographies has been studied (using e.g., model
checking) while compositionality is not considered in this line of
research.
Moreover, as shown in~\cite{fl17} some of the main decidability
results were flawed.

While the main aim of c-automata is to provide a choreography model
based on FSAs, we remark here that it is rather expressive and
complements existing models of choreographies or multiparty session
types.

In particular, the expressive power of c-automata is not comparable
with the one of the multiparty session types in~\cite{ScalasY19},
which subsumes most systems in the literature. More precisely, the
c-automaton
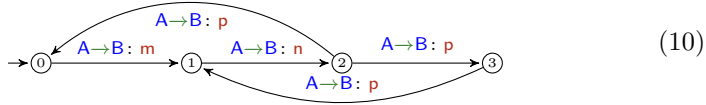
\begin{equation}\label{eq:sy19}
  \begin{tikzpicture}[node distance=2cm]
    \tikzstyle{every state}=[cnode]
    \tikzstyle{every edge}=[carrow]
\node[draw=none,fill=none] (start) {};
    \node[state, right = .3cm of start] (0) {$0$};
    \node[state, right of = 0]     (1) {$1$};
    \node[state, right of = 1]     (2) {$2$};
    \node[state, right of = 2]     (3) {$3$};
    \path  (start) edge node {} (0)
           (0)     edge node [above] {$\gint[]$} (1)
           (1)     edge node [above] {$\gint[][a][n][b]$} (2)
           (2)     edge node [above] {$\gint[][a][p][b]$} (3)
           (2)     edge[bend right=40] node [below] {$\gint[][a][p][b]$} (0)
           (3)     edge[bend left=25] node [above] {$\gint[][a][p][b]$} (1)
    ;
 \end{tikzpicture}
\end{equation}
cannot be syntactically written in~\cite{ScalasY19} due to the two
entangled loops.
The example \eqref{eq:sy19} cannot be expressed in global
graphs~\cite{gt18} either, again due to the intersecting loops.
We note that the infinite unfolding of the c-automaton \eqref{eq:sy19}
is regular and therefore it fits in the session type system considered
in~\cite{sptd17}.
However, this type system has not been conceived for choreographies (it
is binary session type system) and does not allow non-determinism.

On the other side, examples such as~\cite[Ex. 2, Fig. 4]{ScalasY19}
cannot be written in our model (since we expect the same roles to
occur in branches which are coinitial, branches inside loops require
that all participants in a loop are notified when the loop ends).
We conjecture that a refinement of well-branchedness is possible to
address this limitation.
An advantage of global graphs is that they feature parallel
composition of global specifications, which c-automata lack.
We note however that the product of c-automata allows one to model
parallel composition in the case where the two branches have disjoint
sets of participants (as typically assumed in multyparty session
types with parallel composition).
Mapping global graphs without parallel composition into c-automata is
trivial.
The same considerations apply to choreography languages where possible
behaviours are defined by a suitable process algebra with parallel
composition such as~\cite{lgmz08,BravettiZ07}.

\subsubsection{Future work}\label{sec:fw}
We leave as future work the extension of our theory to
\emph{asynchronous} communications. While the general approach should
apply,\linebreak well-formedness and the conditions ensuring a safe composition
should be refined.
\ifhideproofs
\else
Indeed, our blending operation does not work for asynchronous
semantics as it is, as the following example shows.
Consider
\[
  \begin{array}{c@{\hspace{22mm}}c}
    \quad
    \begin{tikzpicture}[node distance=1.6cm,]
     \tikzstyle{every state}=[cnode]
     \tikzstyle{every edge}=[carrow]
\node[state]           (zero)                        {$0$};
     \node[state]           (one)  [right  of=zero]  {$1$};
     \node[draw=none,fill=none] (start) [left = 0.3cm of zero]{$\chora$};
     \node[state]           (two)  [right  of=one]  {$2$};
\path  (start) edge node {} (zero) 
     (zero)    edge                                  node [above] {$\gint[][A][a][I]$} (one)
     (one)    edge                                  node [above] {$\gint[][K][b][A]$} (two)
     ;
   \end{tikzpicture}
   \qand
   \begin{tikzpicture}[node distance=1.6cm,]
     \tikzstyle{every state}=[cnode]
     \tikzstyle{every edge}=[carrow]
\node[state]           (zero)                        {$0$};
     \node[state]           (one)  [right  of=zero]  {$1$};
     \node[draw=none,fill=none] (start) [left = 0.3cm  of zero]{$\chora[B]$};
     \node[state]           (two)  [right  of=one]  {$2$};
\path  (start) edge node {} (zero) 
     (zero)    edge                                  node [above] {$\gint[][B][b][H]$} (one)
     (one)    edge                                  node [above] {$\gint[][J][a][B]$} (two)
     ;
      \end{tikzpicture}
  \end{array}
\]
whose produce is the automaton in \cref{ex:reflective}.
The interfaces $\HH$ and $\KK$ are compatible, so let us compute
$\sync{{(\chora{\times}\chora[B])}}{H}{K}$
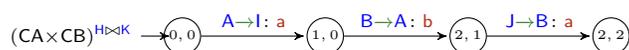
\[
  \begin{tikzpicture}[node distance=1.9cm,]
    \tikzstyle{every state}=[cnode]
    \tikzstyle{every edge}=[carrow]
\node[state]           (00)                        {$0,0$};
    \node[state]           (10)  [right of=00]  {$1,0$};
    \node[draw=none,fill=none] (start) [left = 0.3cm of  00]{$\sync{{(\chora{\times}\chora[B])}}{H}{K}$};
    \node[state]            (21) [right of =10,] {$2,1$};
    \node[state]            (22) [right of=21] {$2,2$};
\path  (start) edge node {} (zero) 
    (00)    edge                                  node [above] {$\gint[][A][a][I]$} (10)
    (10)    
    edge                                   node [above] {$\gint[][B][b][A]$} (21)
    (21)    edge                                   node [above] {$\gint[][J][a][B]$} (22)
    ;
  \end{tikzpicture}
\]
If we now compute $\sync{{(\sync{(\chora{\times}\chora[B])}{H}{K}})}{I}{J}$, we obtain
\[
  \begin{tikzpicture}[node distance=1.9cm,]
     \tikzstyle{every state}=[cnode]
     \tikzstyle{every edge}=[carrow]
\node[state]           (00)                        {$0,0$};
     \node[draw=none,fill=none] (start) [left = 0.3cm of 00]{$\sync{{(\sync{(\chora{\times}\chora[B])}{H}{K}})}{I}{J}$};
\path  (start) edge node {} (00)  
     ;
   \end{tikzpicture}
 \]
which is precisely the expected choreography under the
 synchronous semantics.
If, instead, the asynchronous semantics was considered, we
 should have obtained
 
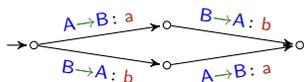
\[
   \begin{tikzpicture}[node distance=2.5cm]
     \tikzstyle{every state}=[cnode]
     \tikzstyle{every edge}=[carrow]
\node[draw=none,fill=none] (start){};
     \node[state,right = .3cm of start] (0) {};
     \node[state] (1) [above right of=0,yshift=-1.5cm] {};
     \node[state] (2) [below right of=0,yshift= 1.5cm] {};
     \node[state] (3) [above right of=2,yshift=-1.5cm] {};
\path
     (start) edge (0) 
     (0) edge node [above] {$\gint[][A][a][B]$} (1)
     edge node [below] {$\gint[][B][b][A]$} (2)
     (1) edge node [above] {$\gint[][B][b][A]$} (3)
     (2) edge node [below] {$\gint[][A][a][B]$} (3)
     ;
   \end{tikzpicture}
   \]
\fi   

An interesting future development is to adopt B\"uchi automata as
c-automata.
This extension is technically straightforward (we just add a set of
final states in the definition of c-automata and define
$\omega$-languages accordingly), but it probably impacts greatly
on our underlying theory.
For instance, it would be interesting, yet not trivial, to devise
well-formedness conditions on this generalised class of c-automata
that guarantee a precise correspondence with the $\omega$-languages of
the projections.
In this new setting, one can then explore milder forms of
liveness where, for instance, it is not the case that all participants
have to terminate (as the \emph{termination awareness} condition of
\cite{gt18}).

The interplay between FSAs and formal languages could lead to a theory
of projection and composition of choreographies based on languages
instead of automata.
For instance, one could try to characterise the languages accepted by
well-formed c-automata, similarly to what done in~\cite{aey03,loh02,gt18}.
In those approaches global specifications are rendered as partial
orders and the distributed realisability is characterised in terms of
some closure properties of languages instead of using well-formedness
conditions.

\bibliographystyle{plain}

\newpage
\appendix

\section{On blending}\label{app:aux}
\subsection{Composition in the working example}\label{app:we}
The composition of $\chora[P]$ and $\chora[V]$  on \iH and \iK of \cref{ex:we1} is in \cref{fig:we2}.
\begin{figure}[!ht]\centering
\begin{tikzpicture}[node distance=1.9cm,scale=.9,transform shape]
    \tikzstyle{every state}=[cnode]
    \tikzstyle{every edge}=[carrow]
\node[state] (01)                                                     {$0,1$};
    \node[state] (11)  [below of =01  ]     {$1,1$};
    \node[draw=none,fill=none] (start) [above = 1.5cm  of 01]{$\sync{\chora[PxV]}{H}{K}$};
    \node[state] (22)  [below of =11  ]     {$2,2$};
    \node[state] (15)  [right of =11]     {$1,5$};
    \node[state] (14)  [right of =15 ]     {$1,4$};
    \node[state] (05)  [right of =01  ]     {$0,5$};
    \node[state] (04)  [right of =05  ]     {$0,4$};   
    \node[state] (51)  [above right of =01]     {$5,1$};
    \node[state] (41)  [above right of =51 ]     {$4,1$};
    \node[state] (55)  [ right of =51]     {$5,5$};
    \node[state] (54)  [ right of =55 ]     {$5,4$};
    \node[state] (45)  [ right of =41]     {$4,5$};
    \node[state] (44)  [ right of =45 ]     {$4,4$};
    \node[state] (110)  [ left of =11 ]     {$1,10$};
    \node[state] (19)  [ left of =110]     {$1,9$};
    \node[state] (010)  [ left of =01 ]     {$0,10$};
    \node[state] (09)  [ left of =010]     {$0,9$};
    \node[state] (31)  [above left of =01]     {$3,1$};
    \node[state] (310)  [ left of =31 ]     {$3,10$};
    \node[state] (39)  [ left of =310]     {$3,9$};
    \node[state] (37)  [below left of =22,yshift=-8mm  ]     {$3,7$};
    \node[state] (36)  [left of =37  ]     {$3,6$};
    \node[state] (33)  [ left of =36  ]     {$3,3$};
    \node[state] (03)  [below right of =33  ]     {$0,3$};
    \node[state] (13)  [below  of =03  ]     {$1,3$};
    \node[state] (06)  [ right of =03  ]     {$0,6$};
    \node[state] (16)  [ right of =13  ]     {$1.6$};
    \node[state] (07)  [ right of =06  ]     {$0,7$};
    \node[state] (17)  [ right of =16  ]     {$1,7$};
    \node[state] (28)  [below  left of =17 , yshift=-20mm, xshift=-20mm  ]     {$2,8$};
    \node[state] (16)  [ right of =13  ]     {$1.6$};
    \node[state] (07)  [ right of =06  ]     {$0,7$};
    \node[state] (17)  [ right of =16  ]     {$1,7$};
    \node[state] (112)  [ right of =17 ]     {$1,12$};
    \node[state] (111)  [ right of =112  ]     {$1,11$};
    \node[state] (012)  [ right of =07]     {$0,12$};
    \node[state] (011)  [ right of =012  ]     {$0,11$};
    \node[state] (57)  [ above right of =07  ]     {$5,7$};
\node[state] (512)  [  right of =57  ]     {$5,12$};     
    \node[state] (511)  [  right of =512  ]     {$5,11$};     
    \node[state] (47)  [ above right of =57  ]     {$4,7$};
\node[state] (412)  [  right of =47  ]     {$4,12$};     
    \node[state] (411)  [  right of =412  ]     {$4,11$};     
\path  (start) edge node {} (01) 
    (01)    edge node [above] {$\gint[][C][tick][Q]$} (11)
    (05)    edge node [above] {$\gint[][C][tick][Q]$} (15)
    (04)    edge node [above] {$\gint[][C][tick][Q]$} (14)
    (51)    edge node [above] {$\gint[][I][text][Q]$} (01)
    (55)    edge node [above] {$\gint[][I][text][Q]$} (05)
    (54)    edge node [above] {$\gint[][I][text][Q]$} (04)
    (41)    edge node [above] {$\gint[][Q][text][I]$} (51)
    (45)    edge node [above] {$\gint[][Q][text][I]$} (55)
    (44)    edge node [above] {$\gint[][Q][text][I]$} (54)
    (45)    edge node [above] {$\gint[][A][wait][B]$} (41)
    (55)    edge node [above] {$\gint[][A][wait][B]$} (51)
    (05)    edge node [above] {$\gint[][A][wait][B]$} (01)
    (15)    edge node [above] {$\gint[][A][wait][B]$} (11)             
    (44)    edge node [above] {$\gint[][A][tock][E]$} (45)
    (54)    edge node [above] {$\gint[][A][tock][E]$} (55)
    (04)    edge node [above] {$\gint[][A][tock][E]$} (05)
    (14)    edge node [above] {$\gint[][A][tock][E]$} (15)
(39)    edge node [above] {$\gint[][B][tock][E]$} (310)
    (09)    edge node [above] {$\gint[][B][tock][E]$} (010)           
    (19)    edge node [above] {$\gint[][B][tock][E]$} (110)  
    (310)  edge node [above] {$\gint[][B][go][A]$} (31)      
    (010)  edge node [above] {$\gint[][B][go][A]$} (01)                
    (110)  edge node [above] {$\gint[][B][go][A]$} (11)       
    (39)    edge node [above] {$\gint[][Q][alt][I]$} (09)
    (310)    edge node [above] {$\gint[][Q][alt][I]$} (010)           
    (31)    edge node [above] {$\gint[][Q][alt][I]$} (01)           
    (09)    edge node [above] {$\gint[][C][tick][Q]$} (19) 
    (010)    edge node [above] {$\gint[][C][tick][Q]$} (110)                                   
(11)    edge node [above] {$\gint[][Q][text][A]$} (22)                                   
    (22)    edge     [bend right=50]      node [above] {$\gint[][A][nack][Q]$} (44)                                   
    edge      [bend right=30]       node [above] {$\gint[][A][ack][Q]$} (33)  
(28)    edge     [bend left=25]      node [above] {$\gint[][B][ack][Q]$} (39)  
    edge     [bend right=50]      node [above] {$\gint[][B][nack][Q]$} (411)      
(33)    edge node [above] {$\gint[][A][tock][E]$} (36)
    (03)    edge node [above] {$\gint[][A][tock][E]$} (06)           
    (13)    edge node [above] {$\gint[][A][tock][E]$} (16)  
    (36)  edge node [above] {$\gint[][A][go][B]$} (37)      
    (06)  edge node [above] {$\gint[][A][go][B]$} (07)                
    (16)  edge node [above] {$\gint[][A][go][B]$} (17)       
    (33)    edge node [above] {$\gint[][Q][alt][I]$} (03)
    (36)    edge node [above] {$\gint[][Q][alt][I]$} (06)           
    (37)    edge node [above] {$\gint[][Q][alt][I]$} (07)           
    (03)    edge node [above] {$\gint[][C][tick][Q]$} (13) 
    (06)    edge node [above] {$\gint[][C][tick][Q]$} (16)                                   
    (07)    edge node [above] {$\gint[][C][tick][Q]$} (17)                                                                
(511)    edge node [above] {$\gint[][B][tock][E]$} (512)
    (011)    edge node [above] {$\gint[][B][tock][E]$} (012)
    (111)    edge node [above] {$\gint[][B][tock][E]$} (112)  
    (512)    edge node [above] {$\gint[][B][wait][A]$} (57)
    (012)    edge                                  node [above] {$\gint[][B][wait][A]$} (07)  
(112)    edge node [above] {$\gint[][B][wait][A]$} (17)                                                 
(412)    edge node [above] {$\gint[][B][wait][A]$} (47)                           
    (411)    edge node [above] {$\gint[][B][tock][E]$} (412) 
    (47)    edge node [above] {$\gint[][Q][text][I]$} (57) 
(412)    edge node [above] {$\gint[][Q][text][I]$} (512)                                                    
    (411)    edge node [above] {$\gint[][Q][text][I]$} (511)                                                    
    (57)    edge node [above] {$\gint[][I][text][Q]$} (07)                                                   
(512)    edge node [above] {$\gint[][I][text][Q]$} (012)                                                                  
    (511)    edge node [above] {$\gint[][I][text][Q]$} (011)                                                                                 
    (012)    edge node [above] {$\gint[][C][tick][Q]$} (112)                                                                              
    (011)    edge node [above] {$\gint[][C][tick][Q]$} (111)                                                                               
(17)    edge node [above] {$\gint[][Q][text][B]$} (28)                                                                                             
    ;
  \end{tikzpicture}
\caption{The c-automaton $\compose{\chora[P]}{\chora[V]}H K$}\label{fig:we2}
\end{figure}

\newpage

\subsection{Blending algorithm}\label{app:blending}
The blending algorithm is given by the following pseudo-code.
\\\begin{minipage}[c]{1.0\linewidth}
  \lstset{
    numbers=left,
    numberstyle=\sffamily\linenumfontsize,
    backgroundcolor=\color{black!3},
    basicstyle=\sffamily\scriptsize,
    procnamestyle=\scfamily\scriptsize,
    mathescape=true,
    morekeywords={each},
    commentstyle=\color{blue!80!red!40},
    escapeinside={(*@}{@*)},
    keywordstyle=\textcolor{ForestGreen}
  }
  \renewcommand{\thelstlisting}{\arabic{lstlisting}}
\begin{lstlisting}[language=Python, caption=\label{alg:blending}The blending algorithm]
$\text{\textcolor{ForestGreen}{Input}}$: $\chora$ c-automaton, roles $\iH \text{ and }\iK$
$\text{\textcolor{ForestGreen}{Output}}$: $\sync \chora H K$ (cf. $\mbox{\cref{def:sync}}$)

def forwarding($\chora[X]$,$\p[U]$,$\p[V]$):
  $\chora[aux] = \chora[X]$                                 # $\chora[aux]$ is an auxiliary c-automaton
  for each $t = \cauttr[p][{\gint[][A][m][U]}][q][][]$ in $\chora[X]\label{ln:stepI}$:
    $\chora[aux] = \chora[aux] \setminus \{t\}$                               # remove transitions involving interfaces
    for each $\cauttr[q][{\gint[][V][m][B]}][r][][]$ in $\chora[X]$:
      if $\p \neq \q$:                         # if interfaces are forwarders,
        $\chora[aux] = \chora[aux] \cup \{\cauttr[p][{\gint[][A][m][B]}][r][][]\}$                           # blend the transitions in a unique one
      else: throw("Undefined")     # otherwise blending is not defined
  return $\chora[aux]$

$\chora[tmp]$ = forwarding(forwarding($\chora$, $\p[h]$, $\p[k]$), $\p[k]$, $\p[h]$)$\label{ln:stepI-end}$
$\chora[tmp] = \chora[tmp] \setminus \{\cauttr[p][{\gint[][X][m][Y]}][r][][]
\text{ in } \chora[tmp] \sst \{X,Y\} \cap \{\iH,\iK\} \neq \emptyset\}$$\label{ln:stepII}$
return $\chora[tmp] \setminus \{\text{state and transitions in } \chora[tmp] \text{ not reachable from the initial state}\}$$\label{ln:stepIII}$
\end{lstlisting}
\end{minipage}
Lines-\ref{ln:stepI}-\ref{ln:stepI-end} correspond to Step \ref{stepI}
of the informal presentation of the blending algorithm in
\cref{subsec:blending}, line-\ref{ln:stepII} corresponds to step
\ref{stepII}, and line-\ref{ln:stepIII} corresponds to step
\ref{stepIII}.

\begin{proposition}[Complexity of blending]\label{prop:complexity}
  The complexity of  \cref{alg:blending} is $O(n^2)$ where $n$ is the number
  of transitions of input c-automaton $\chora$.
\end{proposition}
\begin{proof}
  We assume that the c-automaton, seen as a graph, is stored as an
  array of nodes (the states), each one with an unordered list of
  outgoing edges (the transitions).  Step I has complexity
  $O(n^2)$. After the step, the number of nodes is unchanged and the
  number of edges is at most $O(n^2)$.  Step II has linear cost on the
  new size.  Step III requires to visit nodes and edges at most once,
  but since in the worst case the number of edges is larger, it also
  has linear complexity. The thesis follows.
\qed
\end{proof}

\end{document}